\documentclass[usenatbib,usegraphicx]{mn2e}
\usepackage{psfig,epsf}
\newcommand{\kms}{$\,\mbox{km}\,\mbox{s}^{-1}$}

\newcommand{\HI}{H\,{\sc i}\ }

\newcommand{\alfa}{{Kravtsov}}
\newcommand{\mnras}{MNRAS}
\newcommand{\apj}{ApJ}
\newcommand{\aj}{AJ}
\newcommand{\aap}{A\&A}
\newcommand{\aaps}{A\&A Sup.{}}
\newcommand{\apjl}{ApJL}

\newcommand{\nat}{Nature}

\begin{document}
\title[LSB halo profiles]{Simulating Observations of Dark Matter Dominated Galaxies: Towards the Optimal Halo Profile}
\author[de Blok, Bosma \& McGaugh]{W.J.G.~de~Blok$^{1,2}$, Albert Bosma$^3$, Stacy McGaugh$^4$\\
$^1$ Australia Telescope National Facility,
PO Box 76, Epping NSW 1710, Australia\\
$^2$ Department of Physics and Astronomy, Cardiff University, 5 The Parade, Cardiff CF24 3YB, United Kingdom\\
$^3$ Observatoire de Marseille, 2 Place Le Verrier, 13248 Marseille Cedex 4, France\\
$^4$ Department of Astronomy, University of Maryland, College Park, MD 20742-2421, USA}

\maketitle

\begin{abstract}
Low Surface Brightness (LSB) galaxies are dominated by dark matter,
and their rotation curves thus reflect their dark matter
distribution. Recent high-resolution rotation curves suggest that
their dark matter mass-density distributions are dominated by a
constant-density core. This seems inconsistent with the predictions of
Cold Dark Matter (CDM) models which produce halos with compact density
cusps and steep mass-density profiles.  However, the observationally
determined mass profiles may be affected by non-circular motions,
asymmetries and offsets between optical and dynamical centres, all of
which tend to lower the observed slopes.  Here we determine the impact
of each of these effects on a variety of halo models, and compare the
results with observed mass-density profiles.  Our simulations suggest
that no single systematic effect can reconcile the data with the cuspy
CDM halos. The data are best described by a model with a soft core
with an inner power-law mass-density slope $\alpha = -0.2 \pm
0.2$. However, no single universal halo profile provides a completely
adequate description of the data.
\end{abstract}


\section{Introduction}

Low Surface Brightness (LSB) galaxies are dominated by dark matter.
Their stellar populations are dynamically unimportant and the rotation
curves are thus direct tracers of their dark matter distributions
\citep{edb_rot}. This is important, as in brighter galaxies it is
difficult to disentangle the dynamical contributions of the dark and
visible matter, and unambiguously determine the distribution of the
dark matter. LSB galaxies give us a unique opportunity to compare
observations with cosmological dark matter simulations.

These numerical models, based on the Cold Dark Matter (CDM) paradigm,
make very specific predictions about the distribution of dark matter
in galaxies \citep{dubinskicdm}. In CDM halos the mass density
distribution in the inner parts is characterized by a steep density
``cusp'', usually described by a power-law $\rho(r)\sim r^{\alpha}$ with
slopes varying from $\alpha = -1$ \citep{NFW96,NFW97} to $\alpha =
-1.5$ \citep[e.g.][]{moore98,moore99,bul99}. This cusp manifests
itself in a steeply rising rotation curve with a specific shape.
Observations of dwarf and LSB galaxies \citep{moore94, optcur_data,
optcur_pap2, blais01, lsbopt_bosma, marchesini02} show that real
rotation curves rise less steeply than predicted, and do not have the
required CDM shape. Similar conclusions have been reached by
\citet{salucci01} and \citet{salbor} for high surface brightness disk
galaxies.  At small radii, the empirically determined mass
distribution can be well described by a central kpc-sized
constant-density core (i.e.\ $\alpha \sim 0$ in the inner parts)
(\citealt{optcur_letter,lsbopt_bosma} [hereafter dBMBR and dBB02
respectively]).

dBMBR and dBB02 base their conclusions on observations of rotation
curves of over 60 LSB galaxies. They determine the mass density
profiles that give rise to the observed rotation curves and fit the
inner parts with a single power-law. The resulting distribution of
inner mass-density slopes shows a strong peak at $\alpha = -0.2 \pm
0.2$ (see Fig.~2 in dBMBR), but with a large spread, varying from
steep and cuspy $\alpha=-2$ profiles to flat and even moderately
positive slopes.  This was interpreted as a result of resolution
effects: dBMBR find that the best resolved curves show the flattest
profiles (see Fig.~3 in dBMBR and Fig.~14 in dBB02), contrary to what
is expected for CDM.  At lower resolutions the CDM and the ISO models
both predict the same slopes, leading to an ambiguity as to which
model fits lower resolution data best \citep{vdB&S}.

However, resolution is not the only effect that can influence the
measured slopes. Other effects, which dBMBR and dBB02 did not
thoroughly investigate, include non-circular motions, offsets between
dynamical and optical centres, lopsidedness, asymmetries, and
slit-width.  These effects all have the potential to decrease the
measured inner mass-density slopes. It is thus important to
investigate if the observed shallow slopes truly reflect the
underlying mass distribution, or whether they are determined by
systematic effects.  For example, small offsets between the dynamical
centre and the optical centre are sometimes observed in late-type
barred galaxies
\citep{DJ98,deVauc72}. If this were a systematic property of all LSB or
dwarf galaxies then observations aligned on the optical centre would
not necessarily probe the dynamical centre.  Even if the galaxy in
question had a steep CDM density cusp at the dynamical centre, the
derived slope would be shallower, giving the impression of a
core-dominated halo. Hence these systematic effects should be
understood, in order to judge whether LSB galaxies have cuspy or
core-dominated halos.  Furthermore, once the impact (if any) of these
systematic effects is understood, one could hope to determine how much
they contribute to the scatter in the observed slopes, and whether the
range in observed slopes can be attributed to ``cosmic scatter'' or
whether a single halo profile for all galaxies suffices.

In this paper we investigate the importance of these systematic
effects by deriving a large ensemble of rotation curves of both cuspy
CDM and core-dominated (simulated) halos, using observed curves as a
constraint.  We apply the relevant systematic effects to these
simulated curves, and then determine the inner mass-density slopes.
Our main conclusion is that none of the systematic effects
investigated can convincingly wipe out the signature of a CDM halo. In
other words, if LSB galaxies had CDM halos, they would be
unambiguously detectable, even in the presence of systematic effects.
The most likely interpretation is that the observed core-dominated
mass-distribution reflects the true dark matter distribution in LSB
galaxies.

The organization of this paper is as follows.  Section~2 discusses
some aspects of the observational comparison sample. Section~3
describes the simulations.  Section~4 contains the results of
simulations that include various systematic effects. In Sect.~5 we
describe some tests performed with real data, while in Sect.~6 we
attempt to use the simulations to improve on the halo model used. In
Sect.~7 we present our conclusions.

\section{Data}

The model rotation curves we derive will be compared with the results
presented in dBMBR and dBB02.  The focus will be on the histograms of
the inner slope $\alpha$ (dBMBR, their Fig.~2) and the diagrams of
inner slope $\alpha$ versus resolution $r_{\rm in}$ of the rotation
curve (dBMBR, their Fig.~3 and dBB02, their Fig.~14). These diagrams
are based on power-law fits to the inner parts of the mass-density
profiles derived from the $\sim 60$ high-resolution rotation curves
from
\citet{optcur_data}, \citet{optcur_pap2} and dBB02.  These rotation 
curves are measured from long-slit H$\alpha$ spectra, mostly sampling
the kinematics in the inner parts, in many cases combined with
lower-resolution \HI rotation curves sampling the outer parts
\citep{edb_bmh96}.

\subsection{Defining a restricted sample}

These data span a large range in resolution, inclination, and general
quality of the rotation curves. It is therefore worth asking whether
the data themselves may in any way be biased by systematic effects.
For example, if all the core-like profiles found turned out to have
been derived from low-resolution, low-quality rotation curves, one
would have valid reasons to question the usefulness of those data and
conclusions based on them.

\citet{McGaughcosmo} present such an analysis where they 
analyse only those rotation curves that resemble NFW CDM rotation
curves best, and conclude that there is no systematic difference in
the distributions of $c$ for their restricted and total samples.

\begin{figure*} 
\begin{center}
\hbox{\epsfxsize=0.48\hsize 
\epsfbox{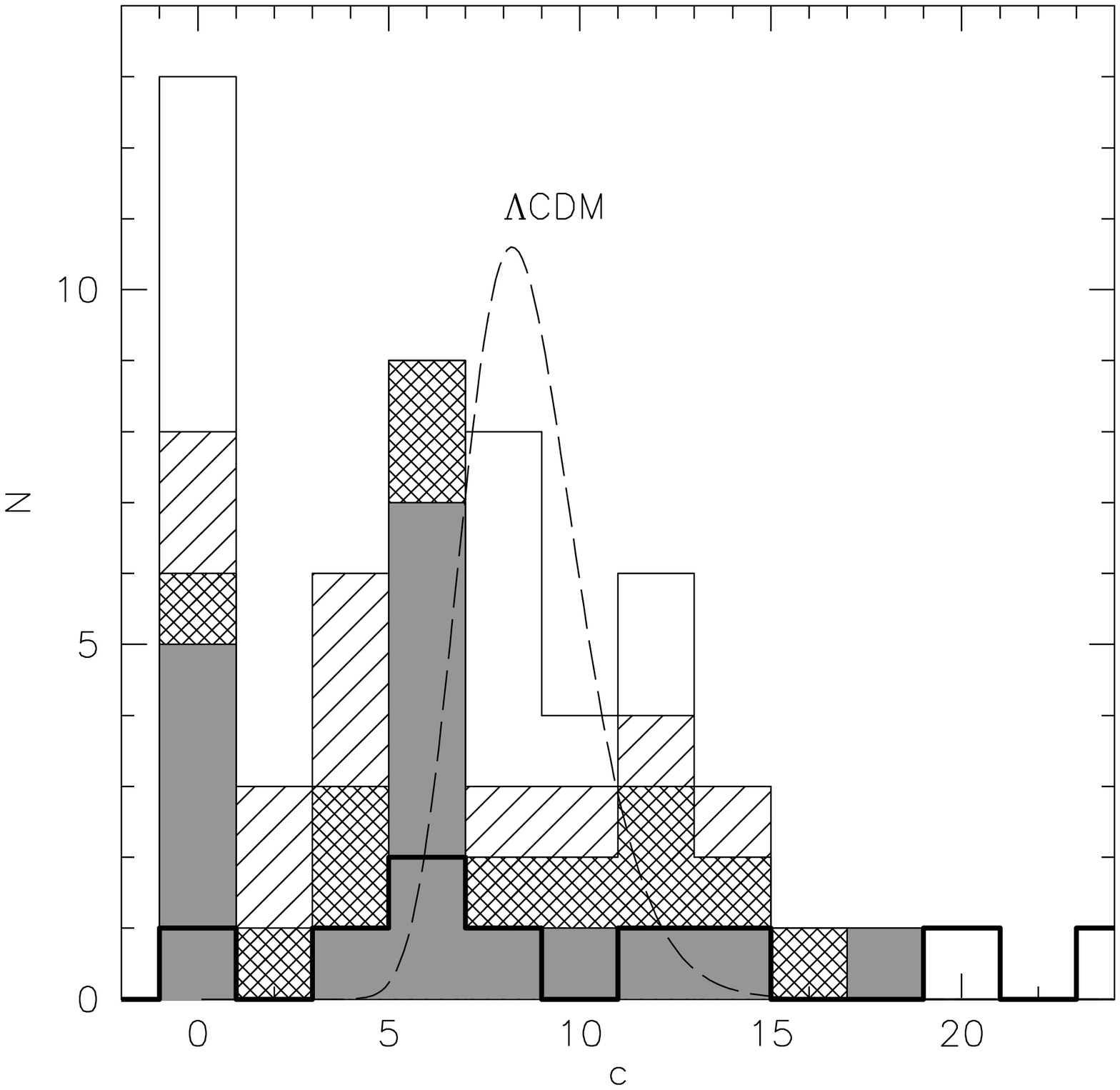}
\epsfxsize=0.48\hsize 
\epsfbox{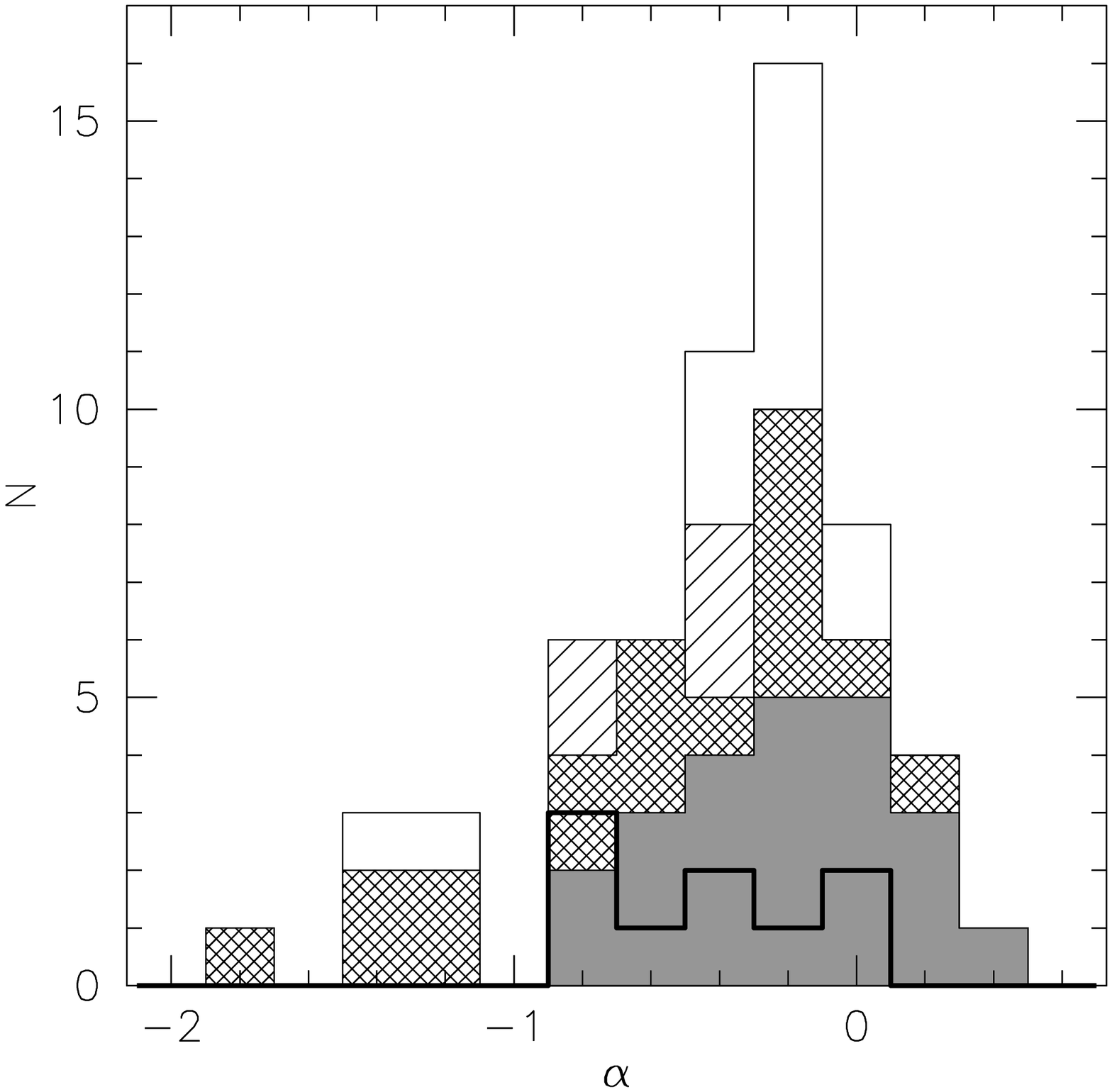}}
\caption[goodslopes.ps]{Distribution of the NFW concentration 
parameter $c$ (left panel) and the inner mass-density slope $\alpha$
(right panel) based on the data from dBMBR and dBB02.  The various
superimposed histograms show the different stages of pruning: open
histogram: full sample from dBB02 and dBMBR; single hatched histogram:
galaxies with $i<30^{\circ}$ and $>85^{\circ}$ have been removed;
double hatched histogram: galaxies with low-quality rotation curves,
asymmetries etc.\ have been removed; grey filled histogram: galaxies
with less than 2 independent data points in the inner 1 kpc are
removed. The thick open histogram in shows the restricted SMBB
sample. The dashed curve in the left panel shows the expected
distribution of $c$ for a $\Lambda$CDM universe (e.g.\
\citealt{McGaughcosmo,bul99}).
\label{goodslope}}
\end{center} 
\end{figure*}

Here we make a similar comparison, but concentrate on the properties
relevant for our comparison with simulations, namely distributions of
$c$ and the inner slope $\alpha$. Starting with the complete sample
from dBMBR and dBB02 (see Fig.~\ref{goodslope}), we apply the
following cuts: firstly, we remove all galaxies with inclinations $i <
30^{\circ}$ and $ i >85^{\circ}$. This leaves 44 galaxies.  The
resulting distribution is insensitive to the choices for the upper and
lower inclination limit: we have tried cuts with lower limits of
$25^{\circ}$ and $40^{\circ}$, and upper limits of $75^{\circ}$ and
$80^{\circ}$, and find almost identical distributions.  Secondly, we
remove all galaxies with low-quality rotation curves, i.e.\ with a
small number of independent data points, with large error-bars, and
large asymmetries.  Galaxies where the minimum-disk assumption is
clearly not valid were removed as well (e.g.\ UGC 6614). The resulting
sample contains 39 galaxies. Finally, as the difference between cusp
and core is most clearly visible in the innermost part of the
galaxies, we demand that the inner part is resolved: we retain only
those galaxies that have at least 2 independent data points in the
inner 1 kpc. This leaves us with a final sample of 19 galaxies, listed
in Table~\ref{sample} and shown in Fig.~\ref{goodslope}.

A comparison of the different stages of pruning shows no systematic
differences between the distributions of the initial full sample and
final restricted sample. If anything, the peak of the distribution of
inner slopes $\alpha$ has shifted towards a more positive value. The
tail towards negative slopes has disappeared.  The peak in the
$c$-distribution still occurs near $c\sim 6$, and the peak at $c\sim
0$ is still present in the final sample.

\subsection{Comparison with literature}

Recently, \citealt{swaters2002} [hereafter SMBB] have made a
similar analysis of a sample of 15 dwarf and LSB galaxies. They reach
the conclusion that the inner mass density slopes in their sample
cover the full range $-1
\la \alpha \la 0$, but claim that though their data prefer a shallow
$\alpha\sim 0$ slope, they cannot rule out steep $\alpha=-1$ slopes
for their sample. The analysis presented in the previous section and
in dBMBR suggests that a large fraction of the steep slopes one finds
are actually due to insufficient resolution.  It would thus be
interesting to see if these effects are also present in the SMBB
data. We subjected their data to the same quality criteria that we
used for ours. In practice this meant that 5 galaxies with less
than 2 independent points in the inner kpc were rejected (F563-V2,
F568-1, F568-3, F568-V1, F574-1), as well as one galaxy (U5721)
where the minimum disk assumption is invalid.  {A high-resolution
optical velocity field of U5721 \citep{ghasp} clearly shows the
presence of a kinematical minor axis not perpendicular to the major
axis. This is a classical indication of an oval distortion
\citep{bosmathesis}. For this galaxy, there are images in the HST
archive that show the presence of a weak bar} (see also dBB02, their
Sec.~9.2.2., noting that U5721 is also known as N3274).

We are thus left 
with 9 well-resolved rotation curves. Our restricted sample
has three galaxies in common with this restricted SMBB sample. We find
that for two of these galaxies the measured slopes are consistent at
better than $1\sigma$ (UGC 731 and UGC 4325). For one galaxy
(UGC 11557) the slopes differ significantly: SMBB find
$\alpha=-0.84 \pm 0.27$, while dBMBR find $\alpha=-0.08 \pm 0.23$. An
inspection of the density profiles in Fig.~1 of dBMBR and Fig.~5 in
SMBB shows that this difference in slope is entirely due to a
different choice of break radius, where SMBB have chosen a larger
radius than dBMBR, resulting in a much steeper slope. Choosing a smaller
break radius will give a value entirely consistent with that in dBMBR.
An inspection of the break radii of the 8 
other high-resolution SMBB profiles shows that UGC 11557 is the only
case where we would claim that the choice in break radius is
ambiguous.
We have overplotted the distribution of slopes of the restricted SMBB
sample in Fig.~\ref{goodslope}. The number of galaxies is small, so it
is difficult to say anything about the intrinsic distribution of
slopes, but it is clear that the SMBB sample is entirely consistent
with our larger restricted sample (and even more so 
when the ambiguous slope for UGC 11577 is removed).

Contrary to the conclusions reached by SMBB we do not find that
edge-on or barred galaxies skew the distribution of slopes. A
comparison between our full and restricted samples shows the inclusion
or omission of edge-on galaxies does not shift the average value of
the slope significantly from $\alpha\sim -0.2$. We also note that
edge-on late-type galaxies are expected to be transparent
\citep{B92,matthews_edgeon}; furthermore the H$\alpha$ spectra from
which the rotation curves are derived do not show the broad velocity
wings expected when projection effects play a large role.

Conclusions regarding the importance of bars are more ambiguous as
these depend on what one defines as a bar. Certainly none of the
galaxies investigated here contain the dominant straight bars found in
earlier-type galaxies. Here we are mostly dealing with Magellanic-type
oval components, and it is obviously a matter of judgement to decide
whether one deals with an intrinsically oval component, or a projected
more circular one.

SMBB list which of the galaxies in their sample they consider barred.
In order to make a consistent barred/non-barred division of the
galaxies in our restricted sample, we use their results to gauge and
calibrate our classification.  That any such classification remains
uncertain is illustrated by the fact that none of the galaxies in the
SMBB sample are classified in NED as unambiguously barred (Hubble type
SB), \emph{except} UGC 2259 [Hubble type SB(s)dm]. SMBB classify this
galaxy as \emph{unbarred}, even though the bar with perpendicular
spiral arms is easily visible on the Digital Sky Survey image. The
inner mass-density slope found for this galaxy is steep at $\alpha =
-0.86 \pm 0.18$.  As bars are disk dynamical features (e.g.,
\citealt{lia2002}), they imply a significant disk mass.  Such heavy disks
more than compensate for any underestimate of the slope which might be
caused by the noncircular motions they induce.  The minimum disk
assumption fails, and with it the steep halo slopes inferred in this
extreme limit.

Despite this UGC 2259 discrepancy, we have used the SMBB results to
gauge which of the galaxies in our restricted sample should be
classified as barred.  We list our classification in Table~1; we find
13 non-barred galaxies, 5 barred galaxies and 1 ambigious case. 

The
galaxies in our restricted sample have been selected not to suffer
from inclination effects or projection effects.  The majority of them
is unbarred, and it is not obvious at all that the presence of a bar
influences any of the results. This comparison of restricted samples
shows that once a proper pruning is made to only include minimum-disk,
well-resolved galaxies, most of the differences are merely a matter of
semantics. The observed \emph{shallow} inner mass-density slopes of
LSB and dwarf galaxies do therefore reflect the intrinsic properties
of their potentials.

\section{Models}

\subsection{Model rotation curves}

\begin{figure*} 
\begin{center}
\epsfxsize=\hsize 
\epsfbox{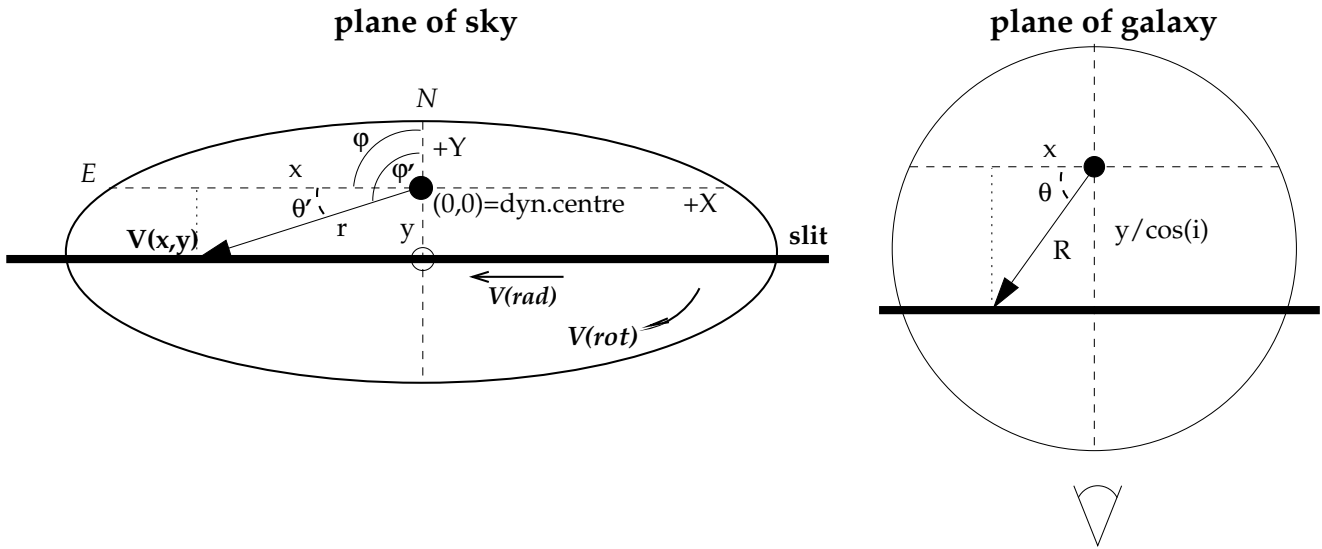}
\caption[geometry.ps]{Sketch of the geometry of an inclined rotating disk
projected on the sky. All angles and distances are measured in the
plane of the sky, except $R$ and $\theta$ which are measured in the
plane of the galaxy. See text for more details.
\label{geometry}}
\end{center} 
\end{figure*}

In order to model the various effects that can affect rotation curves
we use the full expression for the observed
line-of-sight velocity at any position in a galaxy velocity field.
This is given by (see Fig.~\ref{geometry}):
\begin{equation}
V(x,y) = V_0 + V_C(R) \sin i \cos\, \theta +  V_{\rm exp}(R)  \sin i  \sin\, \theta
\label{eq:rotcur}
\end{equation}
where $V(x,y)$ is the observed velocity at position $(x,y)$, where we
define the dynamical centre of the galaxy to be at $(0,0)$; $V_0$ is
the systemic velocity, which is not relevant in this analysis and
which will be ignored; $V_C(R)$ is the actual rotation curve velocity
at radius $R$, where $R$ is measured in the plane of the
galaxy. $V_{\rm exp}(R)$ is an expansion velocity describing
non-circular motions (which can be used to describe e.g.\ streaming
motions). Position $(x,y)$ is measured in the plane of the sky. We
additionally define the radius $r$ as the projection of $R$ on the
plane of the sky. As usual, $i$ is the inclination.  Following
\citet{beeg} we define a coordinate system with the $+X$ direction
towards decreasing right ascension, the $+Y$ direction towards the
north, and the dynamical centre as origin.  Position angles in the
plane of the sky are measured from N towards E.  The major axis of the
galaxy has sky position angle $\phi$.  Position $(x,y)$ then has
position angle $\theta$ with respect to the major axis as measured in
the plane of the galaxy given by:

\[
\cos \theta = \frac{-x\sin \phi + y \cos \phi}{R}
\]
\begin{equation}
\sin \theta = \frac{-x\cos \phi - y \sin \phi}{R\cos i}.
\end{equation}

We now introduce the angle $\phi'$ which is the sky position angle of
position $(x,y)$.  We can express $x$ and $y$ as 

\[
 x = -r\, \cos(\frac{\pi}{2} - \phi') = -r\, \sin \phi' 
\]
\begin{equation}
 y = r\, \sin(\frac{\pi}{2} - \phi') = r\, \cos \phi'.
\end{equation}

Combining this with Eqs.~(2) gives
\[ \cos \theta = \frac{r}{R}(\sin \phi' \sin \phi + \cos \phi' \cos \phi) \]
\begin{equation}
\sin \theta = \frac{r}{R\cos i}(\sin \phi' \cos \phi - \cos \phi' \sin \phi).
\end{equation}

Without loss of generalisation, we can make the simplifying assumption
that the major axis coincides with the $X$-axis, so that $\phi =
\pi/2$, as sketched in Fig.~\ref{geometry}.  Equation~(4) then reduces
to $\cos\, \theta = (r/R) \sin \phi'$ and $\sin\,\theta = - (r/(R\cos
i))\sin \phi'$.

We are now in a position to derive the line-of-sight velocity at any
position $(x,y)$, given any rotation curve $V_C$ and non-circular
motion component $V_{\rm exp}$. For example, Fig.~\ref{geometry} shows a
slit positioned parallel to the major axis, but offset by an amount
$y$. Assuming that $V_{\rm exp}=0$ everywhere, the observed line-of-sight
velocities along the slit are derived by varying $x$ at a constant
$y$, for a given input $V_C(R)$ and known $i$. The radii $R =
\sqrt{x^2+(y/\cos i)^2}$ and $r=\sqrt{x^2+y^2}$ can be 
calculated, as can the position angle $\phi' = \arctan(y/x)$, so that
$\theta$ can be derived. This then trivially gives the line-of-sight
velocities. One last step we need to take, in order to ``simulate''
our ignorance regarding the systematic effects in this particular
example, is to associate the positions along the slit $x$ (rather than
the radii $r$) with the derived line-of-sight velocities, because, as
with real data, these curves will be analysed under the assumption that
no systematic effects are present.  Other situations, including
non-circular motions, can be modeled in a similar way, as shown
below.

\subsection{Halo models}

For $V_C(R)$ we will initially investigate two halo models, apply
identical systematic effects to both, and determine the resulting
rotation curves and mass-density profiles. The first model is the
so-called NFW halo, favoured by the CDM simulations, but seemingly not
by the observations. The other model is the pseudo-isothermal halo,
preferred by observations, but without any basis in cosmology. The
properties of these models are summarised below.

\subsubsection{NFW halos}
The internal structure of DM halos formed in a CDM universe was
investigated in detail by \citet{NFW96,NFW97}. They found that the
halos modeled in their N-body simulations could be well described by
\begin{equation}
\rho(R) = \frac{\rho_i}{\left(R/R_s\right)
\left(1+ R/R_s\right)^2}
\end{equation}
where $R_s$ is the characteristic radius of the halo and $\rho_i$ is
related to the density of the universe at the time of collapse.  The
density and radius are strongly correlated with the mass of the halo.
At small radii the density diverges as $r^{-1}$.  This mass
distribution gives rise to a halo rotation curve
\begin{equation}
V_C(R) = V_{200} \left[\frac{\ln(1+cx)-cx/(1+cx)}
{x[\ln(1+c)-c/(1+c)]}\right]^{1/2},
\end{equation}
where $x = R/R_{200}$.
The radius $R_{200}$ is the radius
where the density contrast exceeds 200, roughly the virial radius
\citep{NFW96}. The characteristic velocity $V_{200}$ of the
halo is defined in the same way as $R_{200}$.  The curve is
characterised by two parameters: a concentration $c$ and a velocity
$V_{200}$. These are not independent, but related through the assumed
cosmology. For a standard $\Lambda$CDM this relation can be well
described as 
\begin{equation}
\log c = 1.191 - 0.064\log V_{200} - 0.032 \log^2 V_{200}
\label{eq:NFW}
\end{equation}
(cf.\ \citealt{NFW97}).  This relationship has an associated scatter,
but estimates in the literature differ. Here we adopt a conservative
logarithmic scatter of $\sigma({\log c}) = 0.18$ \citep{bul99}. Note
that in a more recent paper \citet{wechsler} advocate a scatter of
$\sigma({\log c}) = 0.14$.  As the precise value of the scatter is not
critical for our results, we use the larger value in order to give the
maximum amount of leeway to the CDM models.  See
\citet{optcur_pap2} for more details.

Recent N-body simulations have found even steeper inner
mass-density slopes of $\alpha=-1.5$ \citep{moore98}. 
We will not
consider these steeper slopes here; the NFW model can 
be regarded as a limiting case. If the data do not admit
$\alpha=-1$ slopes, they certainly will not allow $\alpha=-1.5$
values.

\begin{figure*} 
\begin{center}
\epsfxsize=\hsize 
\epsfbox{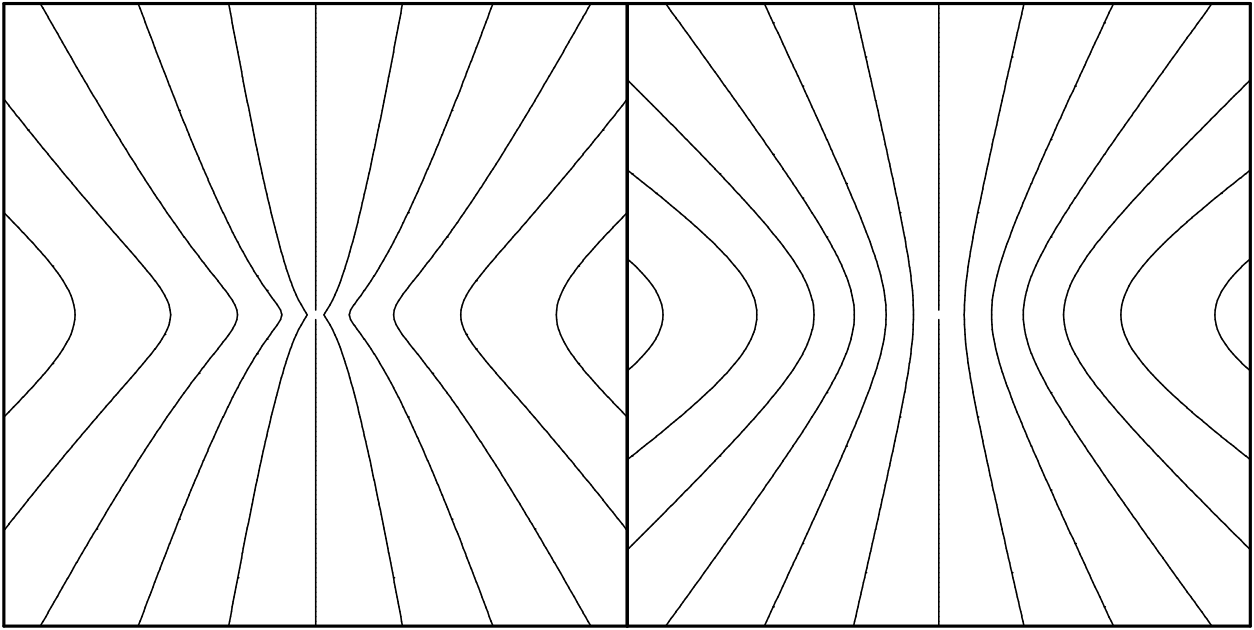}
\caption[velif.ps]{Velocity fields of the inner parts of massless 
disks embedded in a NFW halo (left panel) and an ISO halo (right
panel). The velocity field is seen under an inclination angle of
60\degr, and a position angle of 90\degr. The boxes measure $5 \times
5$ kpc. The vertical minor axis contour is 0
\kms, increasing in steps of 10 \kms outwards. The NFW halo parameters are $c=8.6$ and $V_{200} = 100$ \kms,
the ISO parameters are $R_C = 1$ kpc and $V_{\infty} = 100$ \kms. 
\label{velfi}}
\end{center} 
\end{figure*}

In Fig.~\ref{velfi} we show the inner part of the velocity field of a massless
disk in a NFW halo with $c=8.6$ (i.e.\ approximately the typical value 
predicted by $\Lambda$CDM, see \citealt{McGaughcosmo})
and $V_{200}=100$ \kms. The iso-velocity
contours show a characteristic ``pinch'' in the very inner parts,
which is a signature of the NFW profile.

\subsubsection{ISO halos\label{sec:iso}}

The spherical pseudo-isothermal (ISO) halo differs from the  NFW halo
in that it has a central region with a constant density.
Its  density profile is
\begin{equation} 
\rho(R) = \frac{\rho_0}{ 1 + (
{{R}/{R_C}} )^2 }, 
\end{equation} 
where $\rho_0$ is
the central density of the halo, and $R_C$ the core radius of the
halo.  The corresponding rotation curve is given by
\begin{equation} V_C(R) = \sqrt{ 4\pi G\rho_0 R_C^2 [ 1 -
{({R_C}/{R})}\arctan( {{R}/{R_C}}) ] }.
\end{equation}

The ISO halo is characterised by any 2 out of 3
parameters: an asymptotic velocity $V_{\infty}$, a core radius $R_C$
and a central density $\rho_0$. These parameters are related through
\begin{equation} 
  V_{\infty} = \sqrt{ 4 \pi G \rho_0 R_C^2 }.
\label{eq:ISO}
\end{equation}
Fig.~\ref{velfi} shows the inner part of a velocity field of a massless disk
embedded in an ISO halo with $V_{\infty}=100$ \kms\ and $R_C = 1$ kpc.
The iso-velocity contours are much rounder, and do not show the
``pinch'' of  the NFW velocity field.  

\subsection{Simulations\label{sec:simul}}

In order to be able to directly compare the simulations with the data
it is important that the simulated halos are ``realistic'', in the
sense that the range of halo parameters for the NFW model needs to be
similar to that found in the simulations (which are constructed to
resemble the real Universe), whereas the ISO halo parameters need to
resemble those determined from observations. We therefore use the
following boundary conditions in the models.

{\bf (i)} \emph{halo parameters}.  For the NFW halos we take a random
value for $V_{200}$ between between 20 and 490 \kms, and compute the
corresponding $c$-value using Eq.~\ref{eq:NFW}, modified by a random
scatter as described above.  To simulate ISO halos we choose a random
value for $V_{\infty}$ between 30 and 300 \kms, roughly corresponding
to the observed range in galaxy maximum rotation velocities. For
$\rho_0$ we choose a random value from the uniform range $\log
(\rho_0/[M_{\odot}\, {\rm pc}^{-3}]) = -1.7 \pm 0.4$
\citep{firmani}. The core radius $R_C$ is then determined using Eq.~\ref{eq:ISO}.

{\bf (ii)} \emph{inclination.} A random axis-ratio is chosen in the
range from 0.17 to 0.76, corresponding to inclinations between
40\degr\ and 80\degr. The inclination has no direct effect on the
slopes, but serves as a scale factor in the error-bars determined
below, and hence the uncertainty in the slopes.

\begin{figure*} 
\begin{center}
\epsfxsize=0.48\hsize 
\epsfbox{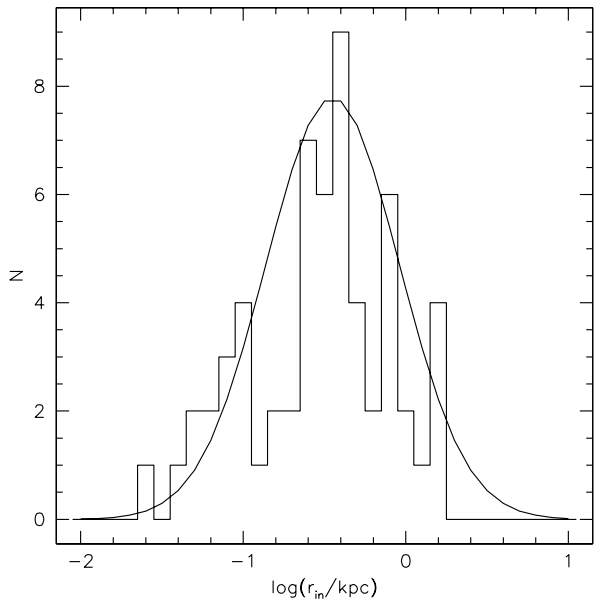}
\caption[distancedist.ps]{Distribution of the resolutions $\log r_{\rm in}$
of the rotation curves in presented in dBMBR and dBB02. Over-plotted
is a Gaussian with average $\mu(\log r_{\rm in}) = -0.45$ and $\sigma(\log
r_{\rm in}) = 0.41$ which was used as probability distribution for the
simulations.
\label{resolution}}
\end{center} 
\end{figure*}

{\bf (iii)} \emph{physical resolution.}  For each simulated galaxy the
resolution in kpc at which it will be ``observed'' is determined as
follows.  Figure~\ref{resolution} shows the logarithmic distribution
of observed resolutions (radii $r_{\rm in}$ of the innermost points of
the rotation curves) from dBMBR and dBB02.  This distribution is well
approximated by a Gaussian with an average $\mu(\log r_{\rm in}) =
-0.45$ and a dispersion $\sigma(\log r_{\rm in})=0.41$. The $2\sigma$
spread in resolution thus varies from 0.05 kpc to 2.3 kpc, reflecting
both the spread in distance as well as the mix of optical and \HI
curves used.  Though the precise parameters of this distribution are
not critical, choosing a resolution distribution that is approximately
equal to the empirical one means we can make a direct comparison
between the simulations and observations.

\begin{figure*} 
\begin{center}
\epsfxsize=0.48\hsize 
\epsfbox{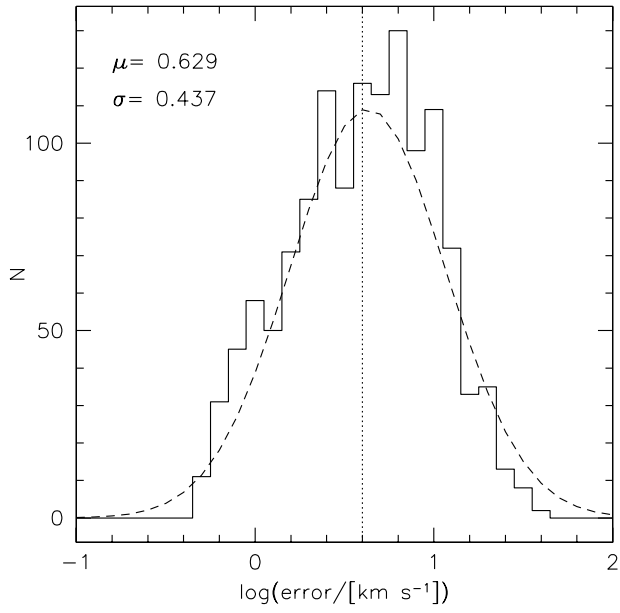}
\caption[errbars.ps]{Distribution of the error-bars in the unsmoothed rotation curves presented in dBB02. Over-plotted is a best fitting Gaussian with $\mu = 0.629$ and $\sigma=0.437$. This was used as the probability distribution for assigning error
s. The vertical line indicates the value of 4 \kms. Errors smaller
than 4 \kms, were set to be 4 \kms, in a similar way as the data.
\label{errors}}
\end{center} 
\end{figure*}

{\bf (iv)} \emph{error-bars.} To simulate the uncertainties in the
data, we de-correct the rotation curve for inclination and add
error-bars.  Figure~\ref{errors} shows the distribution of the
observational uncertainties in the dBB02 data, not corrected for inclination.
The best-fitting Gaussian has an average $\mu(\log \Delta V) = 0.629$
and a dispersion $\sigma(\log \Delta V)=0.437$. As in dBB02 and
\citet{optcur_pap2} we impose a minimum error of 4 \kms\
on the data.  We draw error-bars from this distribution and assign
them to the data points in the simulated rotation curve. The curve is
then corrected back for inclination, thus enlarging the error-bars and
uncertainty accordingly. As the intensity of the H$\alpha$ line is
generally inversely correlated with the magnitude of the error-bar,
this procedure assumes a random distribution of H$\alpha$
intensities. This is likely to be simplistic, but any improvement
involves assuming a specific H$\alpha$ intensity distribution, meaning
the models would loose some of their generality.

{\bf (v)} \emph{slit width.} Equation \ref{eq:rotcur} gives the
line-of-sight velocities along an infinitely thin slit. In reality,
the observations have used a finite slit with a width of $\sim 1''$.
We therefore integrate Eq.~\ref{eq:rotcur} over a width of $1''$ and determine
the average value of $V(x,y)$ at each position along the thick slit.

{\bf (vi)} \emph{mass density profiles.} The mass density profiles are
determined exactly as described in dBMBR, with one difference: dBMBR
choose a ``break radius'' to determine the inner slope. Here we
determine the slope by performing a weighted fit to the innermost 3
points. The uncertainty in the slope is determined in the same manner
as in dBMBR. We have tested fitting to the innermost 2, 4 or 5 points
but this did not change any of the conclusions appreciably.
We assume a spherical halo and minimum disk, i.e.\ the dynamical
significance of the stellar disk is assumed to be negligible and the
stellar mass-to-light is set to zero.  For LSB galaxies minimum disk
is in general a good assumption (e.g.\ \citealt{edb_rot}).  It is
difficult to estimate what effect the stellar disk has on the halo,
but most likely the halo becomes more concentrated under the influence
of the disk.  In the absence of applicable models we will not
explicitly simulate this, except to note that because of the dynamical
insignificance of the disk the effect is likely to be small. The
minimum disk assumption also gives a hard upper limit on the value of
the slope, as taking into account any contribution by the stars to the
mass-density slope reduces the implied dark matter slope, independent
of whether any contraction has taken place or not.

{\bf (vii)} \emph{small-scale deviations.} The model rotation curves
can by necessity be nothing more than pale imitations of those of real
galaxies. One of the features of real rotation curves are the ``bumps
and wiggles'', caused by, among others, streaming motions along spiral
arms, random motions of H\,{\sc ii} regions, and gas motions due to the
effects of star formation. The magnitude and importance of these
effects are of course different for different galaxies. They do
undoubtedly play a role in the observed LSB galaxies rotation curves,
but without modeling individual galaxies in detail it is impossible to
determine and apply these effects to the models in an unbiased and
general manner. We will return to this matter in Sect.~\ref{nosys}. It
should be noted though that the many ways in which rotation curves can
be derived are all designed to remove the effects of these small-scale
deviations: for example, the tilted ring procedure averages velocities
over rings at a certain radius, whereas long-slit rotation curves use
the independent information from radii at the approaching and receding
sides of the galaxy to derive the underlying global rotation
velocities.  Our omission of these deviations means that the
uncertainties we derive in our simulations will be smaller than found
in the data, but this is a small price to pay not having to introduce
ill-constrained random motions that may bear no resemblance with
reality. A specific example will be discussed in Sect.~\ref{nosys}.

\section{Results}

\subsection{Comparison with models}

\begin{figure*} 
\begin{center}
\hbox{\epsfxsize=0.48\hsize 
\epsfbox{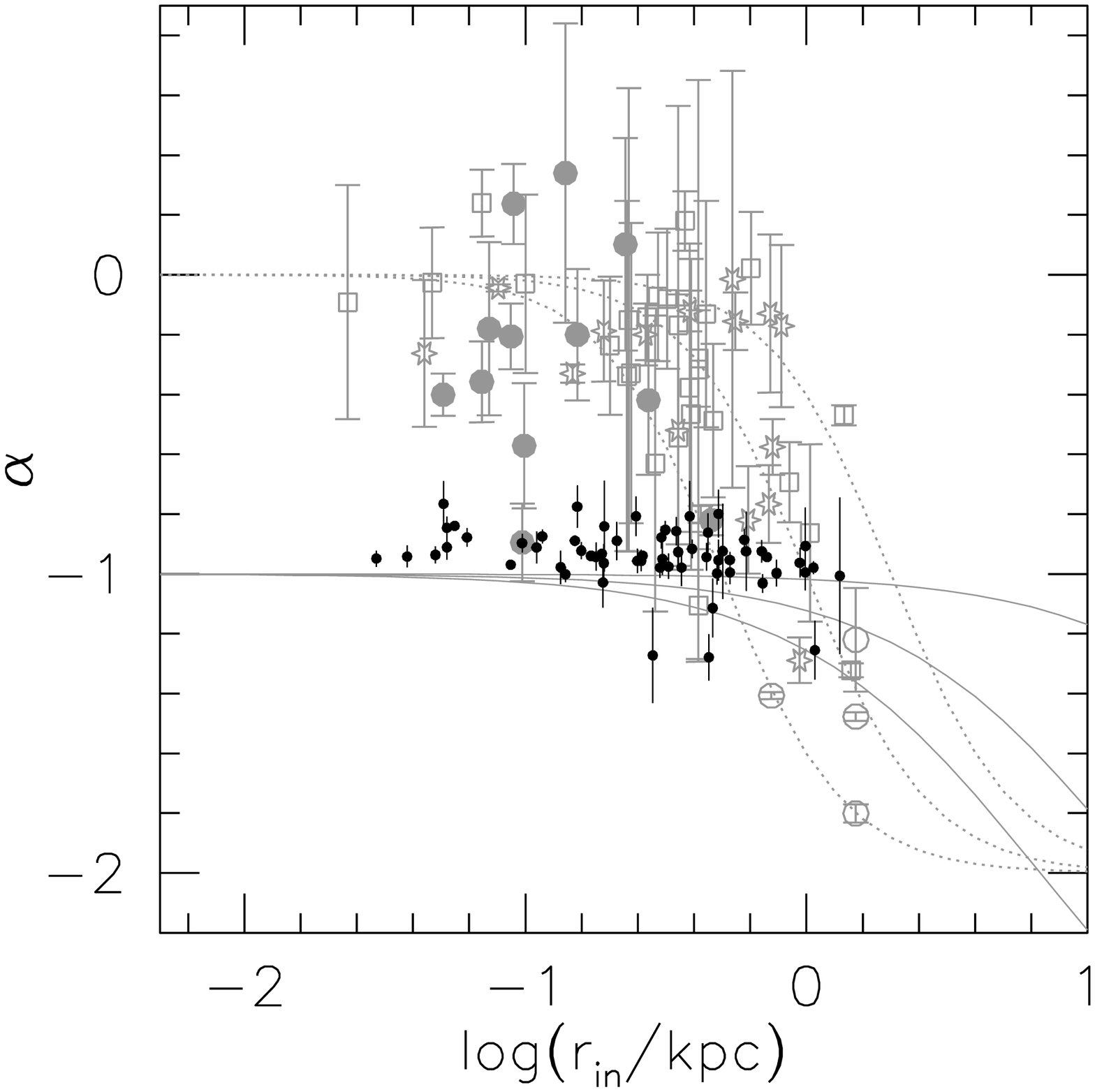}
\epsfxsize=0.48\hsize 
\epsfbox{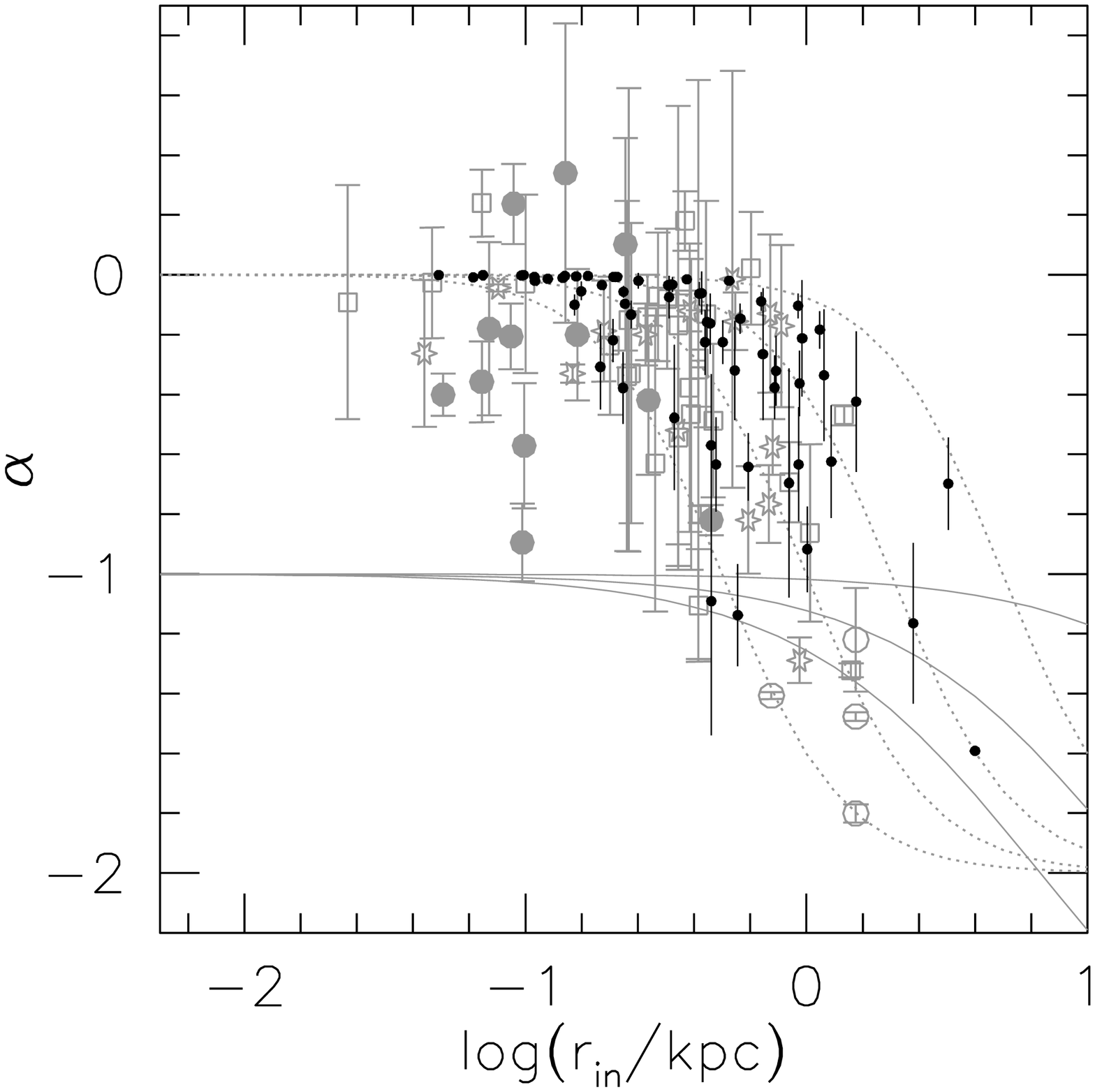}}
\hbox{\epsfxsize=0.48\hsize 
\epsfbox{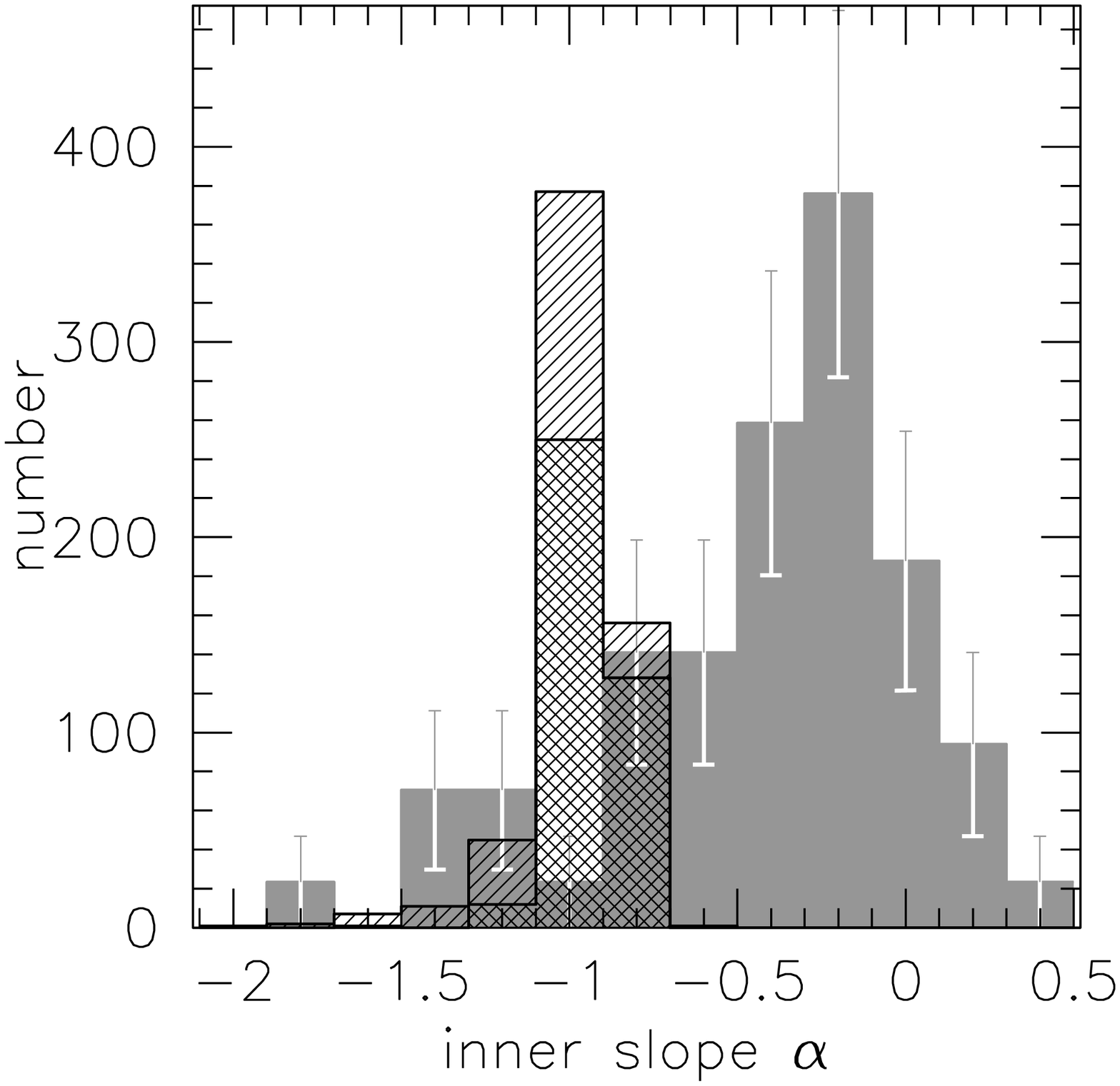}
\epsfxsize=0.48\hsize 
\epsfbox{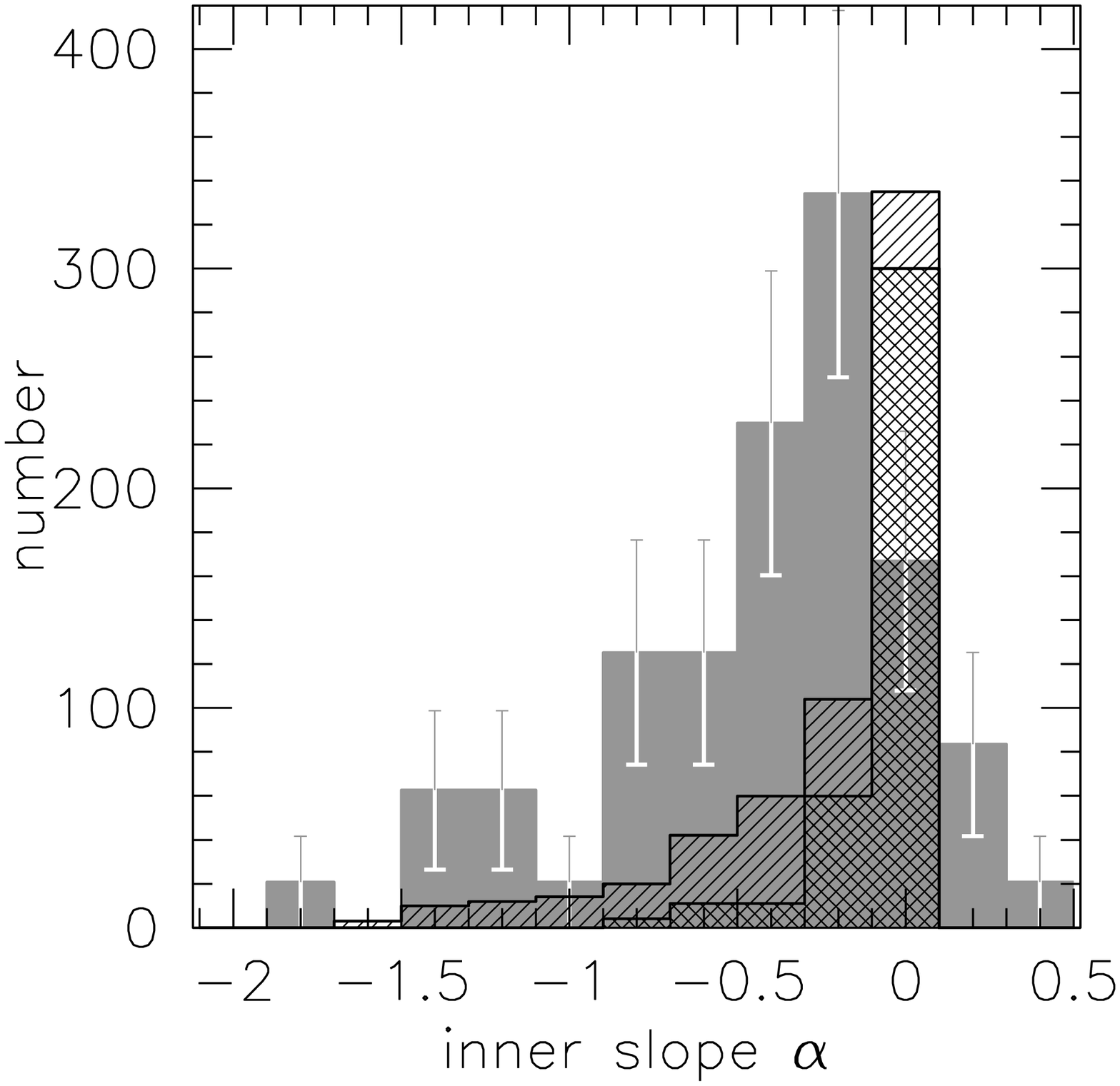}}
\caption[gaussdata]{
Comparison of the simulated NFW (left) and ISO (right) halos.  The top
row shows compares the simulated $r_{\rm in}-\alpha$ data points
(black) with the observed distribution from dBMBR and dBB02 (grey data
points). No systematic effects are assumed. The grey dotted lines
converging on $\alpha=0$ represent the theoretical variations in slope
for ISO halos with $R_C = 0.5, 1, 2$ kpc. The lines converging on
$\alpha = -1$ show the variation in slope for $c/V_{200} = 9.8/50,
8.6/100, 6.1/500$ (the slope only depends on the ratio
$c/V_{200}$). The bottom row shows the histograms of $\alpha$ values
for the two models. The double hatched histogram indicates
well-resolved galaxies with $r_{\rm in} \leq 0.5$ kpc.
\label{NFWISOnosys}}
\end{center} 
\end{figure*}

For each set of input parameters described below we compute 600
rotation curves per halo model.  This is significantly more than the
61 observational data-points available, but avoids small number
statistics affecting the simulated histogram. To compare the data and
the simulations we scale the maximum of the data histogram to that of
the simulation histogram.  To indicate the uncertainties in the data
histogram we over-plot an error-bar on each observed histogram bin
which reflects the uncertainty due to counting statistics in the
observational sample (these error-bars are also scaled) (see
Fig.~\ref{NFWISOnosys} for examples).  In the diagram of inner slope
$\alpha$ versus resolution $r_{\rm in}$ we show observational data
points in the background, with the simulation results over-plotted.
To avoid cluttering the plot we do not always show all 600 galaxies,
but select 61 galaxies (i.e.\ equal to the number of observed data
points) at random (see Fig.~\ref{NFWISOnosys} for examples).  

In order to isolate the investigated systematic effects from
resolution effects, we distinguish in the histograms between
well-resolved galaxies (2 or more independent points per kpc;
cross-hatched histogram) and less well-resolved galaxies (less than 2
independent points per kpc; single-hatched histograms). The
cross-hatched histograms can be directly compared with the
well-resolved sample in Fig.~\ref{goodslope}.

\subsection{The Null-case\label{nosys}}

We first investigate the case of observations without any systematic
effects. That is, we determine the major-axis curves defined by
Eq.~\ref{eq:rotcur} and the ISO and NFW halo models, and follow the
procedure described in Sect.~\ref{sec:simul}. We assume there are no
non-circular motions and no mismatches of any kind.  The results are
shown in Fig.~\ref{NFWISOnosys}. As expected, the derived slopes are
close to the theoretical values. The NFW model shows a pronounced peak
at $\alpha =-1$. The finite width of the slit hardly affects the
measured slopes. Similar conclusions apply to the ISO model. The
high-resolution ISO models show a clear $\alpha=0$ slope, while the
least-resolved models show steeper slopes, comparable to those found
in the NFW model at similar resolutions.

The histograms show only partial agreement with the data. The peak
occurs at the wrong value for the NFW model, and the histogram has the
wrong shape. For the ISO model the peak occurs at roughly the correct
value, but the shape of the histogram differs significantly from the
data. Clearly some sort of effect, whether cosmic or systematic, is
needed to make the simulations more consistent with the data. It is
clear though that this effect needs to be more pronounced for NFW than
for ISO.

\subsubsection{Small-scale deviations}

The simulated error-bars in Fig.~\ref{NFWISOnosys} are smaller than
those of the data. As discussed in item (vii) in
Sect.~\ref{sec:simul}, this is most likely due to the fact that we do
not simulate small-scale features in the rotation curves. This does
however not affect the conclusions we can draw from the
simulations. After all, when averaged over $\sim 60$ rotation curves,
we can regard the ``bumps and wiggles'' as essentially random: they
introduce scatter but are expected to average out and will not affect
the shape in a systematic way.  Another reason not to include the
small-scale deviations is that every galaxy is unique and has features
in the rotation curve that depend on the positions of the H\,{\sc ii} regions,
spiral arms, star formation etc. Given the low star formation rates,
small number of H\,{\sc ii} regions, and lack of pronounced spiral arms, it is
unclear how important any of these effects are in LSB galaxies
compared to ``normal'' galaxies. It is therefore difficult to derive a
universally applicable description short of  modeling each galaxy
individually, which is clearly unfeasible.

\begin{figure*} 
\begin{center}
\hbox{\epsfxsize=0.48\hsize 
\epsfbox{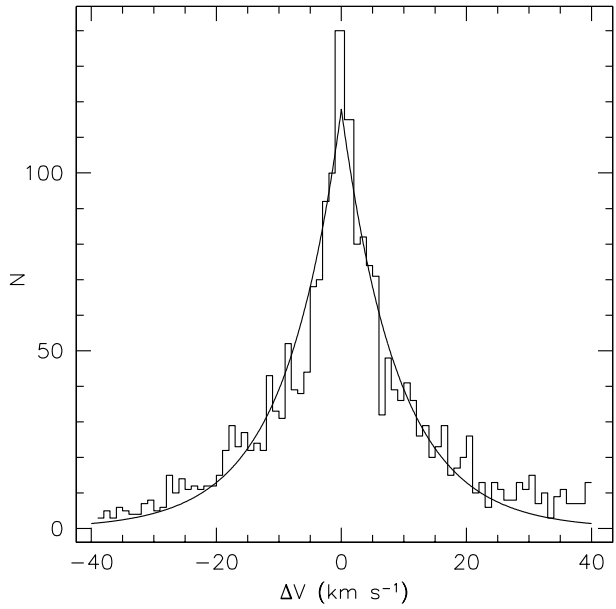}}
\caption[smoothdata]{
Histogram of the deviations of the raw data points from the underlying smooth
rotation curves presented in \citet{optcur_pap2}. The distribution is symmetrical, and can be well described by a function of the form $N = 117.9\, 
e^{-|\Delta V |/9.06}$, as shown by the superimposed curve.
\label{smoothdata}}
\end{center} 
\end{figure*}

We illustrate the fact that small scale deviations most likely do not
affect the conclusions, but merely introduce more scatter, by
investigating a naive (and probably too simplistic) description
for the deviations which we then apply to the simulations presented above.
Fig.~\ref{smoothdata} shows a histogram of the deviations of the raw
observed rotation curves with respect to the smooth rotation curves as
presented in Fig.~7 of dBB02 and Fig.~1 of
\citet{optcur_pap2}. The deviation histogram is well described by a
function of the form $\exp -|\Delta V|$.  If the deviations in the
rotation curves were radially uncorrelated we could add deviations
derived from this distribution to our model curves, use this modified
model curve to determine the underlying smooth rotation curve and
subsequently measure the mass profile.

However, in real galaxies the small-scale deviations are correlated:
if e.g.\ streaming motions along a (resolved) spiral arm cause a data
point to lie above the average curve, then the data point next to it
is more likely to lie above it as well.  Furthermore, the deviations
are defined with respect to the underlying curve, in that the latter
is derived by minimising the former.  Ideally, we would have to
introduce an extra step in the simulations to adjust the model
rotation curves to the effect of the deviations.  On the other hand,
as we would afterwards be fitting to the underlying curve as defined
by a minimisation of deviations, and
\emph{not} to the raw deviations, their impact on the results
would have to be significantly less than suggested by
Fig.~\ref{smoothdata}.  It is thus not clear whether taking small-scale deviations into account would have a significant effect.

\begin{figure*} 
\begin{center}
\hbox{\epsfxsize=0.48\hsize 
\epsfbox{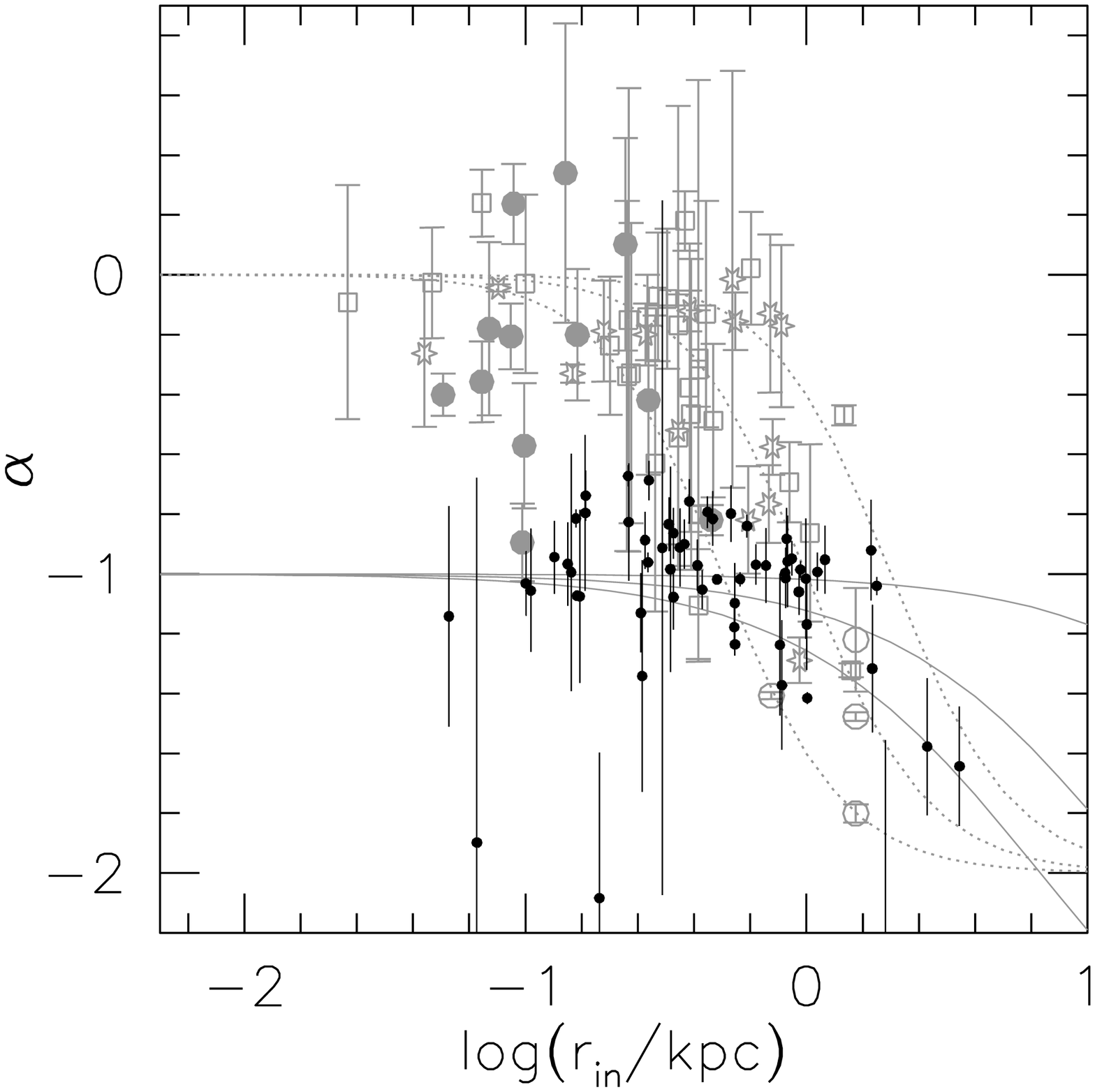}
\epsfxsize=0.48\hsize 
\epsfbox{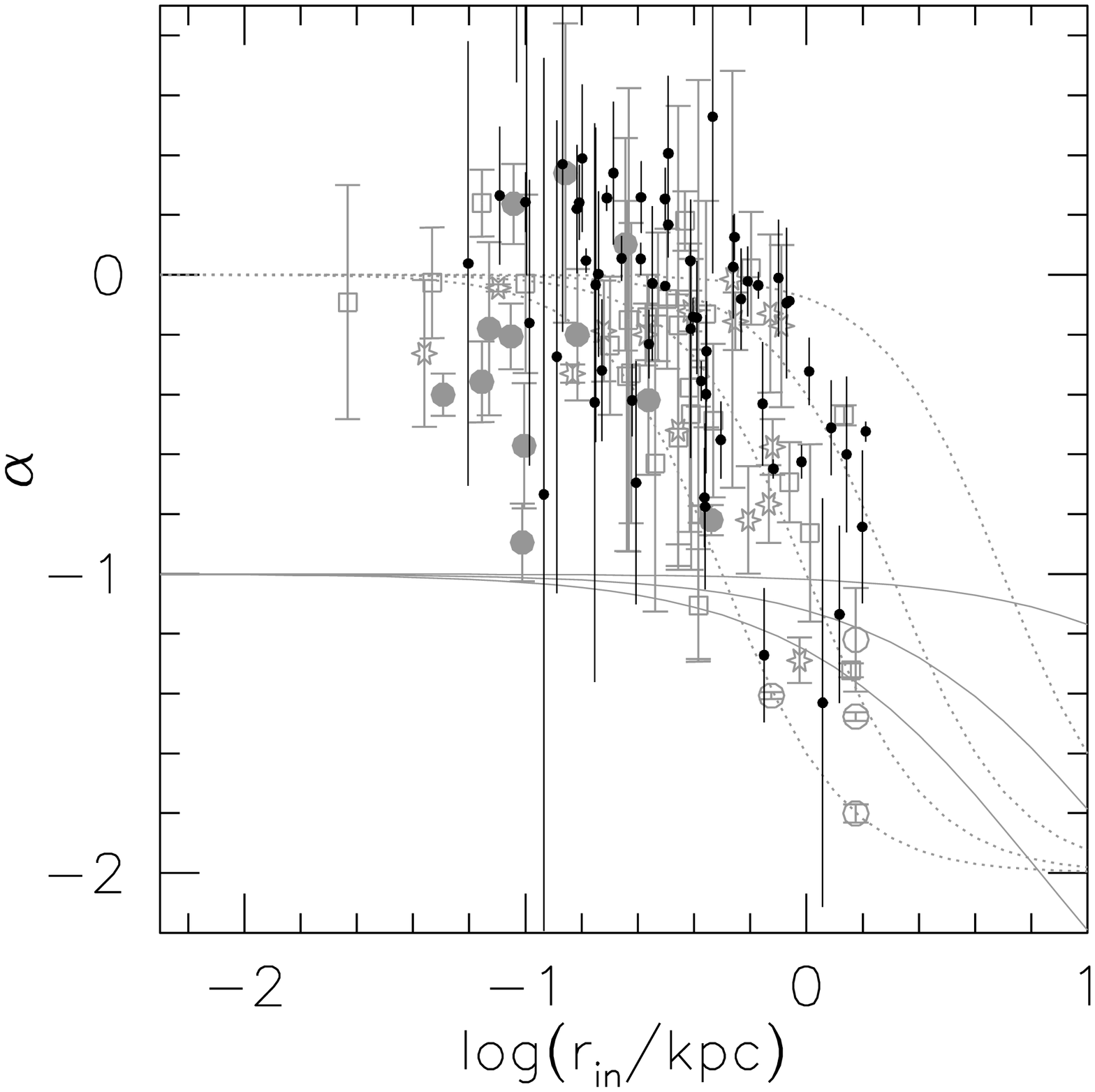}}
\hbox{\epsfxsize=0.48\hsize 
\epsfbox{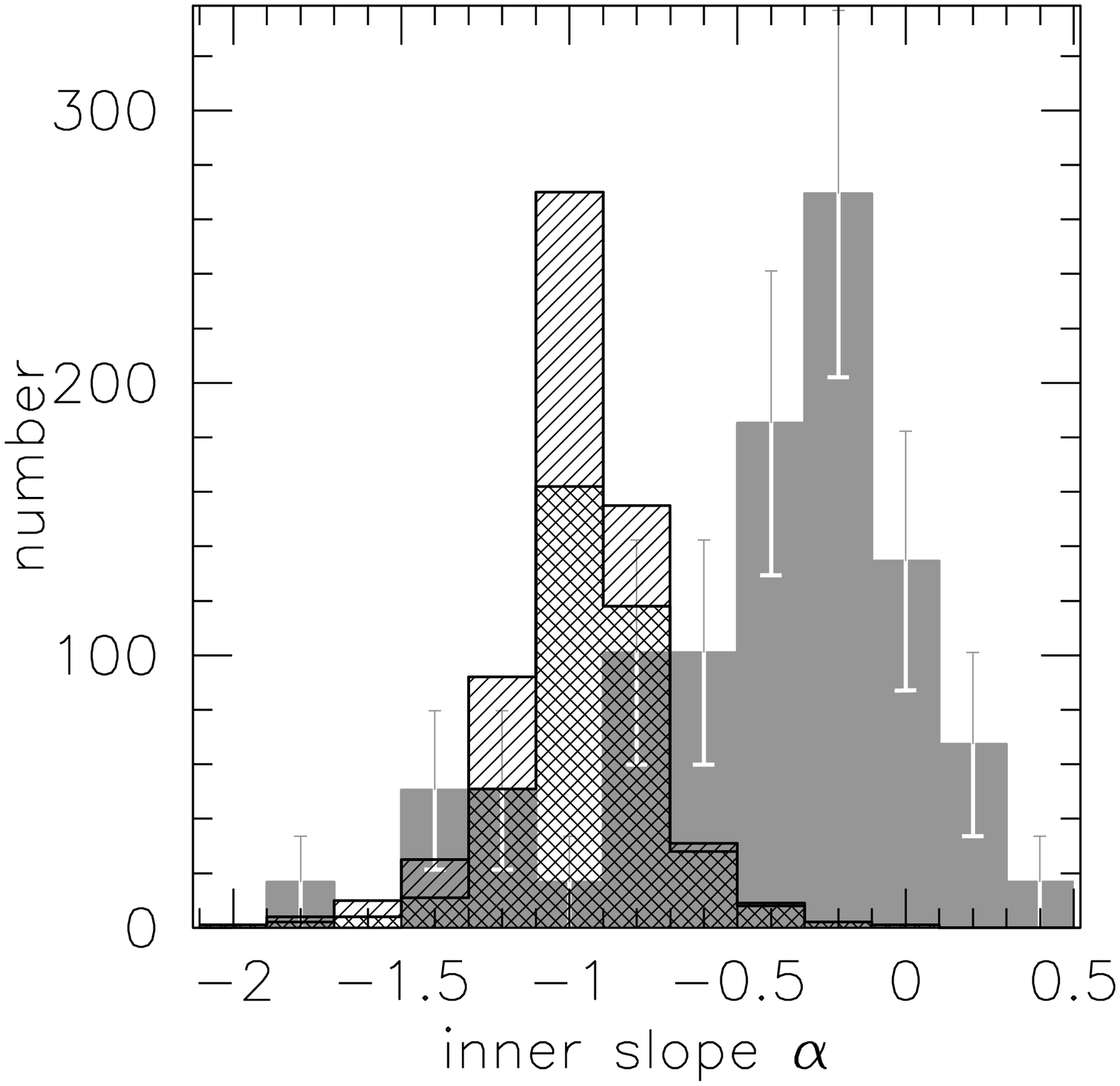}
\epsfxsize=0.48\hsize 
\epsfbox{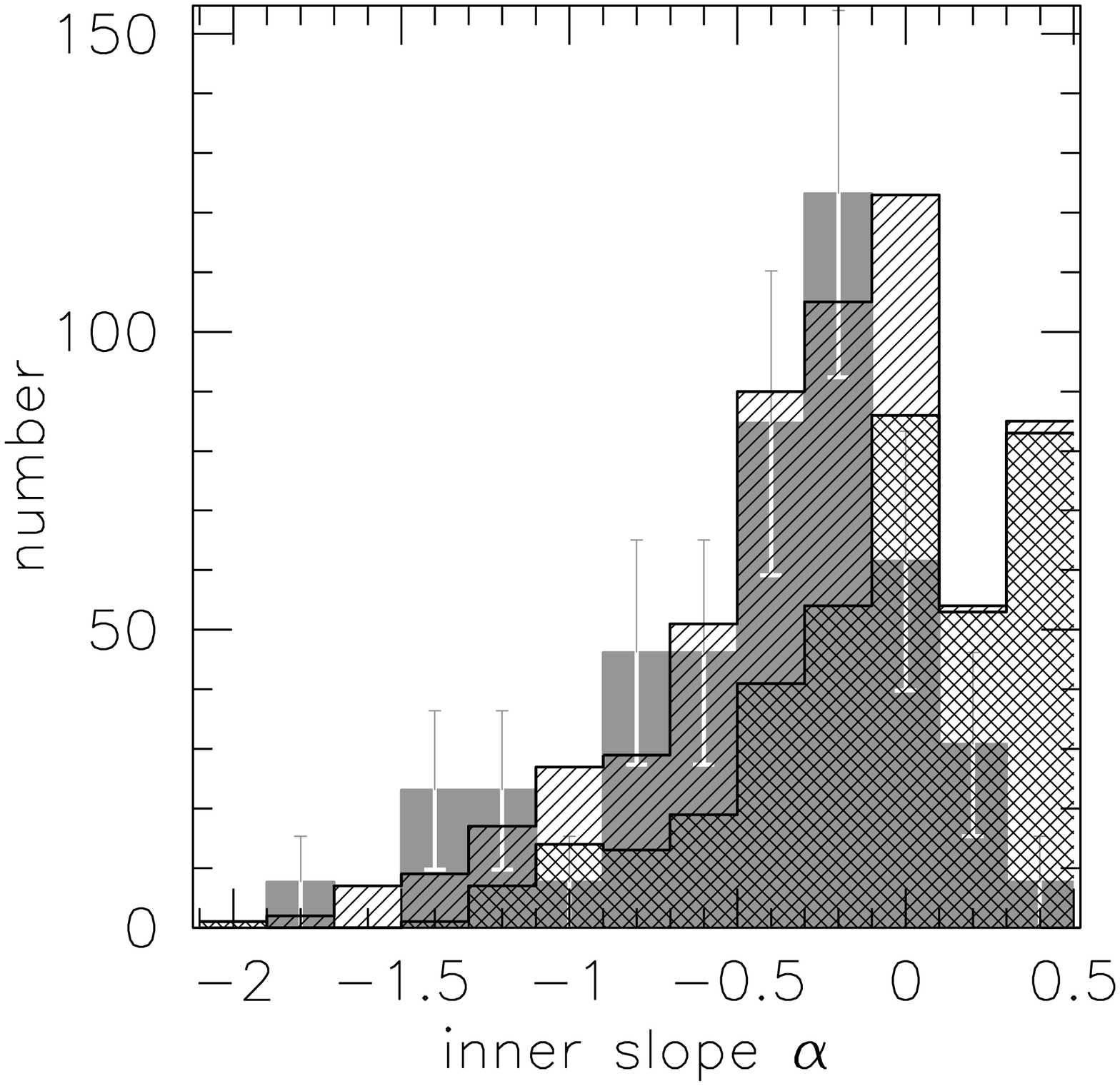}}
\caption[deviatedata]{
Comparison of the simulated NFW (left column) and ISO (right column)
halos with rotation curves modified by small-scale velocity
deviations. The top row compares $r_{\rm in}-\alpha$ data points with
the observed distribution.  The bottom row shows the histograms of
$\alpha$ values for the two models.  Apart from extra scatter, the
addition of small-scale deviations has not changed the general
trend. See Fig.~\ref{NFWISOnosys}.
\label{NFWISOdeviate}}
\end{center} 
\end{figure*}

As a crude attempt to show these effects, we have applied to the two
simulations described above, small-scale deviations drawn from the
distribution shown in Fig.~\ref{smoothdata}. In order to introduce a
degree of correlation between the deviations, we boxcar-smooth over
four data points. With real data the next step would have been to fit
an underlying smooth curve by minimizing the residuals, followed by
the derivation of the mass profile from the smooth curve. To determine
the mass profiles we thus prefer not to use the raw data, since the
mass profile derivation uses the gradient of the rotation
curve. Instead we introduce a bit of binning and smoothing to
determine the underlying gradient.  In practice, we thus have to
soften the impact of the derived (unbinned) deviations on the simulated curves.
Here we (arbitrarily) divide them by a factor of four before adding.
The number of data points to smooth over as well as this softening
factor were found empirically by demanding that the error bars in the
simulated slopes match those of the data.  Fig.~\ref{NFWISOdeviate}
shows the results. The method has introduced scatter, but not changed
the conclusions that were derived from the simulations presented in
Fig.~\ref{NFWISOnosys}.

It is clear that the method described above is unsatisfactory. The
smoothing and softening factors are arbitrary and ill-constrained and
the amount of scatter introduced depends strongly on the values one
assumes for them. Again, the lack of a universally applicable
description of what are essentially stochastic and random processes
makes this an unfeasible exercise.  Furthermore, as the derivation
of rotation curves uses extra information such as continuity and
symmetry between approaching and receding sides it is not clear whether
the net impact would be significant.  Therefore, rather than trying to
fine-tune a clearly unsatisfactory description we prefer to proceed
without it, keeping in mind  that there is this missing ingredient,
but that it is not likely to affect the conclusions in a significant
way.

\subsection{Lopsidedness and asymmetries}

One of the assumptions made in rotation curve analysis is that the
velocity field is symmetrical around the dynamical centre.  This is
not necessarily the case for all galaxies; some seem to have a
``kinematical lopsidedness'' (see e.g.\
\citealt{swaters_lop} for an extensive description), where the
rotation curves of 
the approaching and receding sides each have
different slopes and amplitudes (cf.\ Fig.~2 in
\citealt{swaters_lop}).  The amplitudes of the curves can differ by up
to 20 per cent. Interestingly, kinematical lopsidedness does not seem
to automatically imply a spatial offset between optical and dynamical
centres.  Kinematical lopsidedness can thus affect the interpretation
of mass-density slopes.

Here we investigate the effect of kinematical lopsidedness on the
rotation curves of NFW and ISO halos. We model the lopsidedness by
computing a rotation curve, but choose an incorrect point of symmetry
(or incorrect systemic velocity). After flipping the rotation curve
around this incorrect point of ``symmetry'' we thus end up with
approaching and receding side curves that have different amplitudes
and shapes, and therefore different inner slopes.  We then proceed to
take the average of the inner curves, and determine the mass-density profile.

\begin{figure*} 
\begin{center}
\epsfxsize=\hsize 
\epsfbox{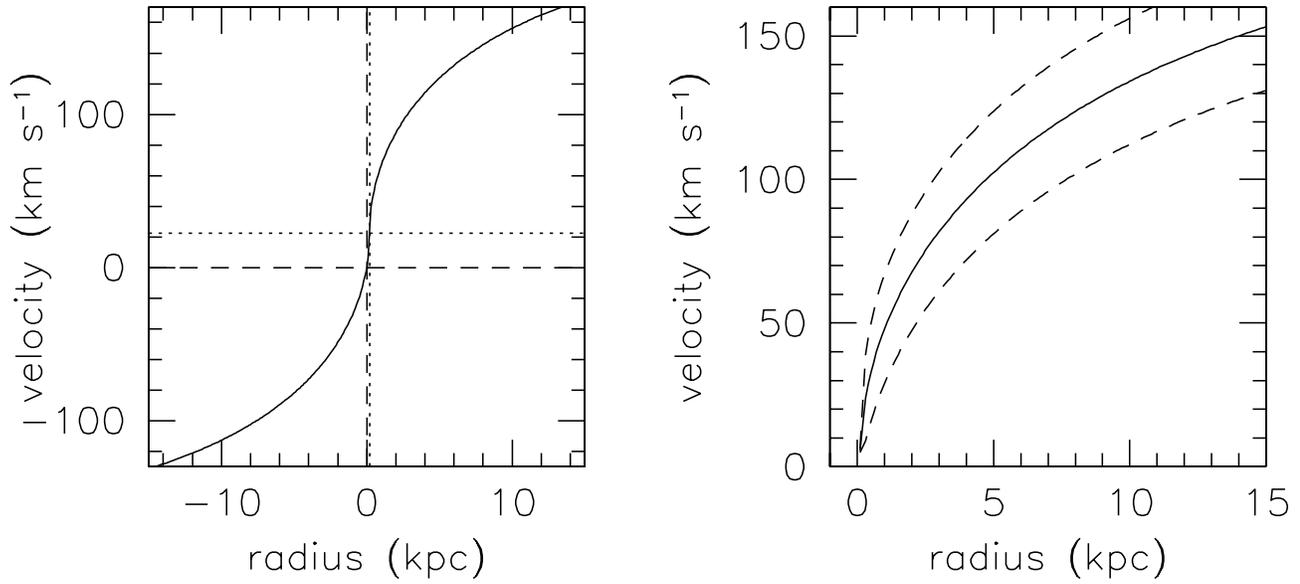}
\caption[kinlop.ps]{Example of the construction of a lopsided rotation curve.
The left panel shows a symmetric NFW halo rotation curve. The dotted
lines indicate the true point of symmetry. The dashed lines indicate
the ``wrong'' centre of symmetry used to construct the lopsided
curve. The offset is 0.2 kpc in radius and 22 \kms\ in velocity. The
right panel shows the lopsided curve (drawn line) which is the average
of the approaching and receding sides' curves (dashed lines).
\label{kinlop}}
\end{center} 
\end{figure*}

As the derived slopes are steepest when the point of symmetry is
chosen correctly, this procedure will flatten the measured
slopes. Figure~\ref{kinlop} shows an example of a ``kinematically
lopsided'' NFW curve.  This procedure does not take into account the
observed behaviour of the curves at large radii (see e.g.\ N4395 in
\citealt{swaters_lop}; here one side of the galaxy exhibits a rising 
rotation curve, while at identical radii on the other side it has
already flattened), but as our interest is in the inner slope, this
will not affect any of the conclusions.

\begin{figure*} 
\begin{center}
\hbox{\epsfxsize=0.32\hsize 
\epsfbox{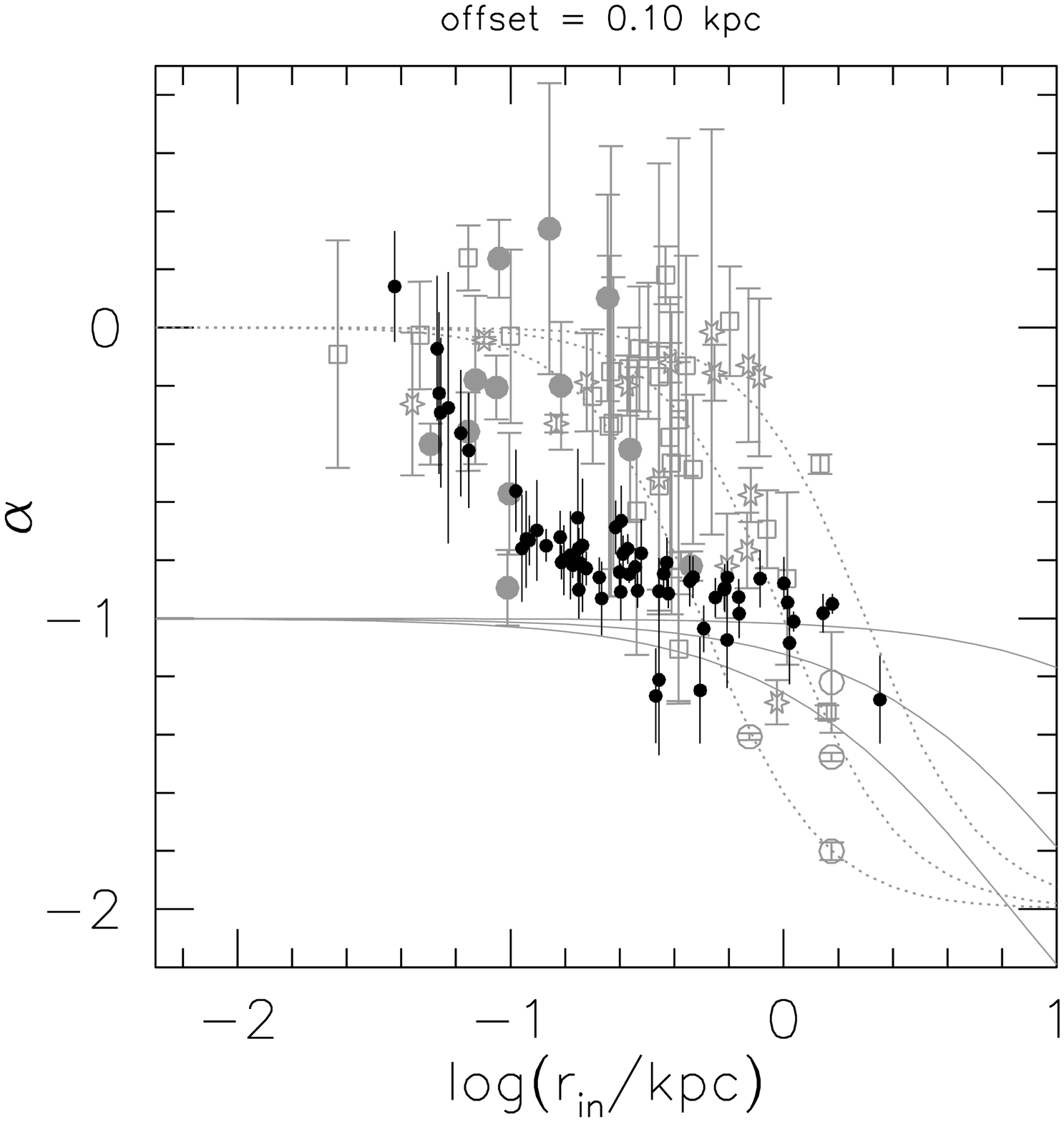}
\epsfxsize=0.32\hsize 
\epsfbox{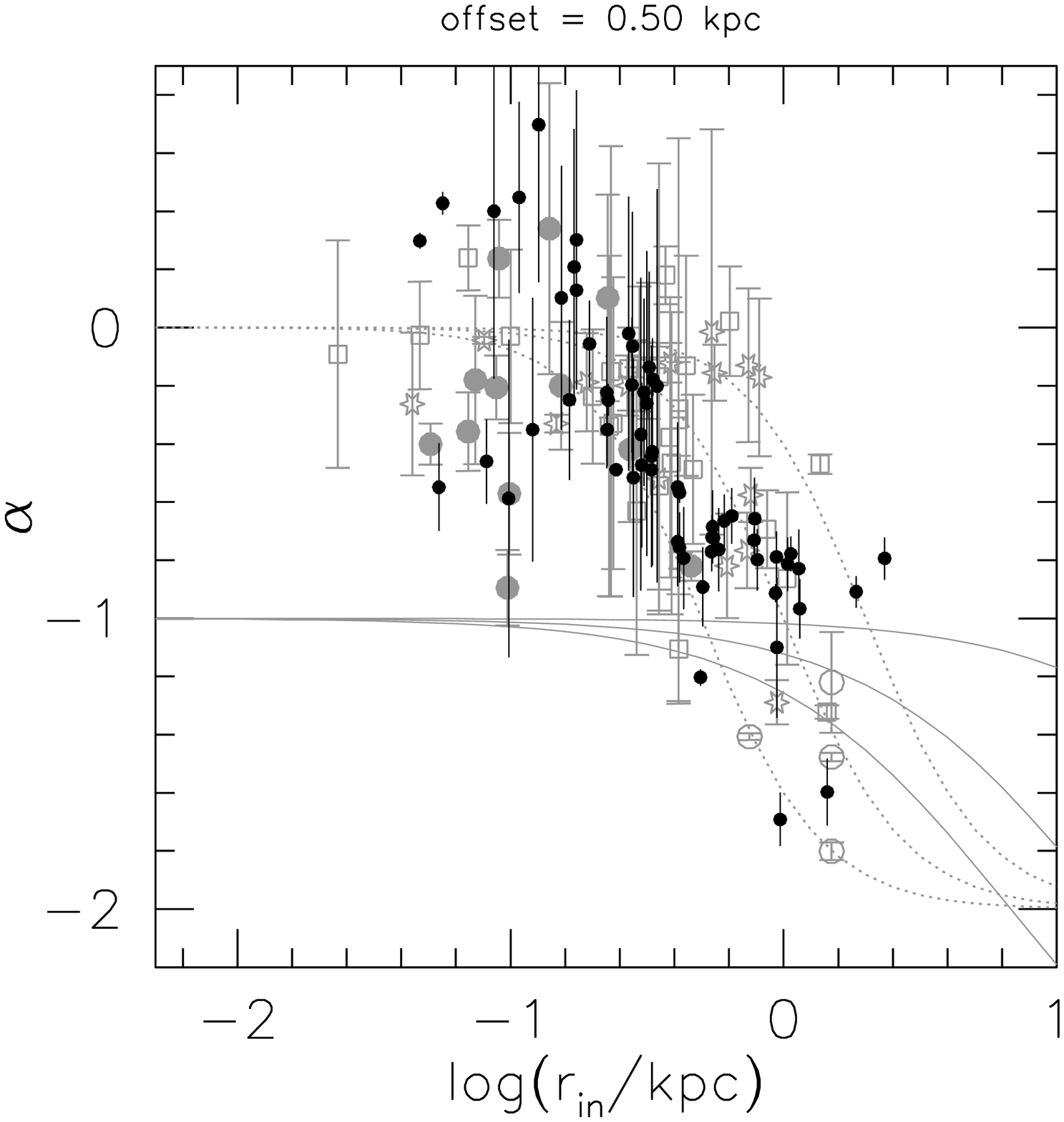}
\epsfxsize=0.32\hsize 
\epsfbox{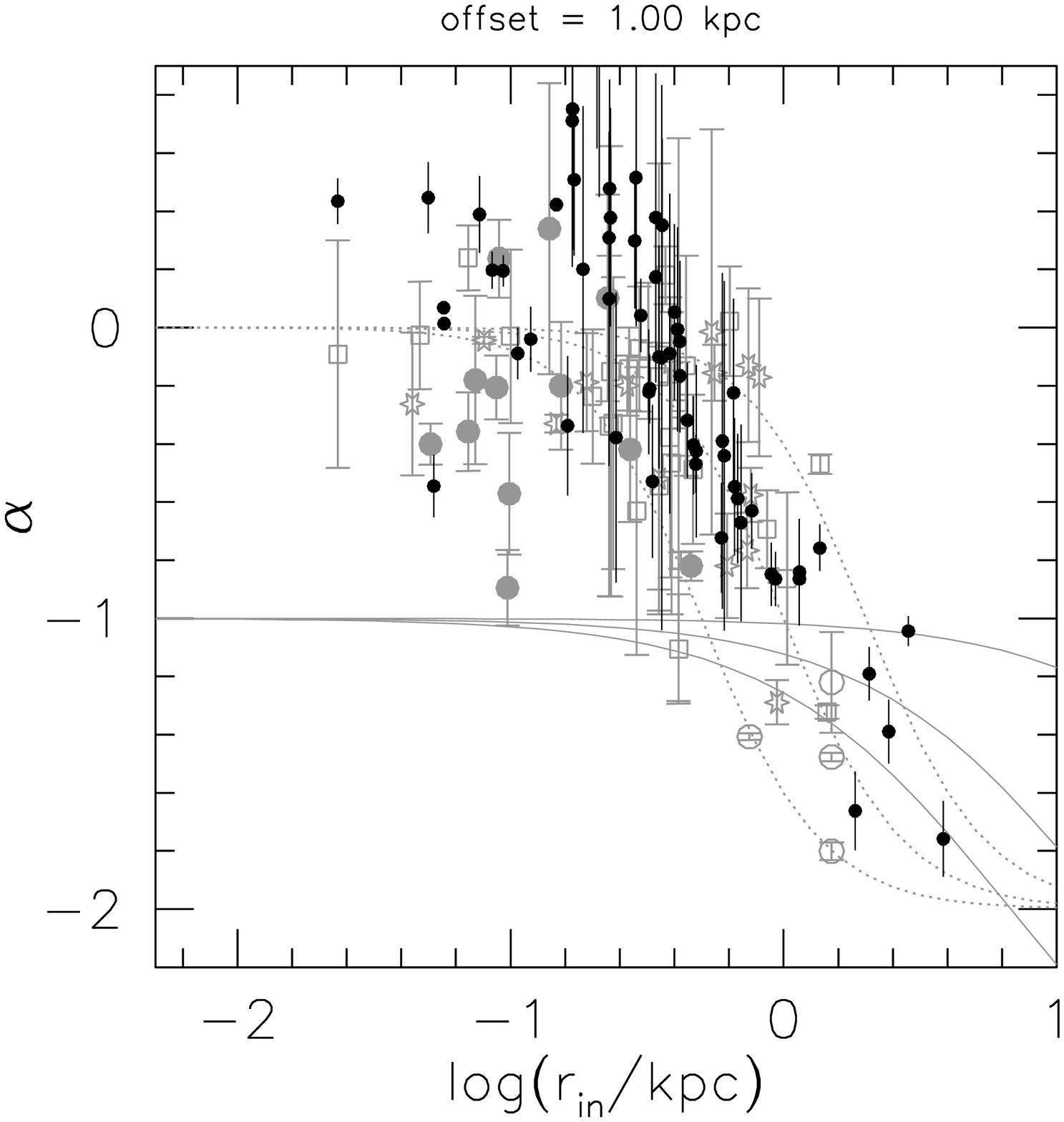}}
\hbox{\epsfxsize=0.32\hsize 
\epsfbox{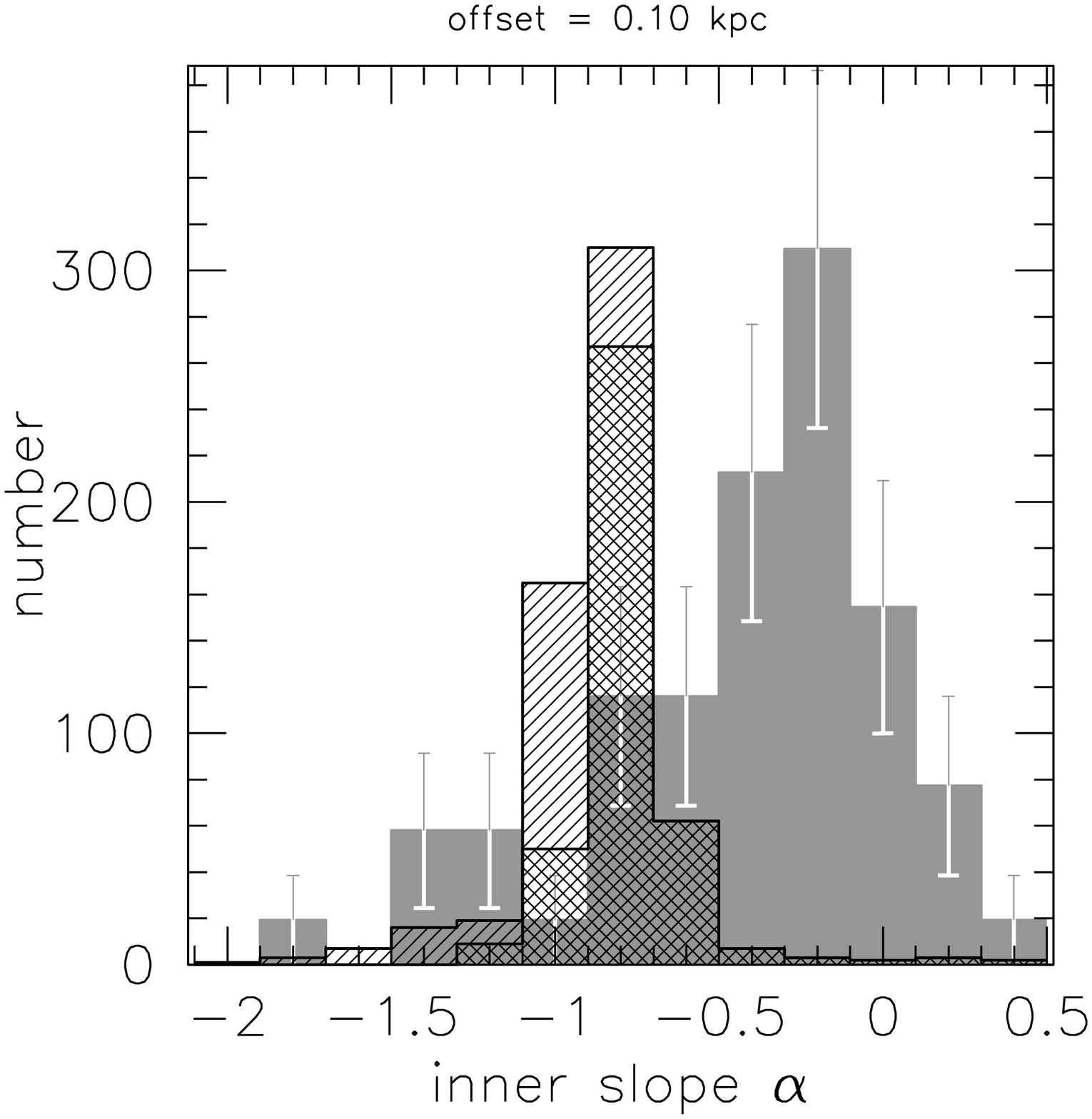}
\epsfxsize=0.32\hsize 
\epsfbox{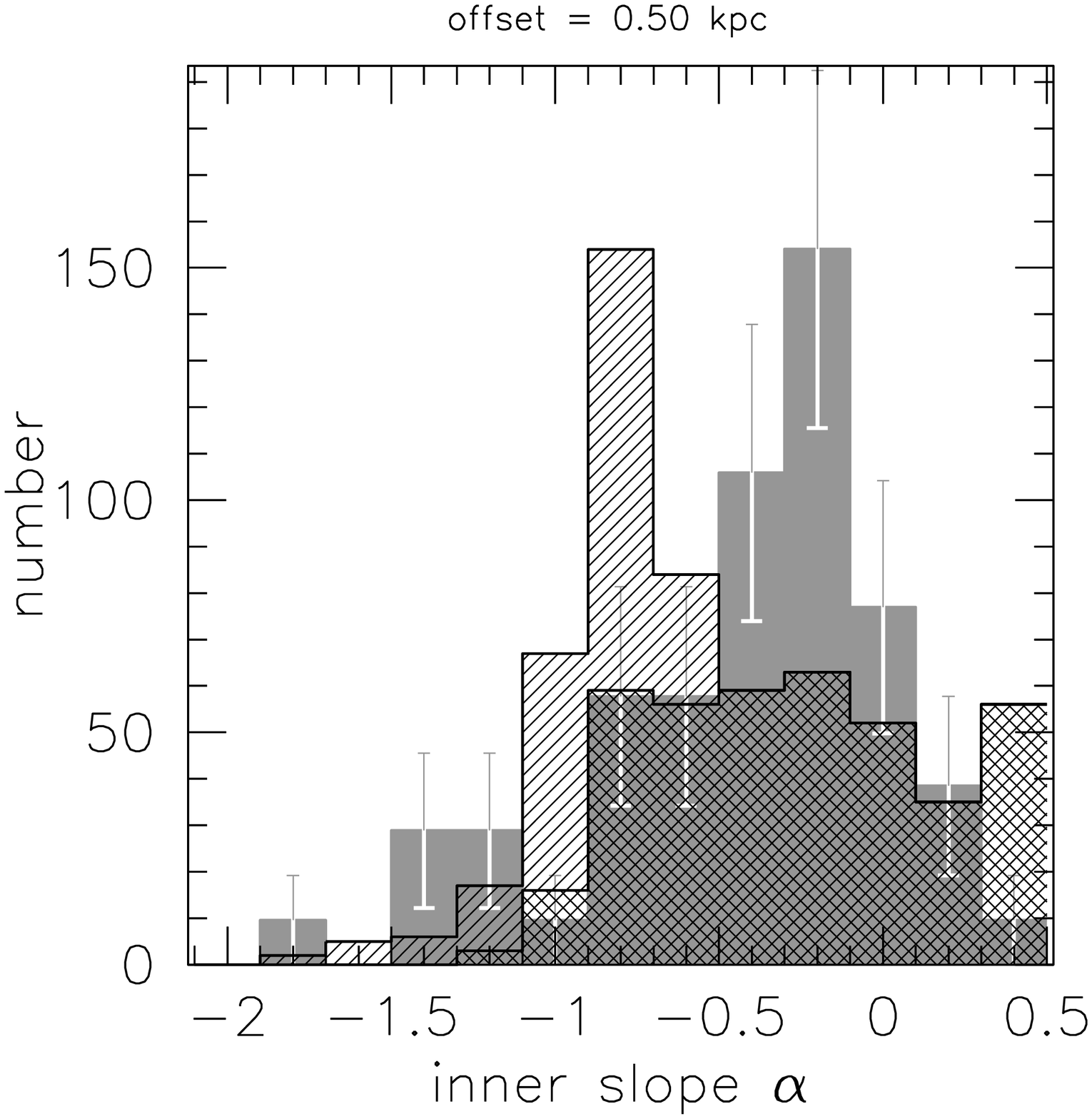}
\epsfxsize=0.32\hsize 
\epsfbox{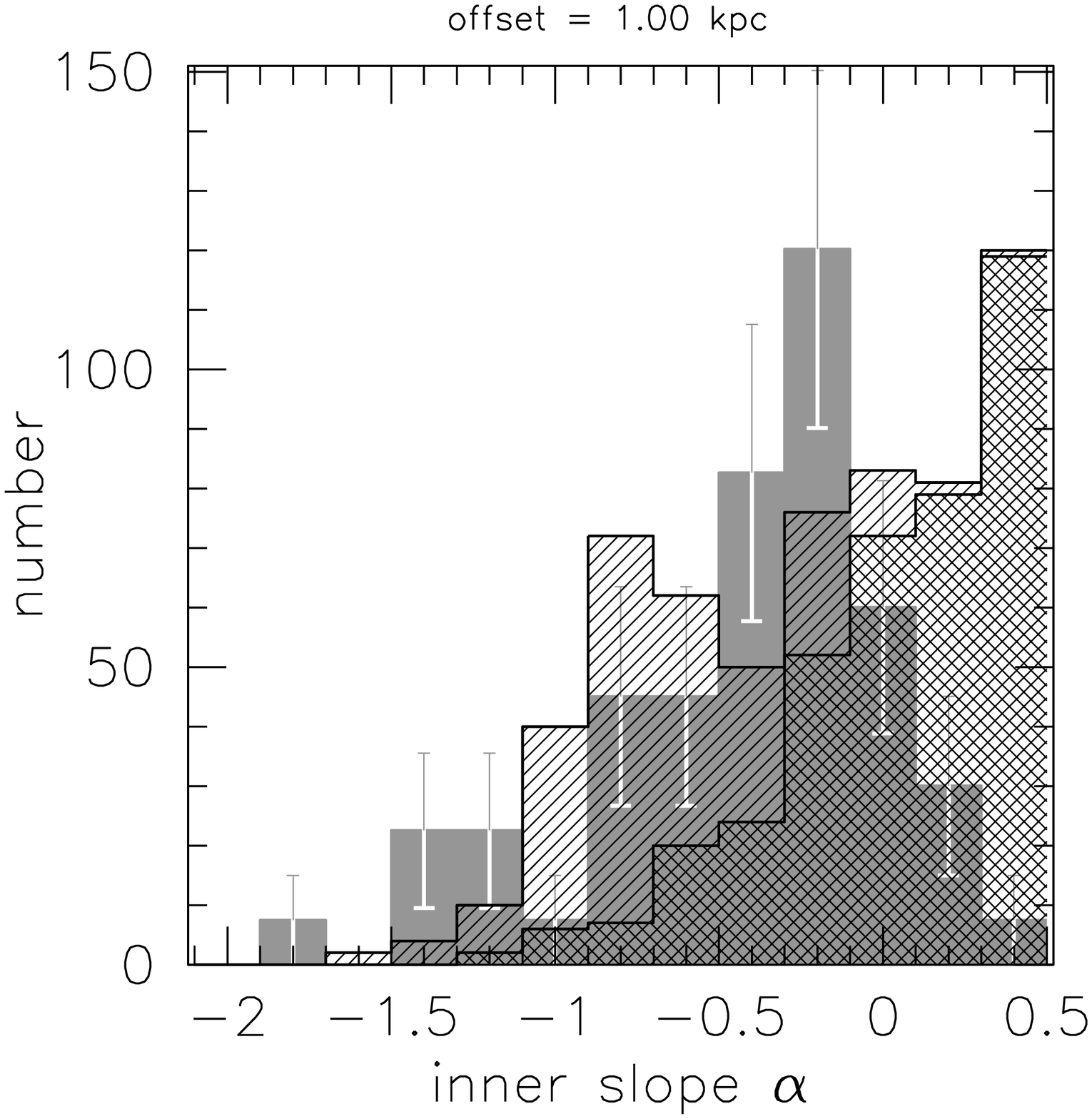}}
\hbox{\epsfxsize=0.32\hsize 
\epsfbox{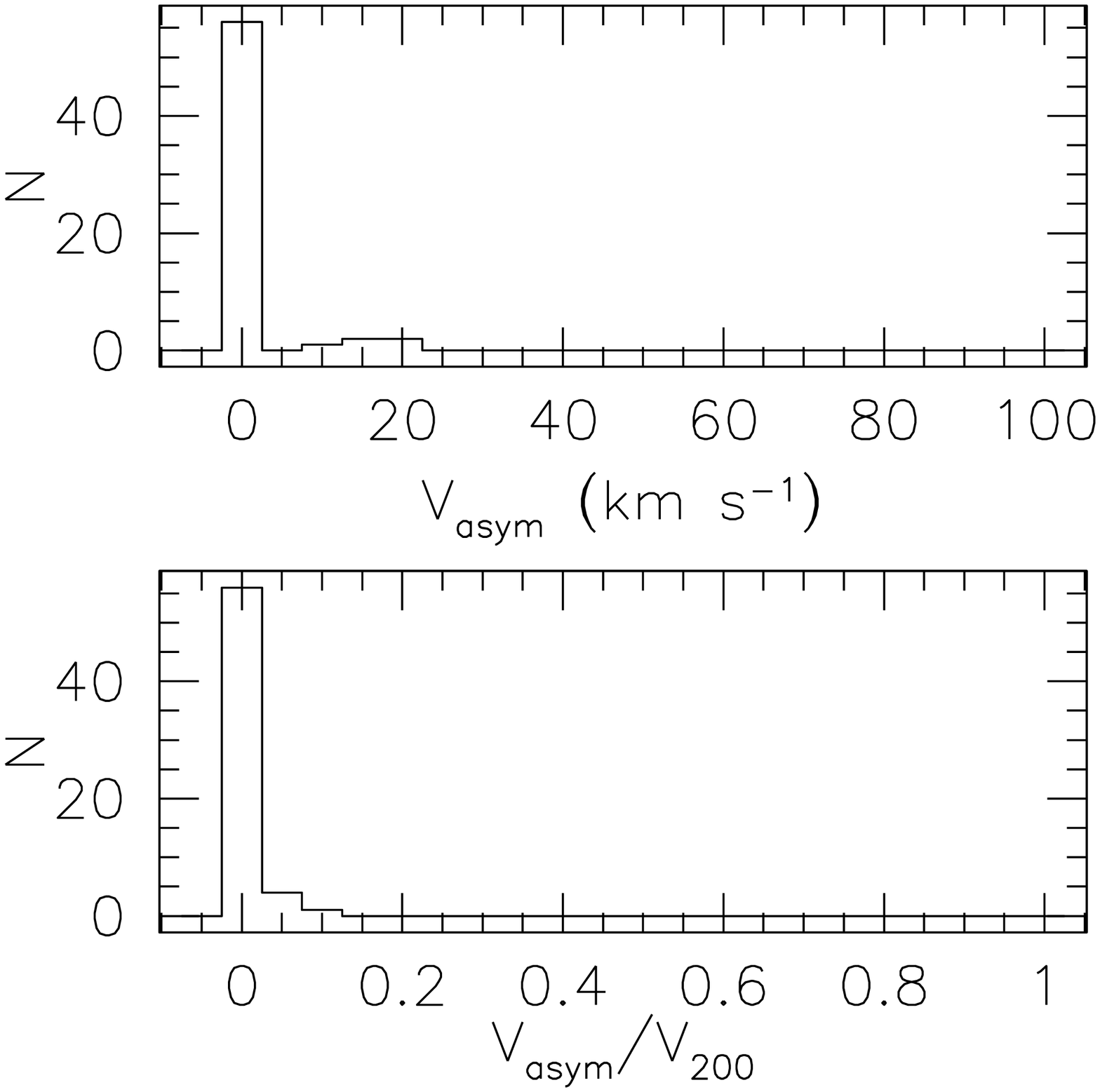}
\epsfxsize=0.32\hsize 
\epsfbox{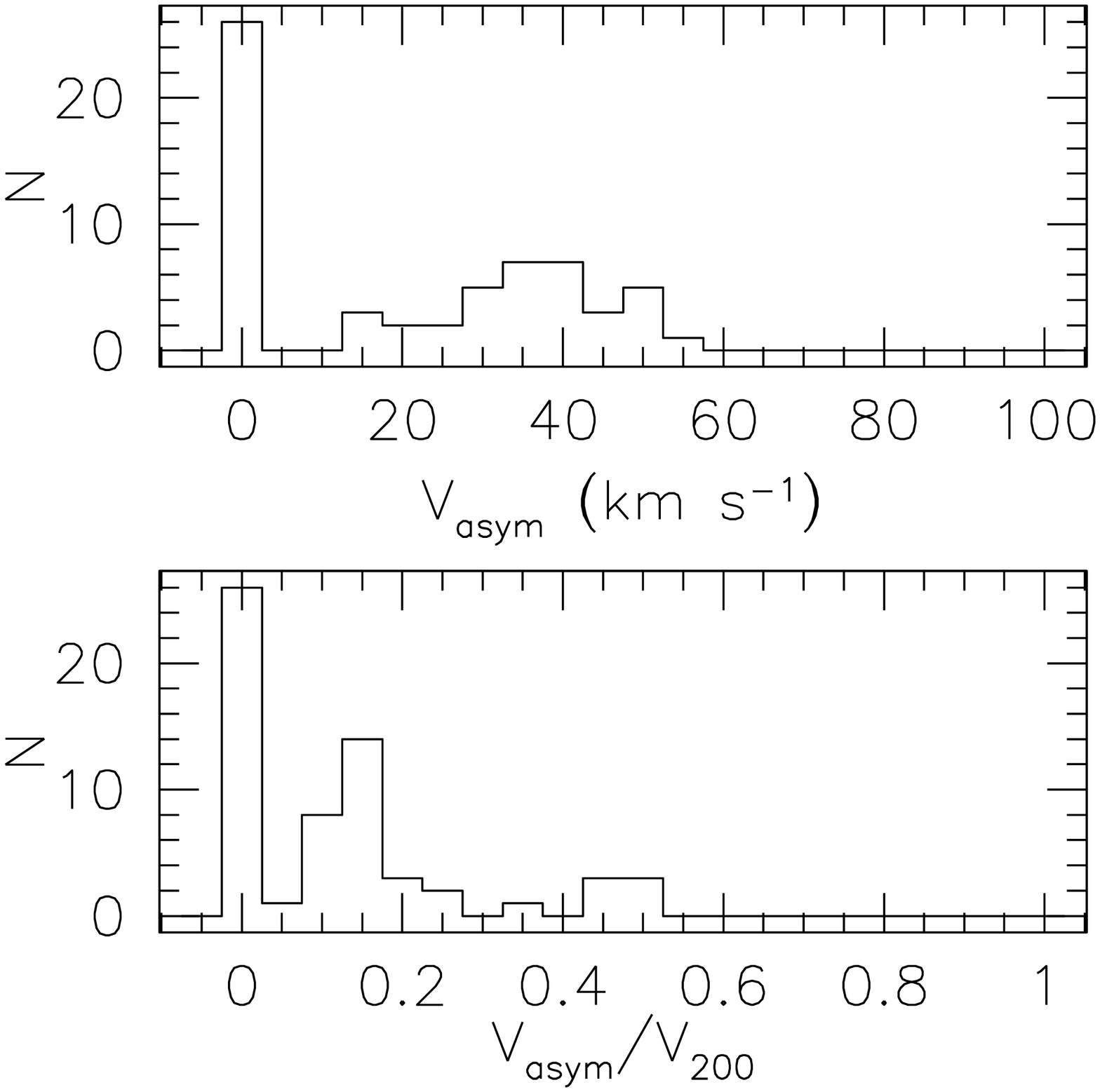}
\epsfxsize=0.32\hsize 
\epsfbox{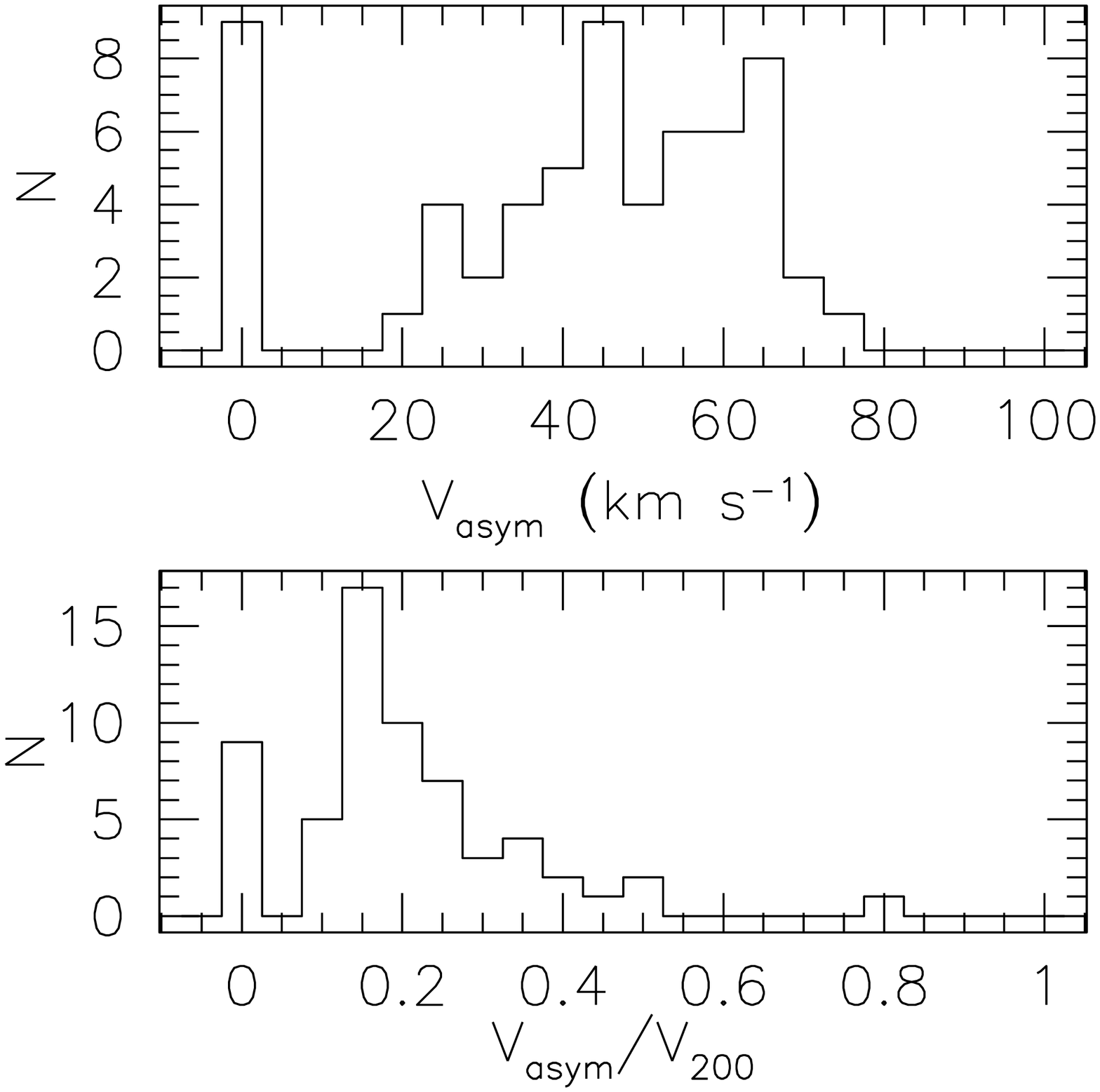}}
\caption[NFW_sym01_data.ps]{As Fig.~\ref{NFWISOnosys}, 
but now showing the results for NFW halos in combination with
kinematical lopsidedness of 0.1 kpc (left column), 0.5 kpc (centre
column) and 1 kpc (right column). The bin at $\alpha=0.5$ also
contains galaxies with $\alpha > 0.5$. The bottom rows shows the
distribution of the resulting velocity differences between approaching
and receding sides, both in absolute terms, and as a fraction of halo
rotation velocity $V_{200}$. The peak at $V=0$ is due to galaxies at
large distance or low inclinations where the kinematical lopsidedness
shift is much less than one resolution element.
\label{NFW_asym}}
\end{center} 
\end{figure*}

We assume offsets of 0.1, 0.5 and 1 kpc.  Note that though the
magnitude of the effect is expressed in kpc, we do not imply that this
involves a shift in dynamical centre. Our notation is merely a
convenient way to express the incorrect choice of systemic velocity
that we use to create the kinematical lopsidedness. The offsets used
here result in average in velocity offsets of $\sim 15$, $\sim 30$ and
$\sim 50$ \kms\ respectively (see bottom row in
Fig.~\ref{NFW_asym}).  We also assume that there are no additional
offsets between optical and dynamical centres.

Figure \ref{NFW_asym} shows the results for the NFW halos.  We see
that lopsidedness is not important as long as the effect occurs on
scales smaller than the resolution. A lopsidedness of between $\sim
0.5$ and $\sim 1$ kpc (which translates into an average velocity shift
of $\sim 40$
\kms) is needed in order to bring most of the NFW points into the
$\alpha=0$ part of the diagram. The peak at $\alpha=-1$ in the 0.1 kpc
case is caused mainly by well-resolved galaxies. Small offsets are
thus inconsistent with NFW models. This peak is still present in the
0.5 kpc case, but is now mainly caused by lower resolution
galaxies. The well-resolved galaxies are distributed more like the
observed histogram, but with an excess of galaxies with positive
slopes. The spread in slopes at these higher resolutions is much
larger than observed.  Large offsets of $\sim 1$ kpc are inconsistent
with the observed distribution as nearly all well-resolved galaxies
exhibit too positive slopes.

The bottom-row in Fig.~\ref{NFW_asym} show the distribution of the
velocity differences between approaching and receding side (not
corrected for inclination) for the various offsets.  Also shown is the
distribution of the same differences but as a fraction of the halo
velocity $V_{200}$.

In order to explain the data with NFW models, one thus needs to assume
a net lopsidedness for \emph{all} galaxies of $\sim 30$ to $\sim 60$
\kms.  Estimates of the fraction of galaxies that are kinematically
lopsided reach up to 30 to 50 per cent
\citep{swaters_lop}, but with much smaller velocity differences. 
It is thus not possible to explain the data with CDM models in
combination with lopsidedness alone.  We do not show the ISO halo
results: ISO halos already show a slope that is slightly more positive
than the data (Fig.~\ref{NFWISOnosys}), adding in extra flattening due
to lopsidedness results in even more positive slopes. ISO halos are
thus not consistent with a large (i.e.\ shift in symmetry point $\ga
1$ kpc) kinematical lopsidedness.

\subsection{Non-circular motions\label{noncirc}}

A frequently made assumption is that the visible matter in a galaxy is
on circular orbits, and therefore that the measured line-of-sight
velocities represent the true rotation velocity.  If, however, part of
the measured velocity is due to non-circular motions, it
is no longer representative indicator of the mass content of a galaxy.
Systematic non-circular motions are most commonly observed as
streaming motions along bars or spiral arms (we do not consider local
non-circular motions due to e.g.\ star formation or supernovae, as explained in Sect.~\ref{sec:simul} and \ref{nosys}).
Bright galaxies can show large streaming motions: e.g.\ $\sim 30$
\kms\ in M81 \citep{ad&westpf} or $\sim 20 $\kms\ in NGC 1365
\citep{joer}. These galaxies are however mostly
grand-design spirals with pronounced spiral structures. Streaming
motions in late-type spirals such as M33, NGC 300, NGC 925 or NGC 1744
are found to be much smaller, and barely distinguishable from random
motions in the disk (i.e.\ $\la 10$ \kms)
\citep{deul,pcb90,DJ98}.  LSB galaxies are similar to these late-type
galaxies and would not be expected to have large circular motions. \HI
velocity fields also show no evidence for large systematic
non-circular motions $\ga 10$ \kms\ \citep{edb_bmh96,Weldrake02}.

We model non-circular streaming motions by introducing a radial
velocity component (see Eq.~\ref{eq:rotcur}). The
effects of streaming motions on a rotation curve  are strongest when
they are oriented parallel to the minor axis. We approximate this by
adding a constant velocity component to the rotation curve over the
inner one-third of its radius (the exact range is not important as we
use only the inner points to determine the mass-density profile). We
thus use Eq.~\ref{eq:rotcur} but change the phase of the radial
component by $\pi/2$.  Again we only show the results for the NFW
halos, as the ISO halos are only consistent with no or very small
non-circular motions, for the same reason as with kinematical
lopsidedness: their profiles are already slightly flatter than the
data, and more flattening is not needed.

\begin{figure*} 
\begin{center}
\hbox{\epsfxsize=0.32\hsize 
\epsfbox{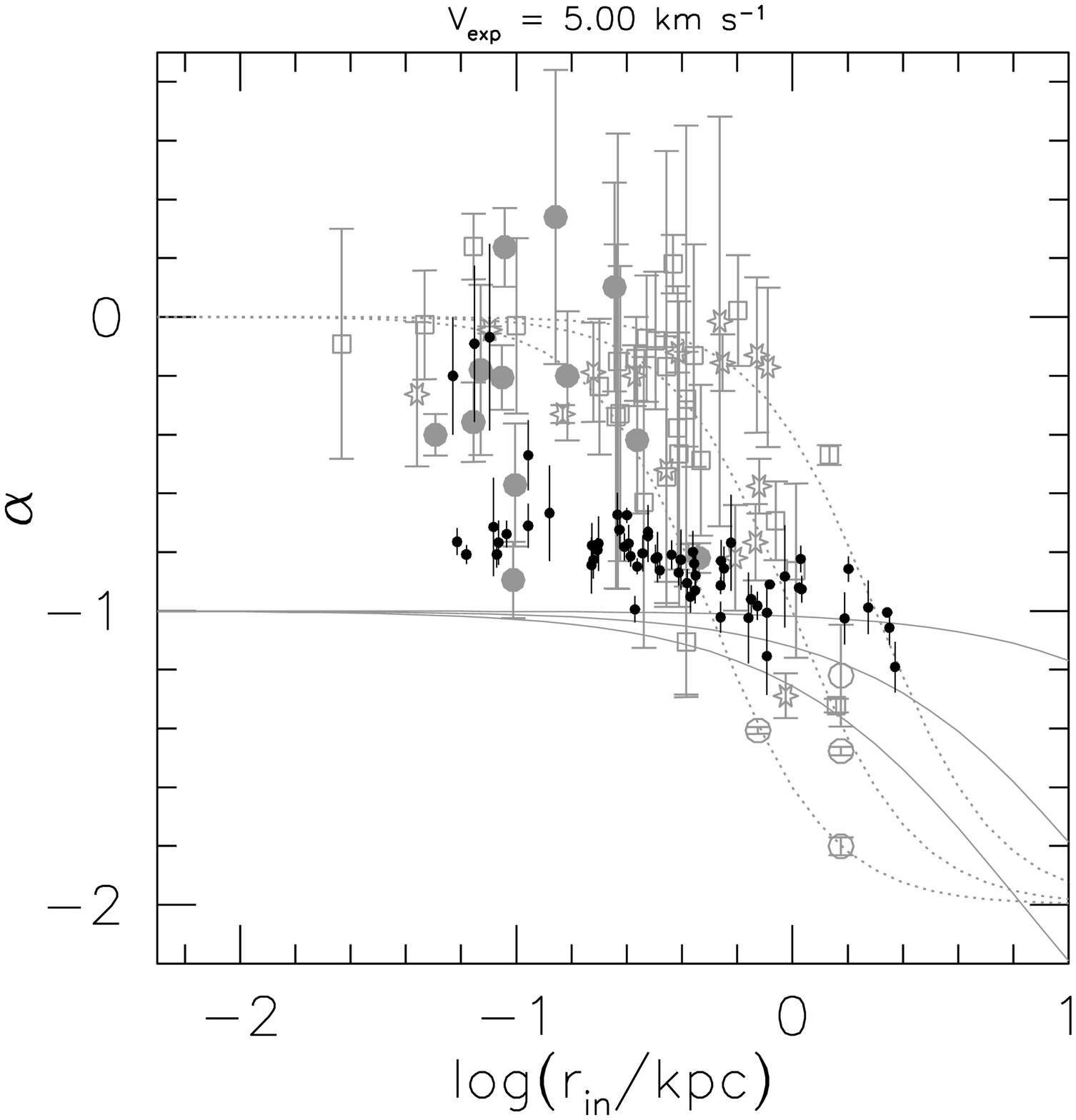}
\epsfxsize=0.32\hsize 
\epsfbox{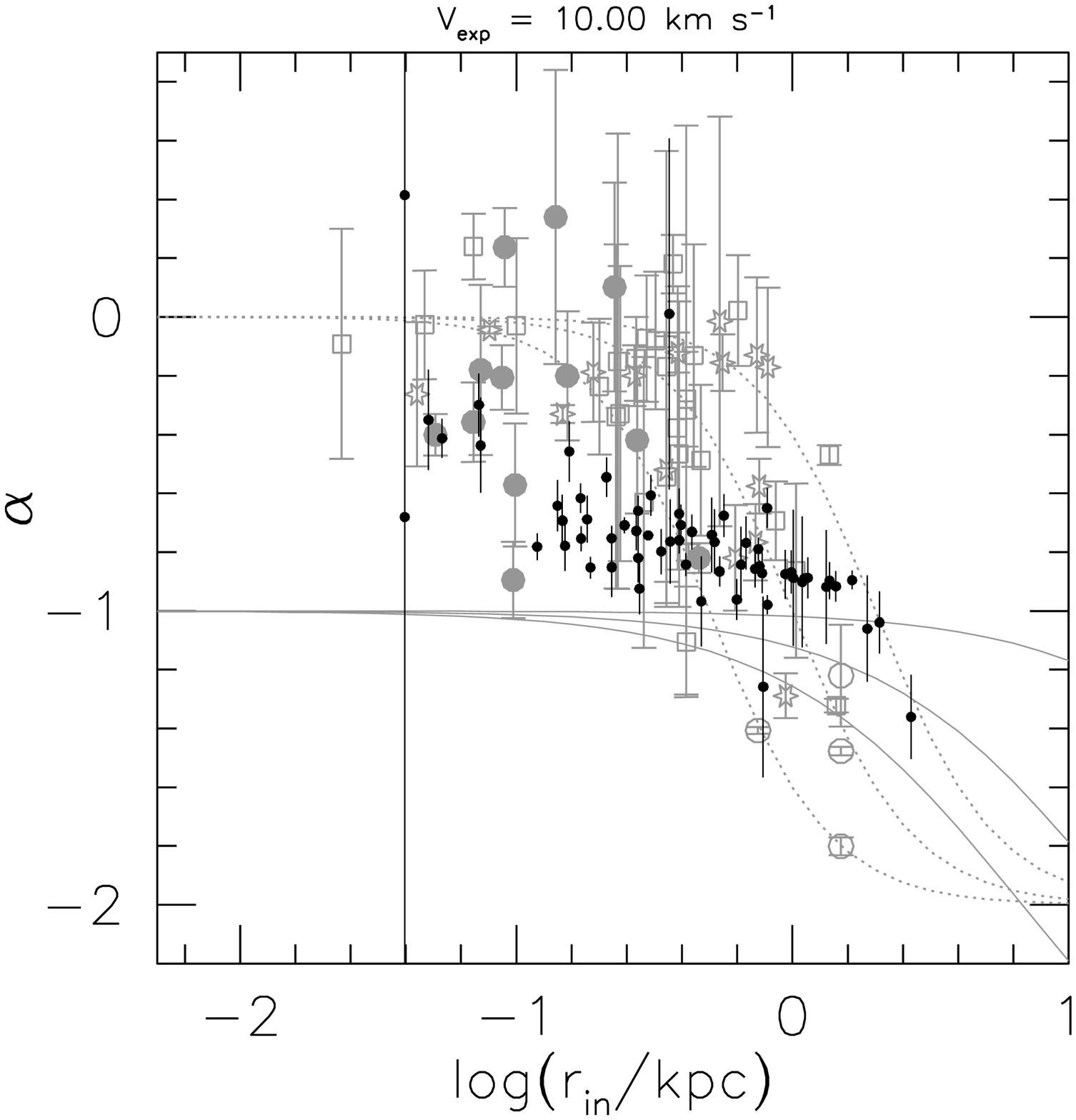}
\epsfxsize=0.32\hsize 
\epsfbox{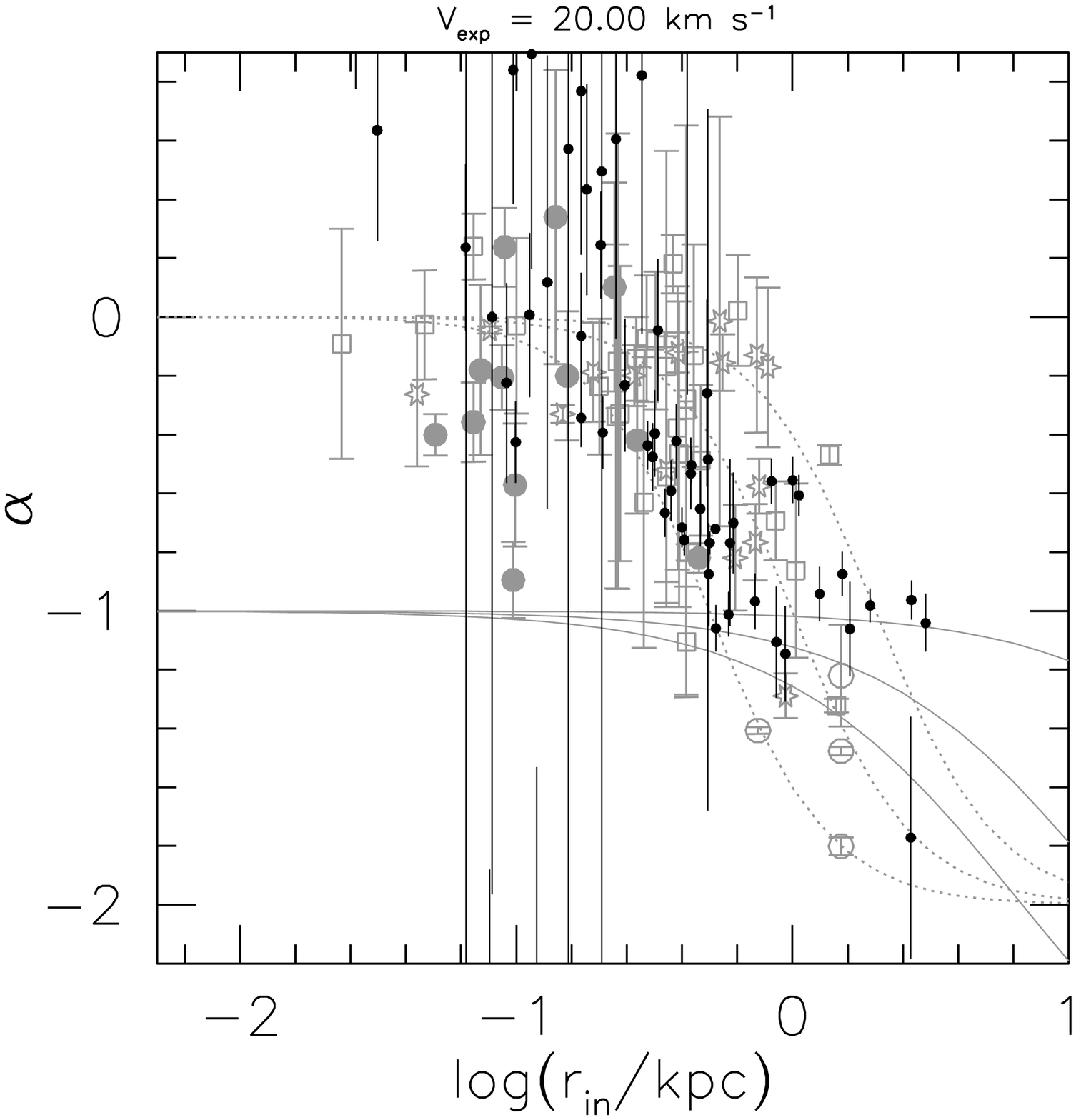}}
\hbox{\epsfxsize=0.32\hsize 
\epsfbox{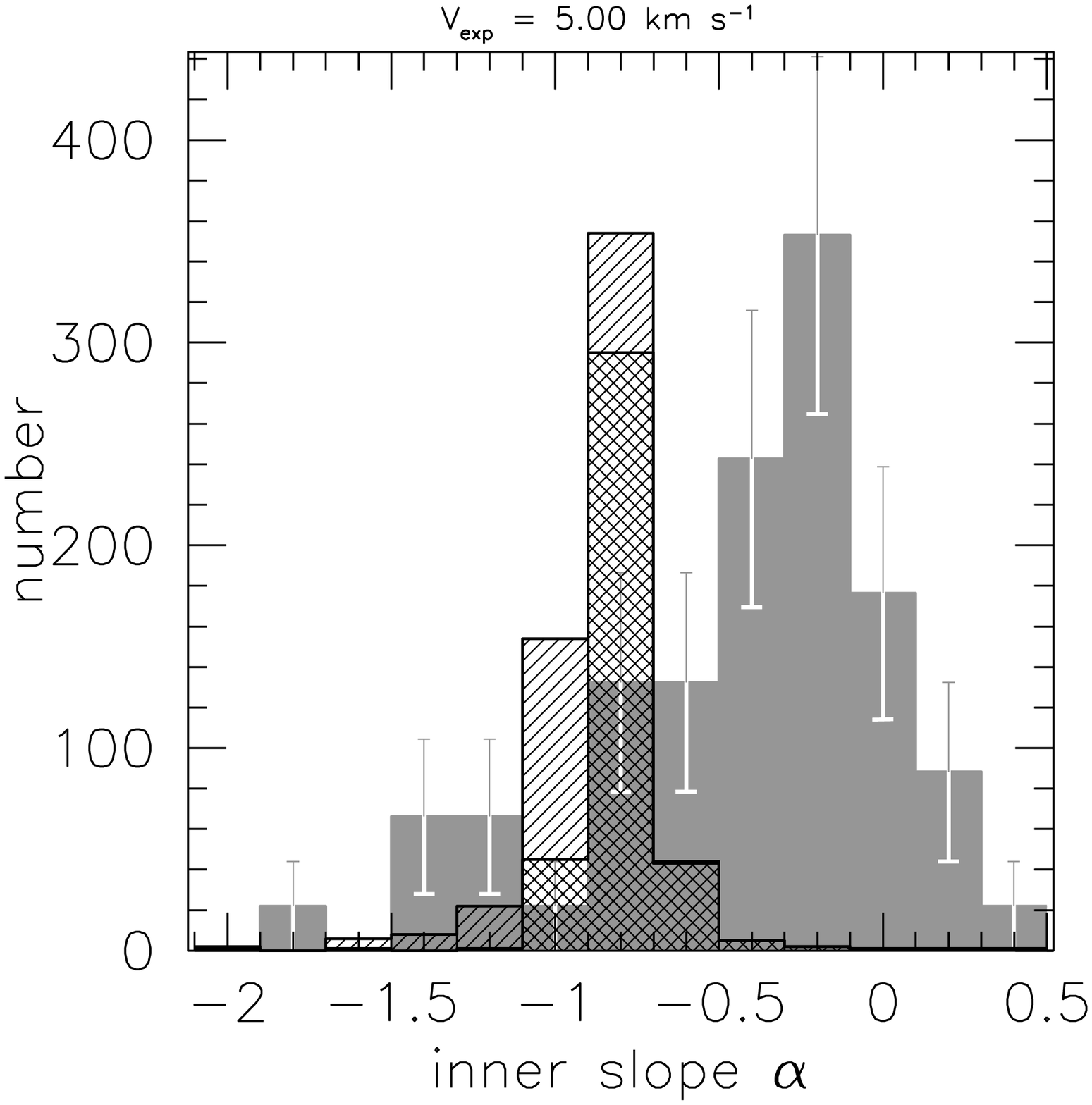}
\epsfxsize=0.32\hsize 
\epsfbox{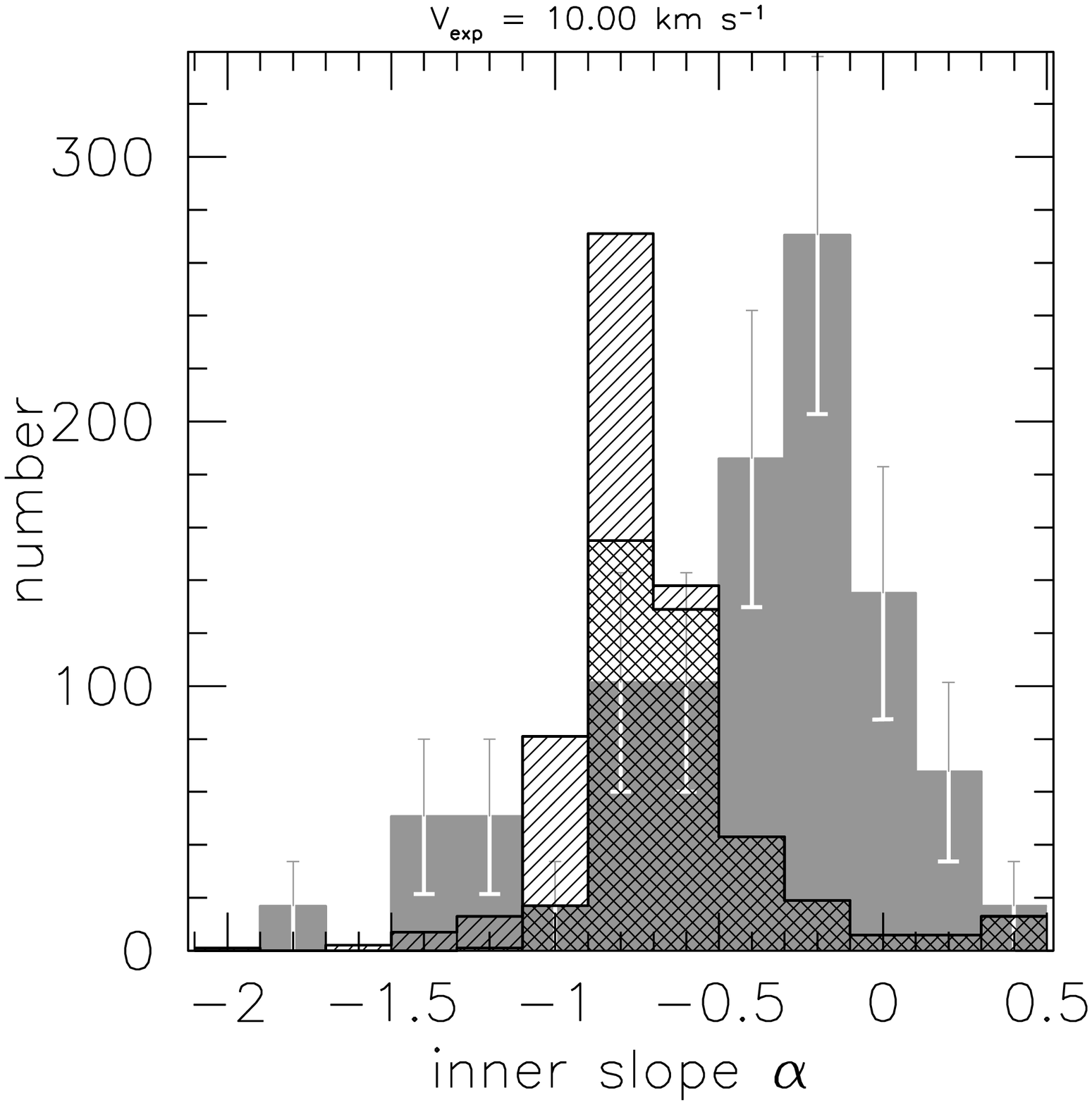}
\epsfxsize=0.32\hsize 
\epsfbox{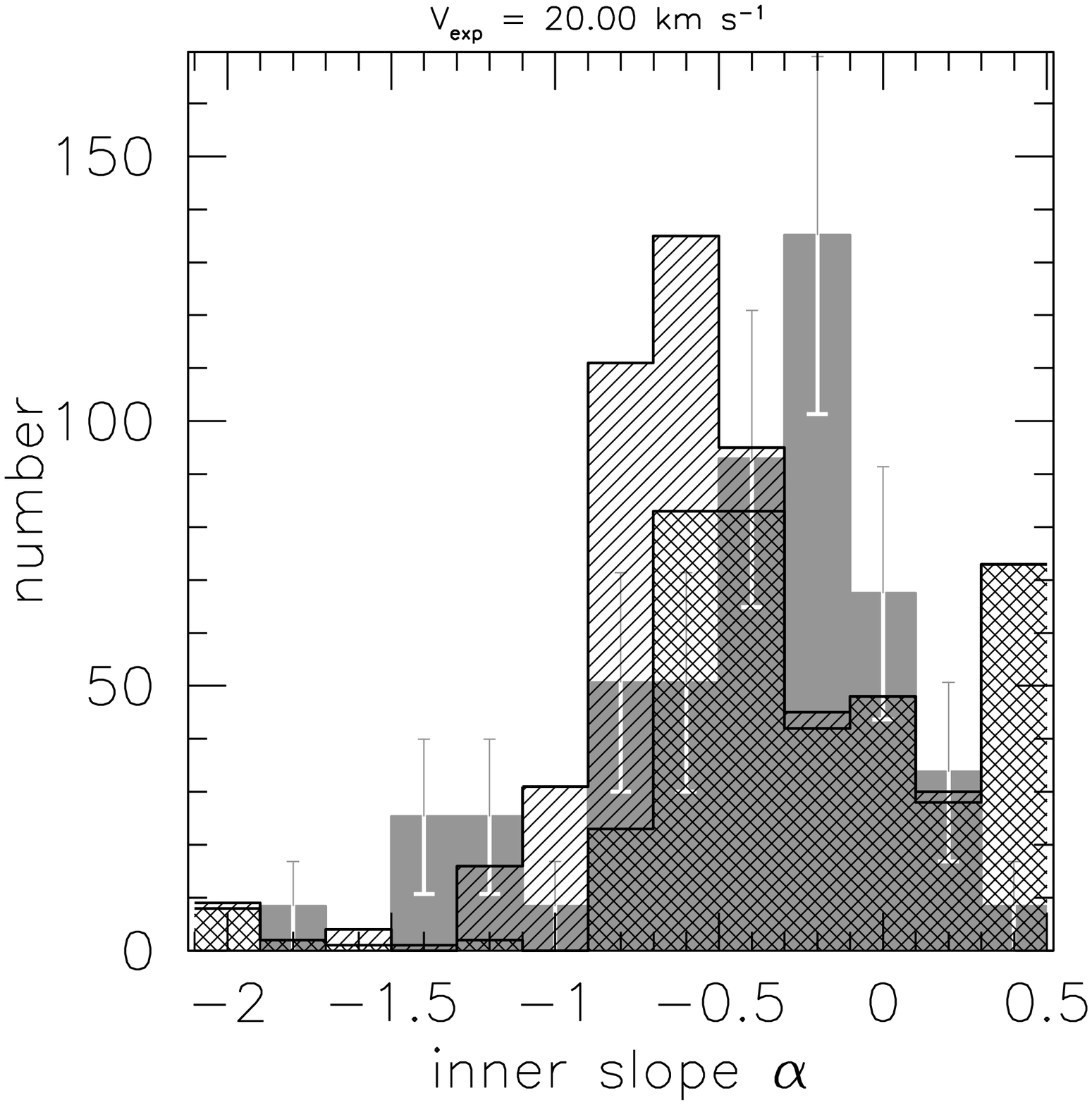}}
\hbox{\epsfxsize=0.32\hsize 
\epsfbox{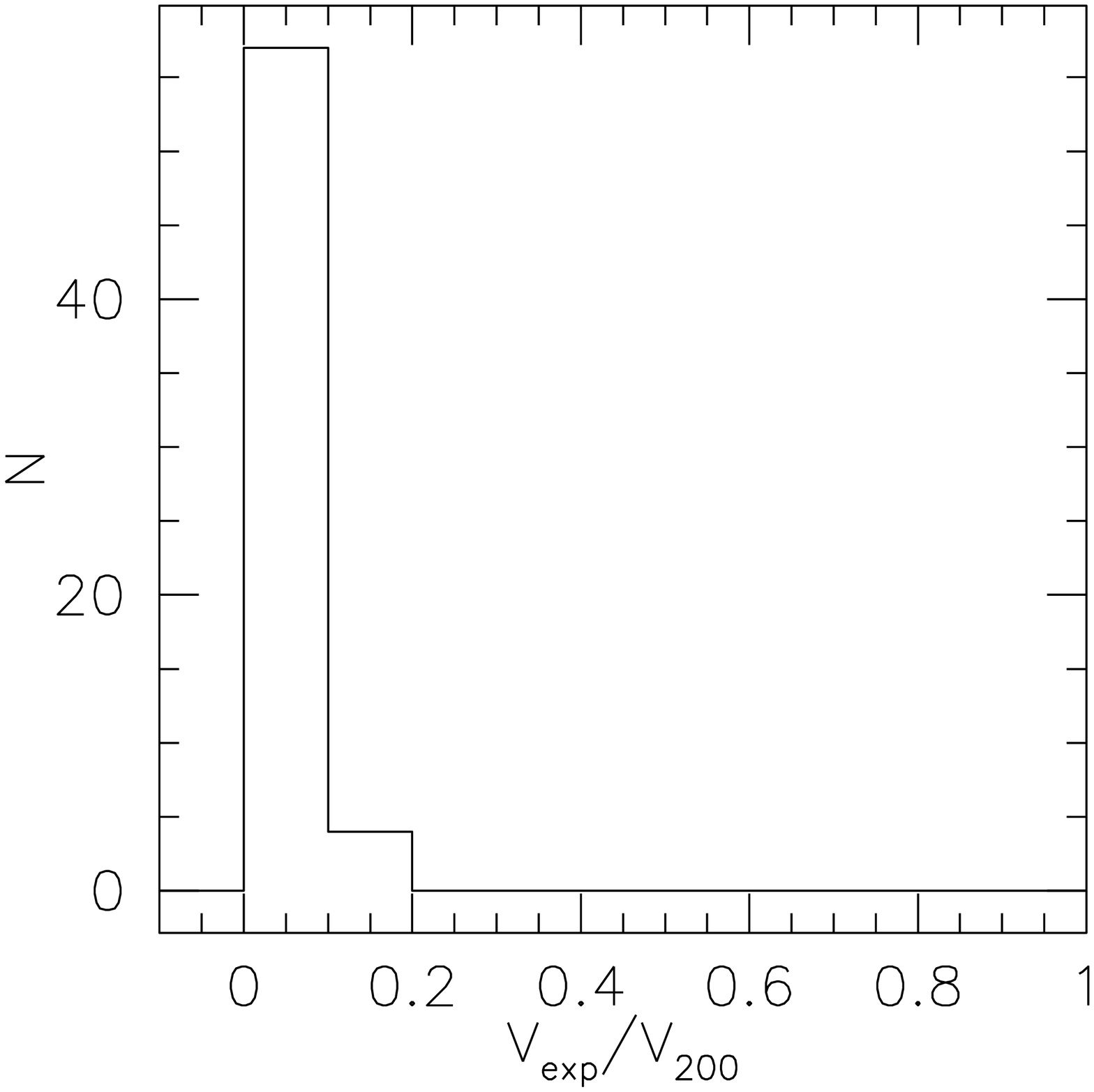}
\epsfxsize=0.32\hsize 
\epsfbox{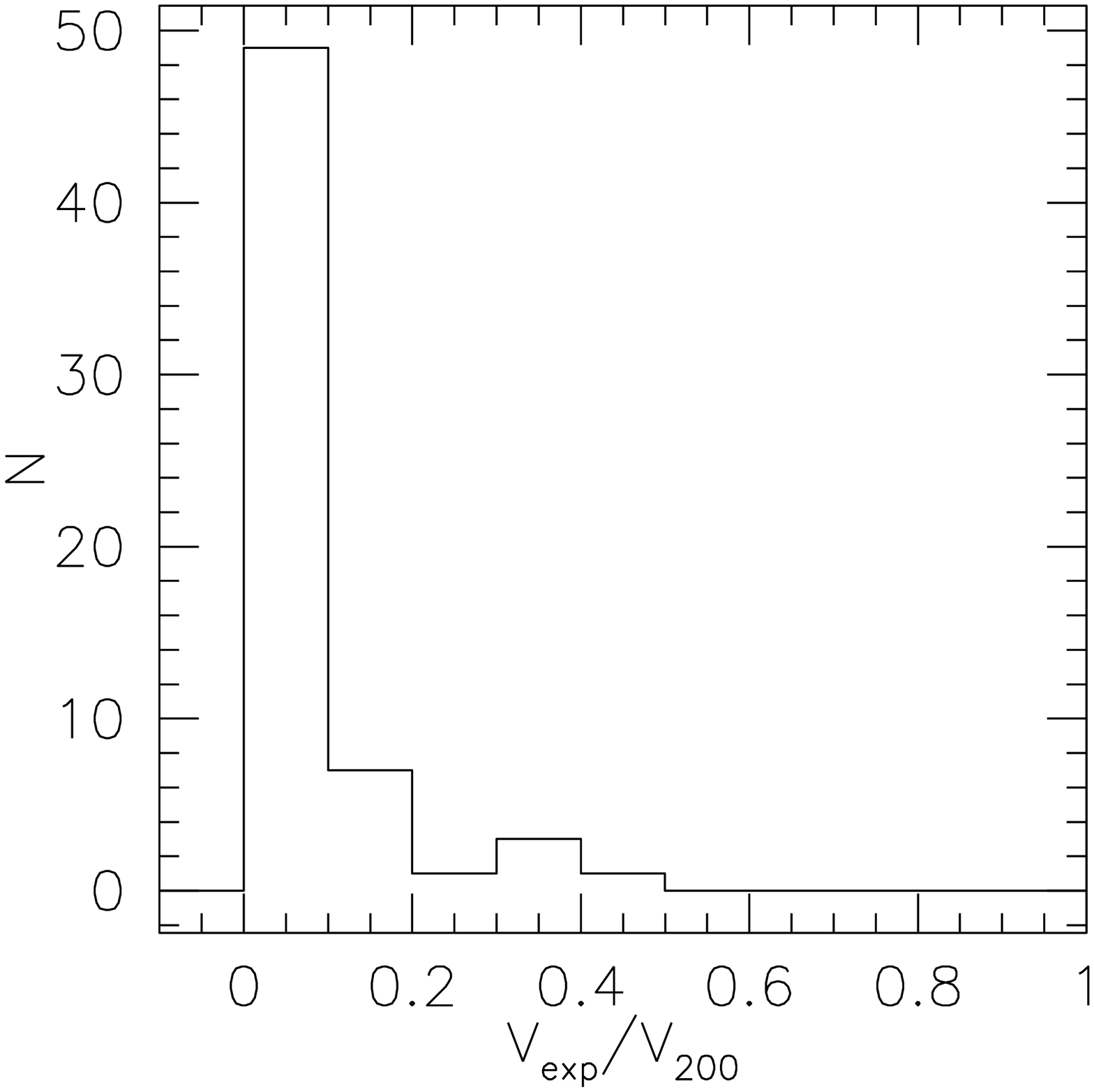}
\epsfxsize=0.32\hsize 
\epsfbox{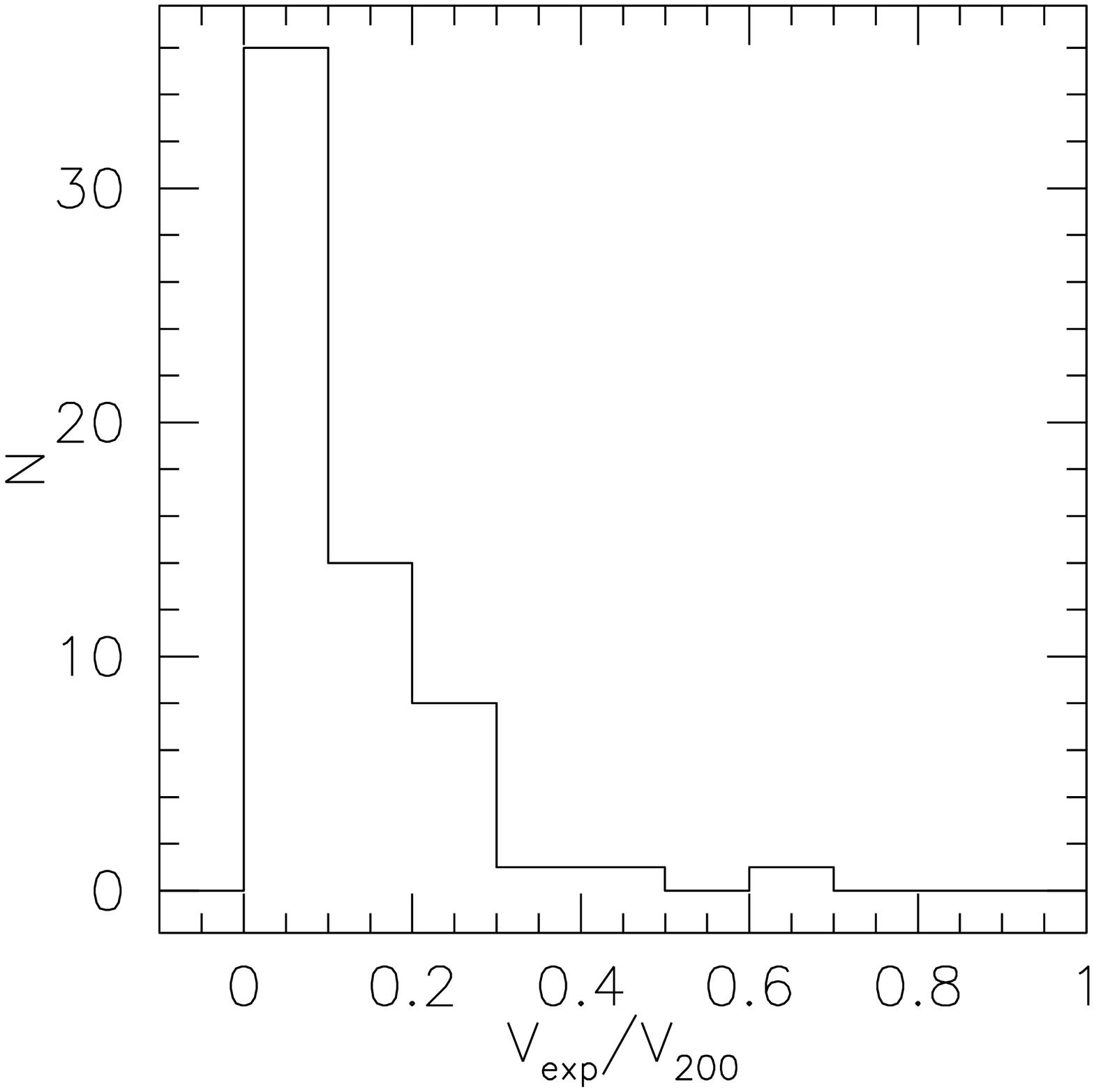}}
\caption[NFW_sym01_data.ps]{As Fig.~\ref{NFWISOnosys}, but now showing the results for NFW halos in
combination with streaming motions parallel to the minor axis. The
left panel shows results for 5 \kms, centre for 10 \kms and right
panels for 20 \kms.  The bin at $\alpha=0.5$ also contains galaxies
with $\alpha > 0.5$.  The bottom row shows the distribution of the
magnitude of the streaming motion as a fraction of the halo velocity
$V_{200}$.
\label{streaming}}
\end{center} 
\end{figure*}

We have computed models for streaming velocities of 5, 10 and 20 \kms.
Figure~\ref{streaming} shows that systematic streaming motions of
order $\sim 20$ \kms\ are needed to at least partially explain the
data for the well-resolved galaxies with a NFW halo. 
There is however still an 
excess of data points with slope $\alpha = -1$ formed of
low-resolution galaxies, as well as an excess of positive slopes due
to well-resolved galaxies. The bottom row in Fig.~\ref{streaming}
shows that for a significant fraction of the galaxies a streaming
motion of this magnitude amounts to $\sim 20$ per cent or more of the
halo velocity $V_{200}$. In order to use streaming motions in
combination with NFW halos, one thus needs to assume that
\emph{all} galaxies have systematic non-circular motions over a large
part of the disk of order $\sim 20$ \kms. This contradicts the
observational data for late-type galaxies.

\subsection{Mismatched position angles}

A mis-match between the true position angle and the observed position
angle does not lead to a change in slope for axisymmetric potentials.
As can be seen from Eq.~\ref{eq:rotcur} it will merely introduce a
scale-factor of order $(\cos \theta/\cos \theta')$ where $\theta$ and
$\theta'$ are the true and mis-matched position angles respectively.
In the absence of non-circular motions and in axisymmetric potentials
one can even correct for a mis-matched position angle. If the
mass-distribution in the central parts is not axisymmetric, then the
precise value of the position angle becomes important. However, in
these cases the inner parts of the galaxy are likely to be dominated
by the stellar component, or the halo is very triaxial. In these cases
the minimum disk and spherical halo assumptions are no longer
valid. One needs to make self-consistent
multi-component galaxy models, and can no longer use the simple
one-component models we have been using.

\subsection{Offsets between dynamical and optical centres}

It is generally assumed that in a galaxy the optical centre and the
dynamical centre coincide. If, however, this is not the case, then, at
least for NFW halos, measuring the kinematics around the optical centre
will lead to the impression that the slope of the inner mass density
profile is shallow rather than steep.

It is known that in some late-type (barred) galaxies the optical centre can
be located up to $\sim 1.5$ kpc away from the dynamical centre
\citep{DJ98,deVauc72,Weldrake02}. This ``morphological lopsidedness'' 
is different from the kinematical lopsidedness discussed before, in
that as long as the stellar population is unimportant, the potential
can still be dominated by the halo and be axisymmetric around the
dynamical centre.  It should be stressed though, that for the majority
of bright galaxies where the kinematics have been investigated in
detail, the optical centre and dynamical centre do coincide
(cf.\ \citealt{beeg}). In these galaxies the slit would sample the
true dynamical centre.

To model the effect of offsets between dynamical and (observed)
optical centers, we run the simulations assuming a distribution of
offsets and observe the halos with a slit offset from the dynamical
center (see Fig.~\ref{geometry}).  We assume Gaussian distributions of offsets with dispersions
$\sigma =$ (0.5, 1, 2, 3, 4, 5, 6) arcseconds. For each of these seven
cases we run the simulations twice, once producing 600 NFW halos and
once producing 600 ISO halos.  As an offset between dynamical and
optical center is equivalent to a slit offset, we draw a random slit
offset from the relevant Gaussian offset distribution for each halo
(i.e.\ the first case produces 600 NFW and 600 ISO halos, each
observed with offset slits, where the distribution of slit offsets was
Gaussian with a dispersion $\sigma=0.5''$,
and so on for all other values of the dispersion).  We have also
modeled uniform distributions using widths identical to the
dispersions listed above, with identical conclusions as the Gaussian
distributions.

\subsubsection{NFW halos}

\begin{figure*} 
\begin{center}
\hbox{\epsfxsize=0.9\hsize 
\epsfbox{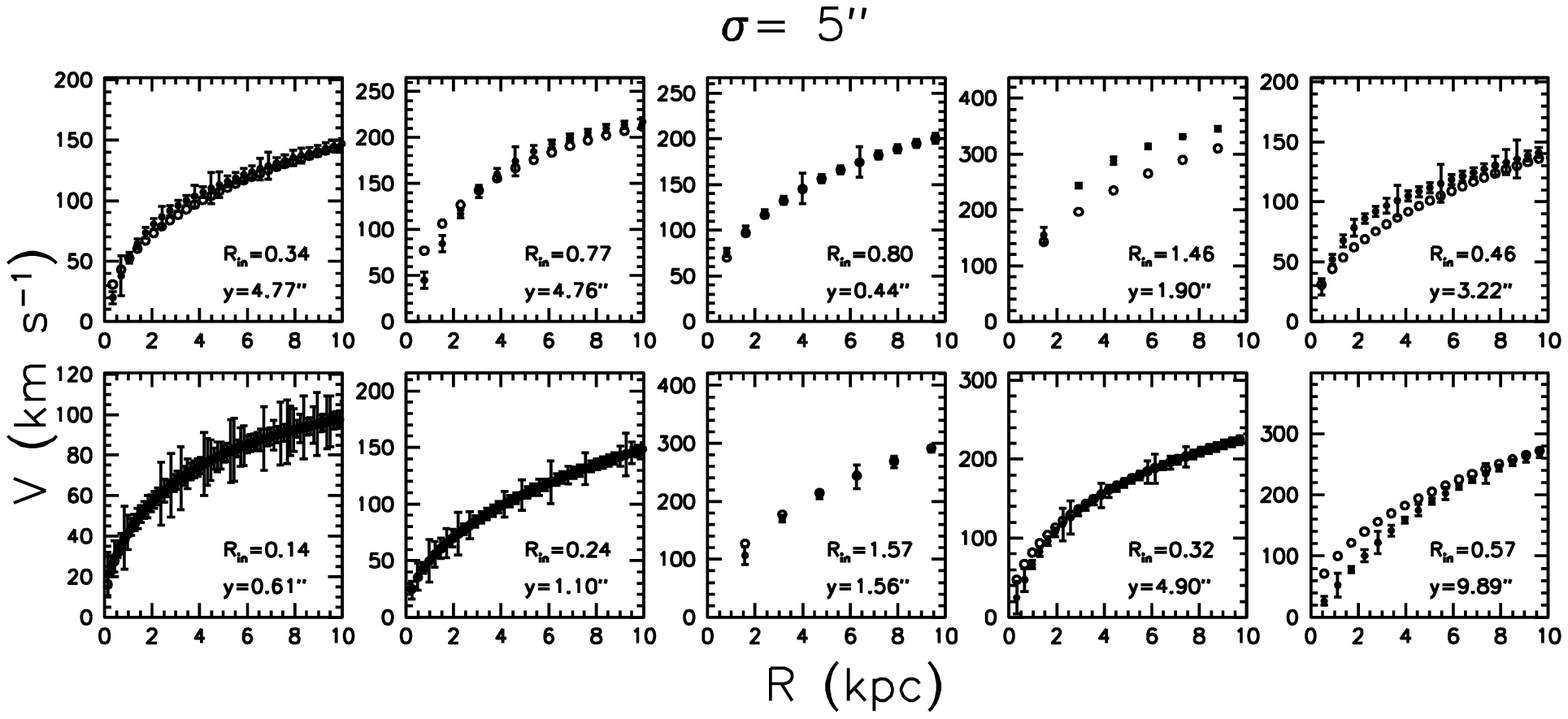}}
\hbox{\epsfxsize=0.9\hsize 
\epsfbox{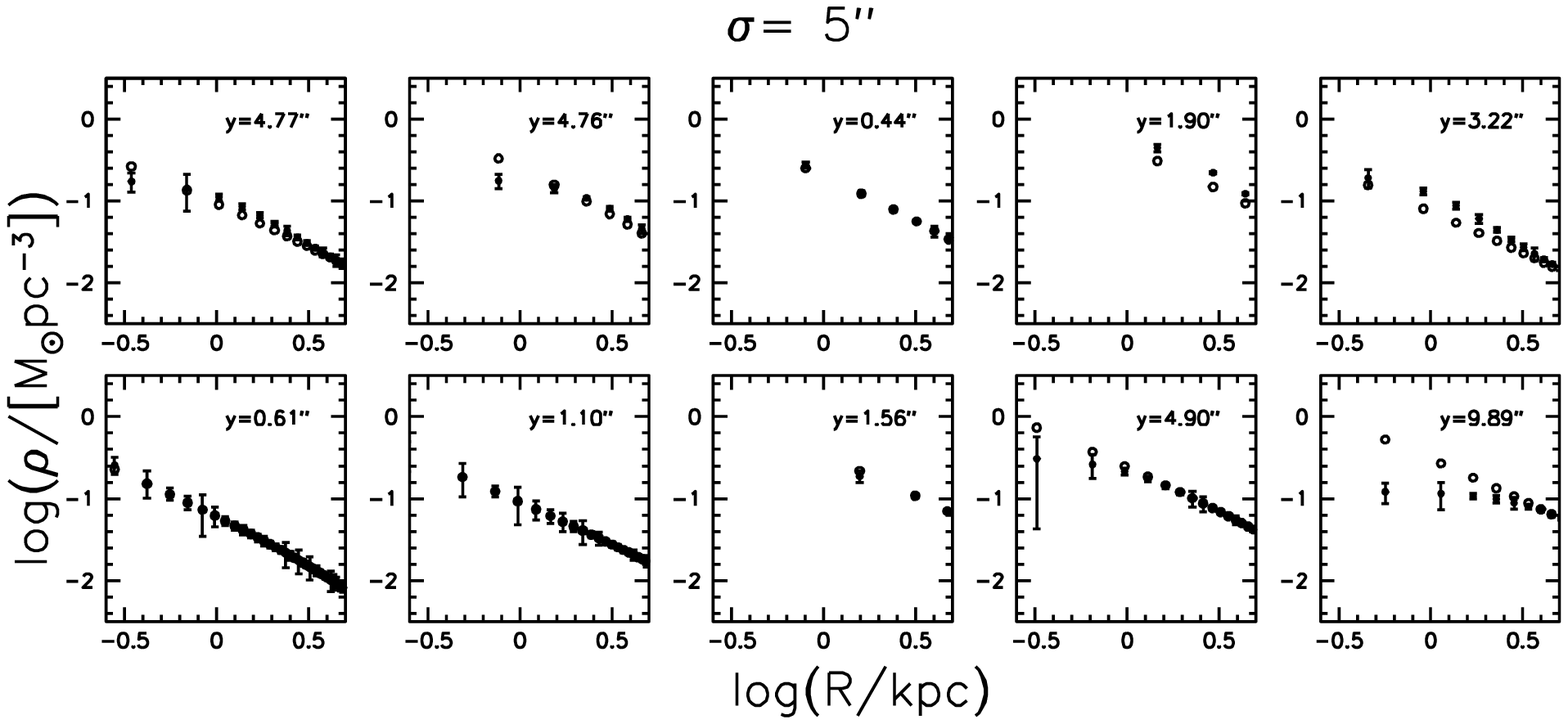}}
\caption[curves.ps]{A random selection of simulated NFW rotation curves 
(top panel) and corresponding density profiles (bottom panel). The
open circles show the original major axis curves and profiles, while
the filled circles with error-bars show the off-axis curves and
profiles. A Gaussian scatter with $\sigma = 5''$ was used for these
curves.  The actual value is given in the sub-panels, along with the
resolution (separation between points) in kpc.
\label{NFWcurves}}
\end{center} 
\end{figure*}

In Fig.~\ref{NFWcurves} a random selection of 10 rotation curves
derived assuming a Gaussian offset distribution with $\sigma=5''$ is
shown in comparison with the corresponding no-offset rotation curves.
Also shown are the corresponding mass density profiles.  For the
average distance of the observed sample of $\sim 73$ Mpc, this
dispersion corresponds to a physical dispersion of $\sim 1.8$ kpc.  We
use such a large offset dispersion to clearly show the differences
between the curves, which would hardly be visible for smaller
dispersions.

\begin{figure*} 
\begin{center}

\hbox{\epsfxsize=0.2\hsize 
\epsfbox{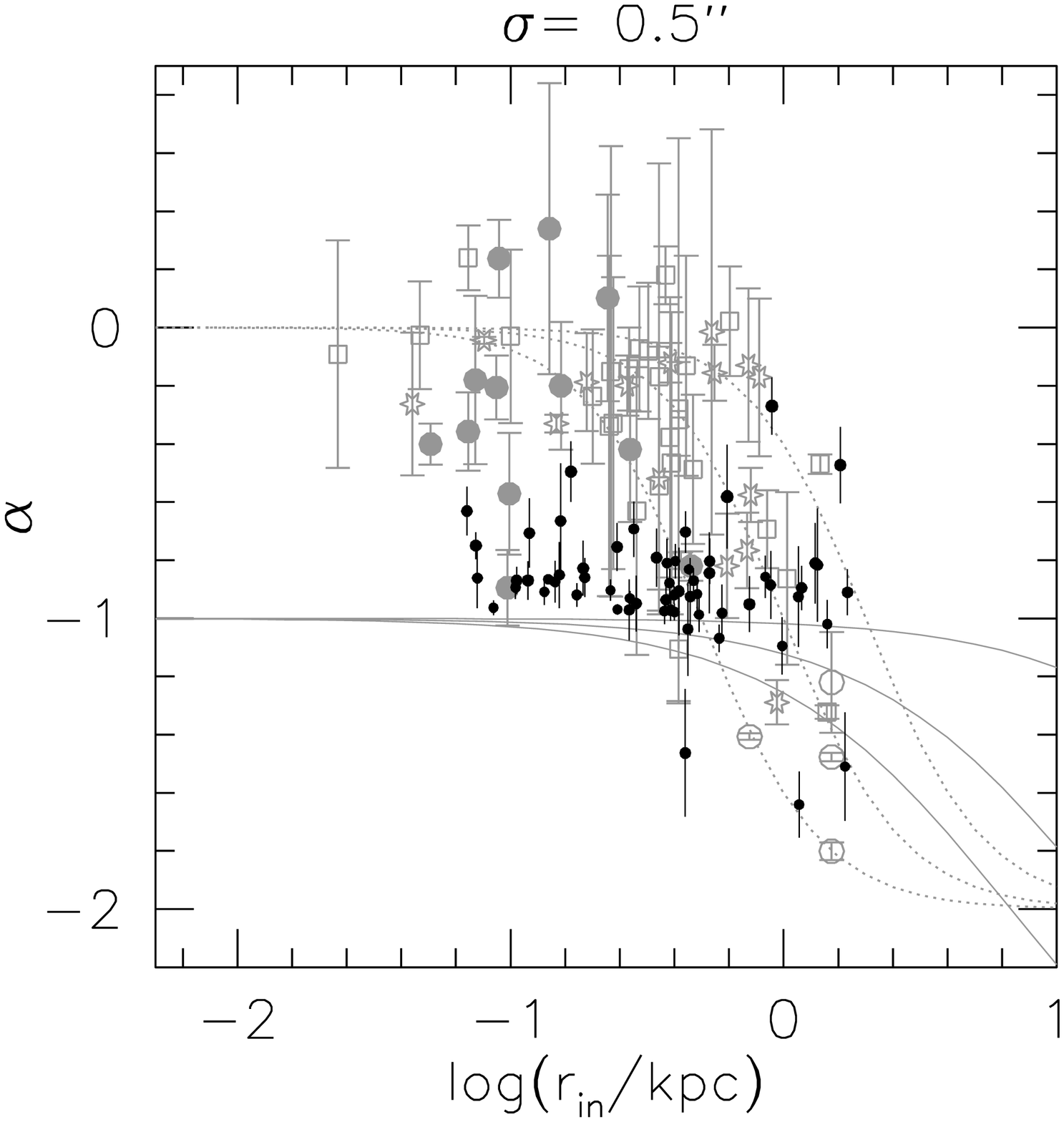}
\epsfxsize=0.2\hsize 
\epsfbox{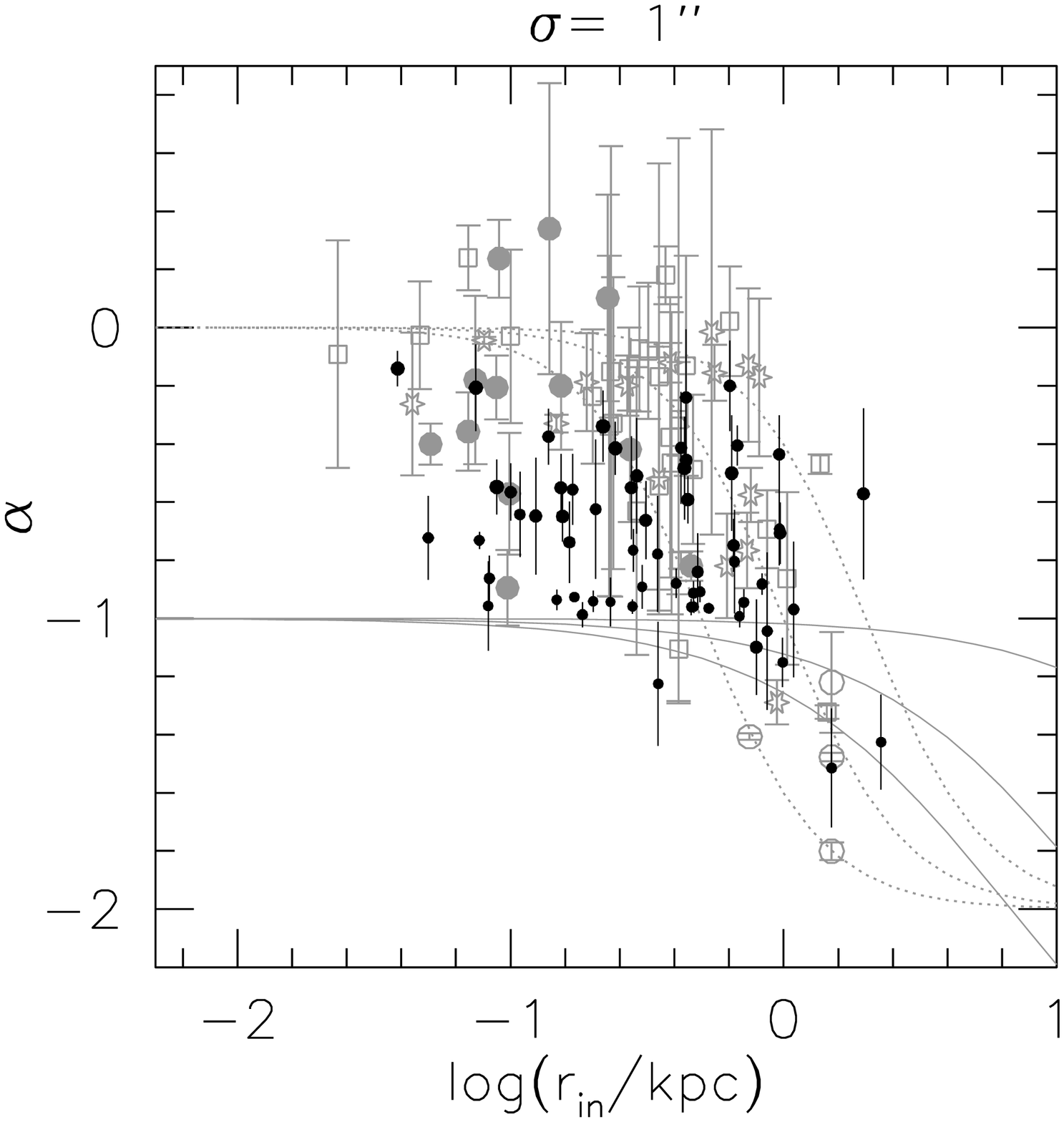}
\epsfxsize=0.2\hsize 
\epsfbox{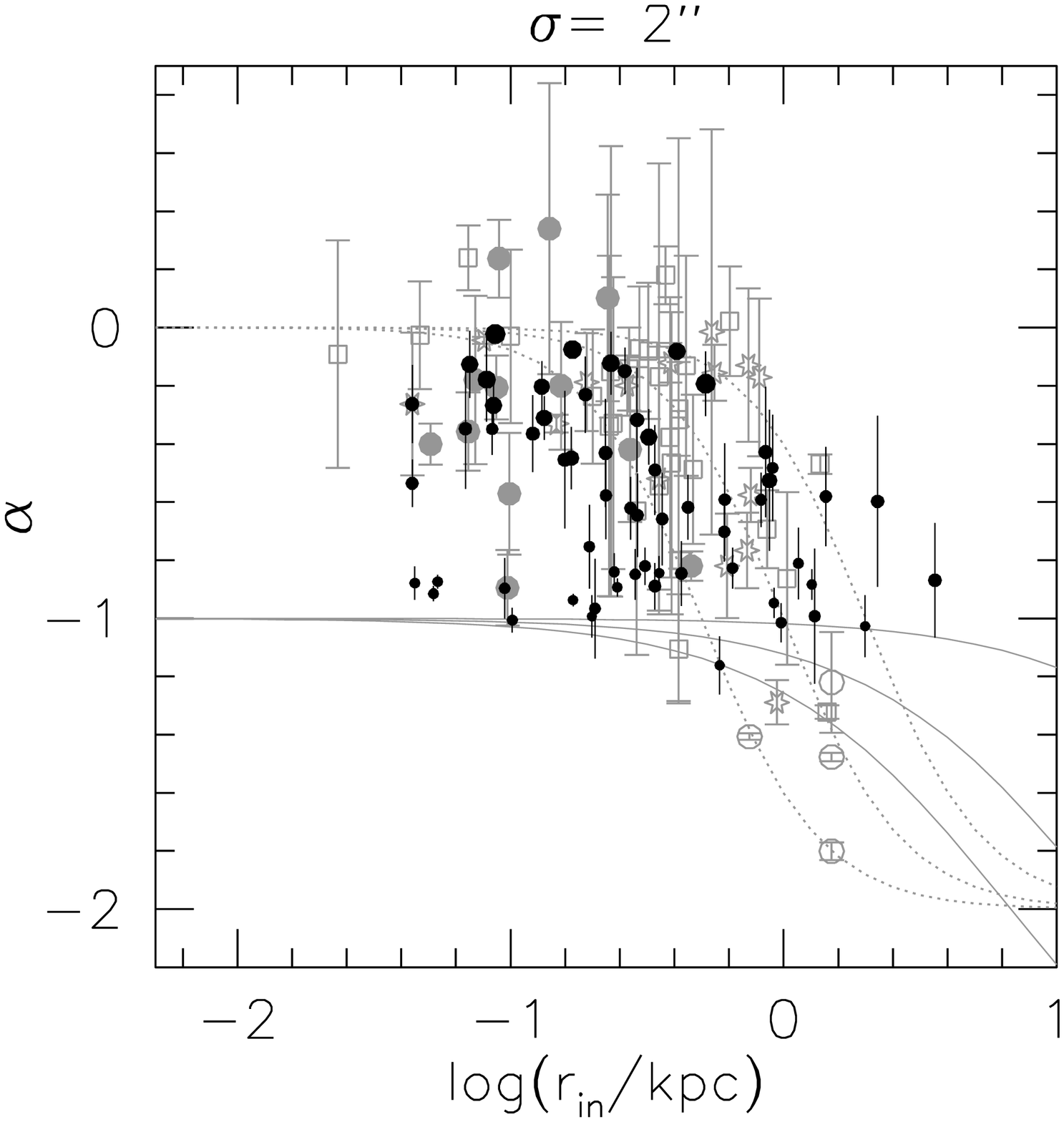}
\epsfxsize=0.2\hsize 
\epsfbox{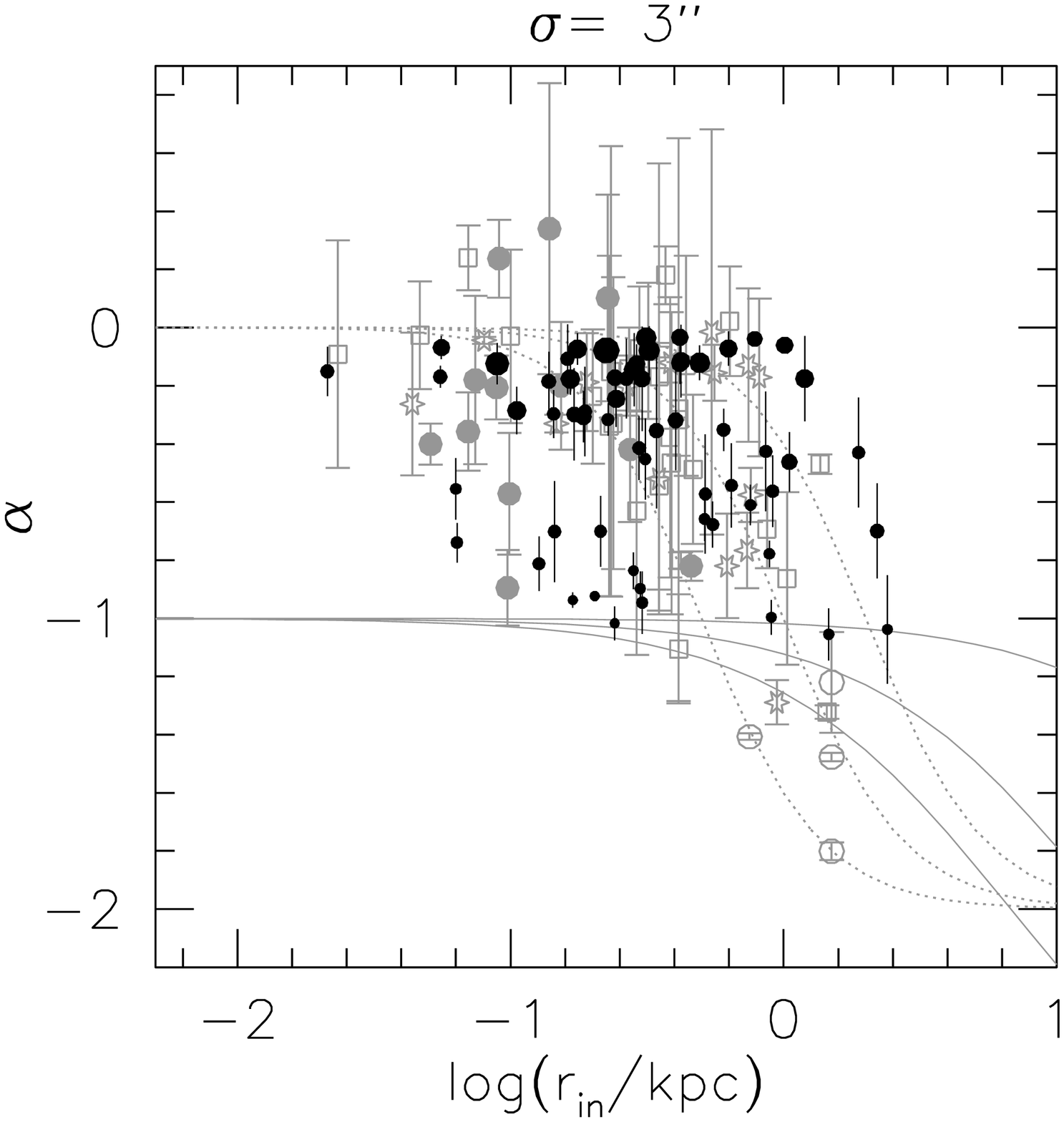}}
\hbox{\epsfxsize=0.2\hsize 
\epsfbox{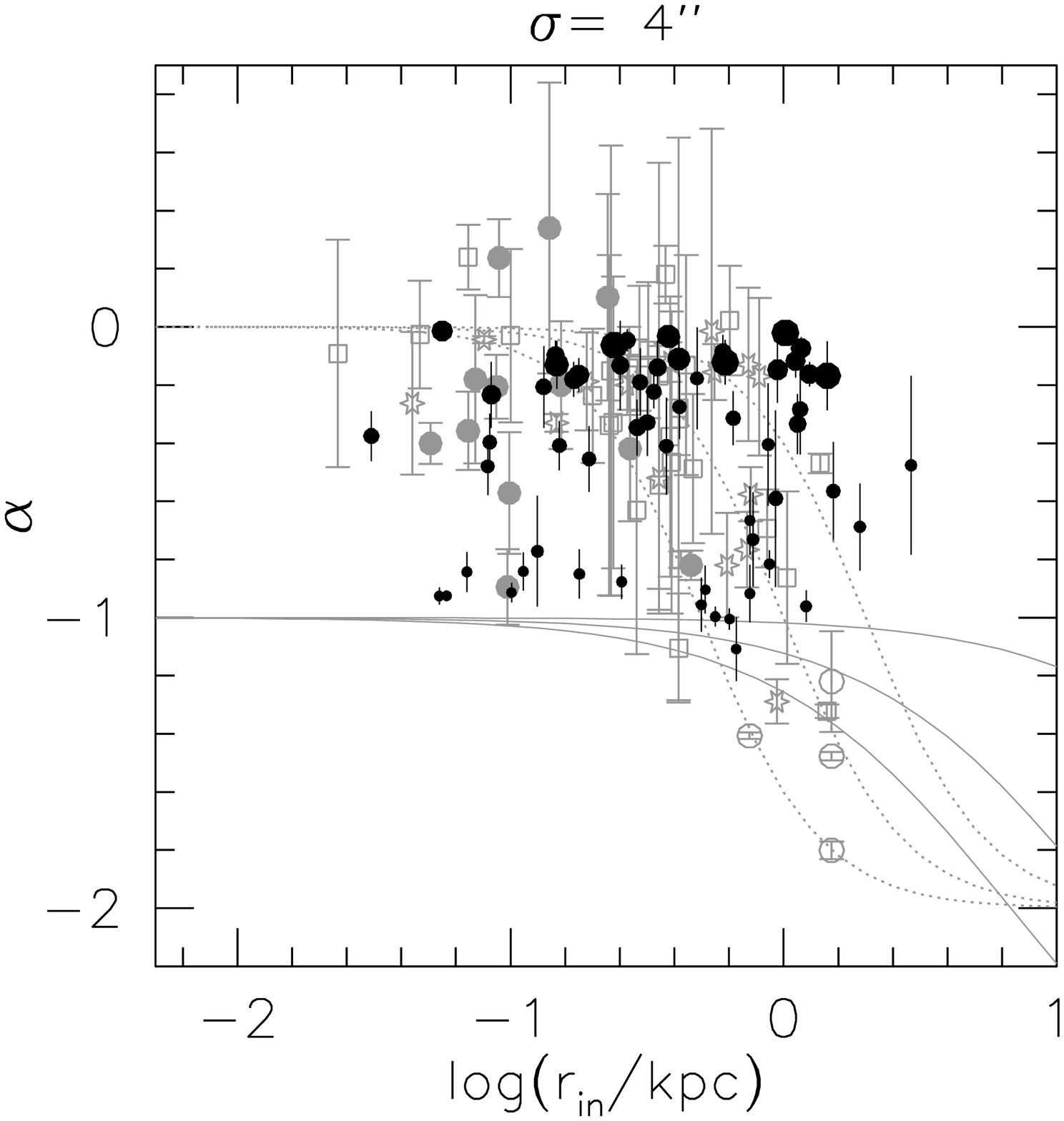}
\epsfxsize=0.2\hsize 
\epsfbox{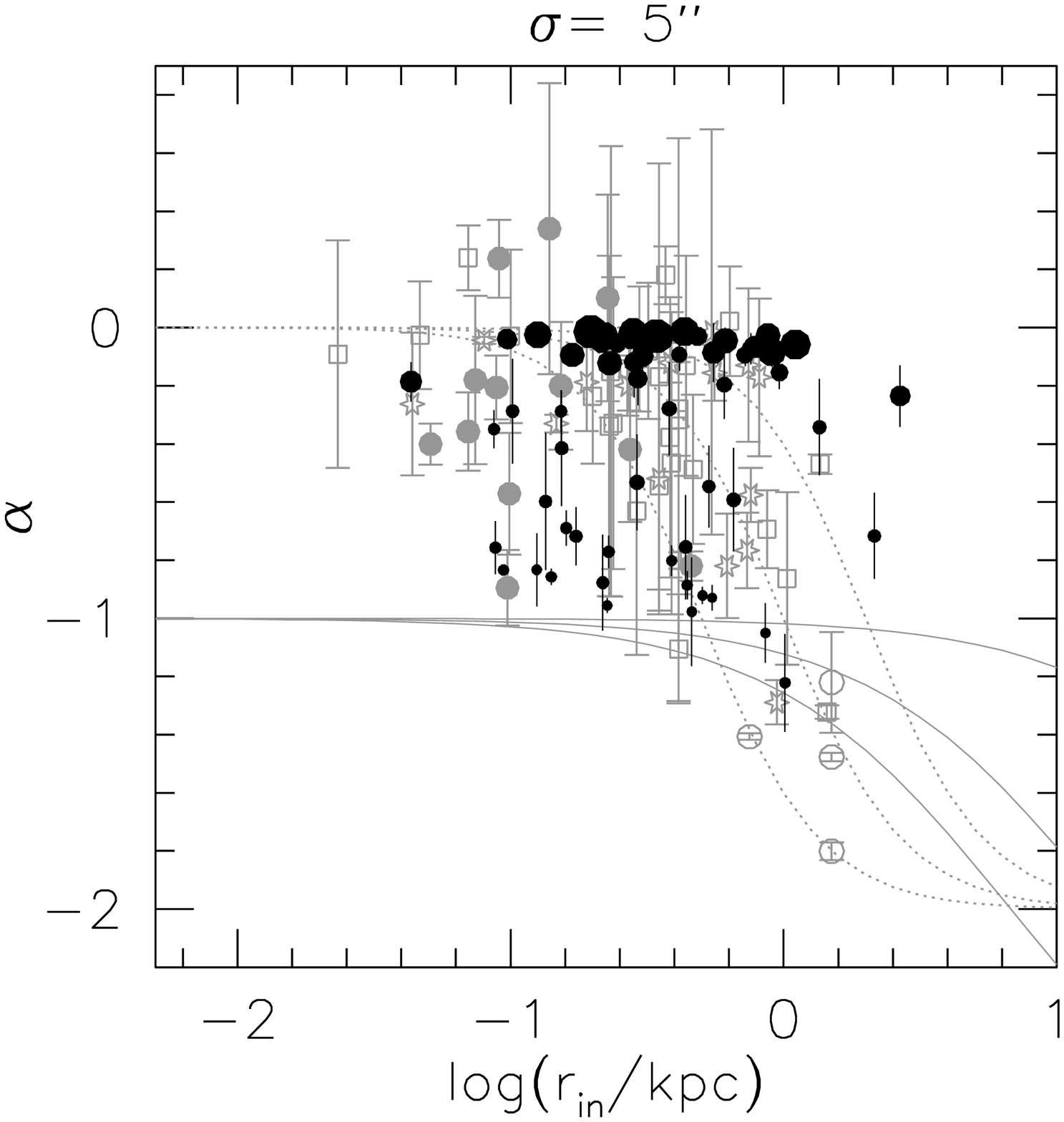}
\epsfxsize=0.2\hsize 
\epsfbox{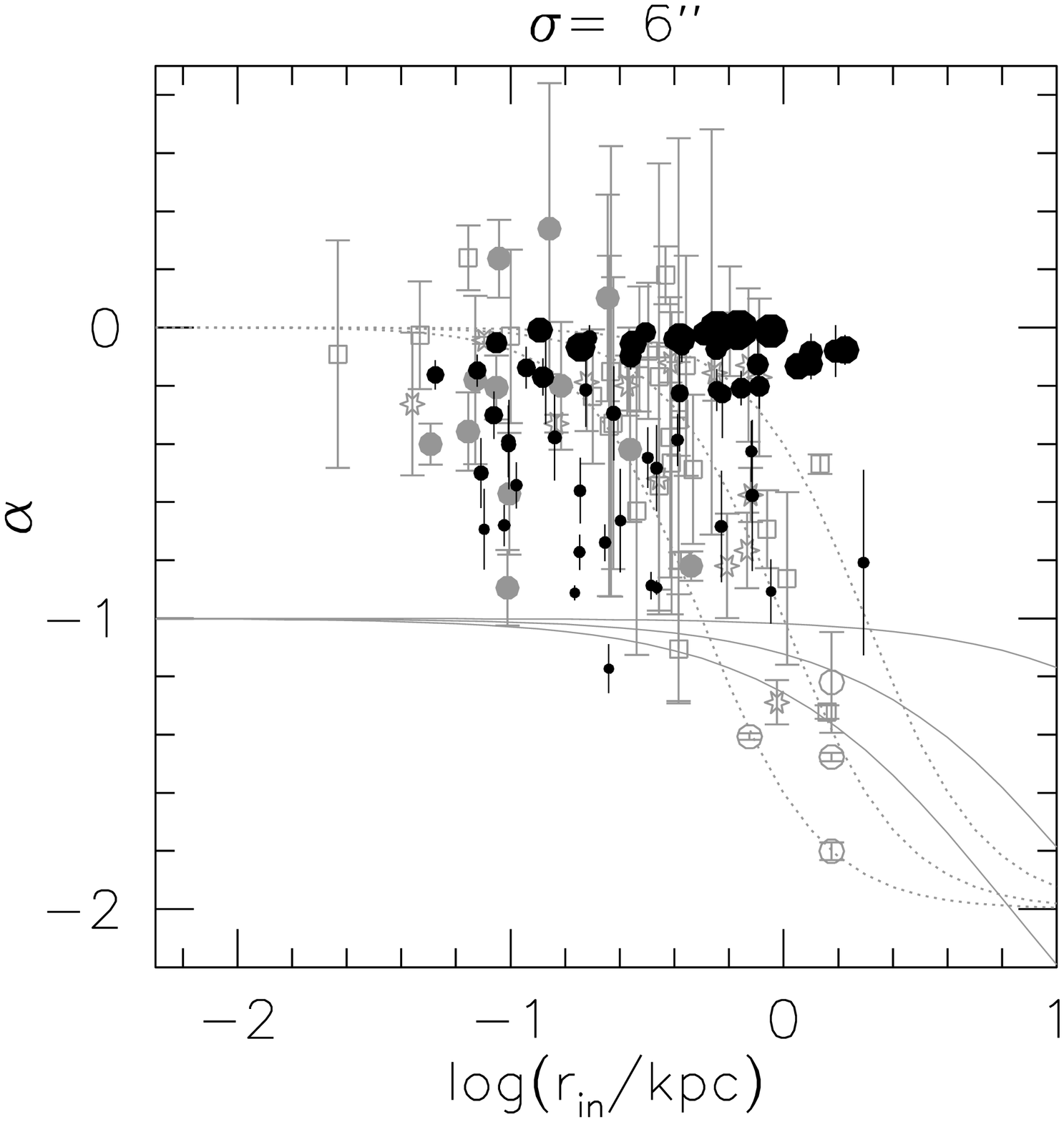}}
\caption[gaussdata]{
Comparison of the simulated NFW halo $r_{\rm in}-\alpha$ data points
with the observed distribution. The simulated data points were derived
using random centre offsets, where the offsets have a Gaussian
distribution with the dispersion given above each sub-panel. As
Fig.~\ref{NFWISOnosys}.
\label{gaussNFWdata}}
\end{center} 
\end{figure*}

\begin{figure*} 
\begin{center}
\hbox{
\epsfxsize=0.2\hsize 
\epsfbox{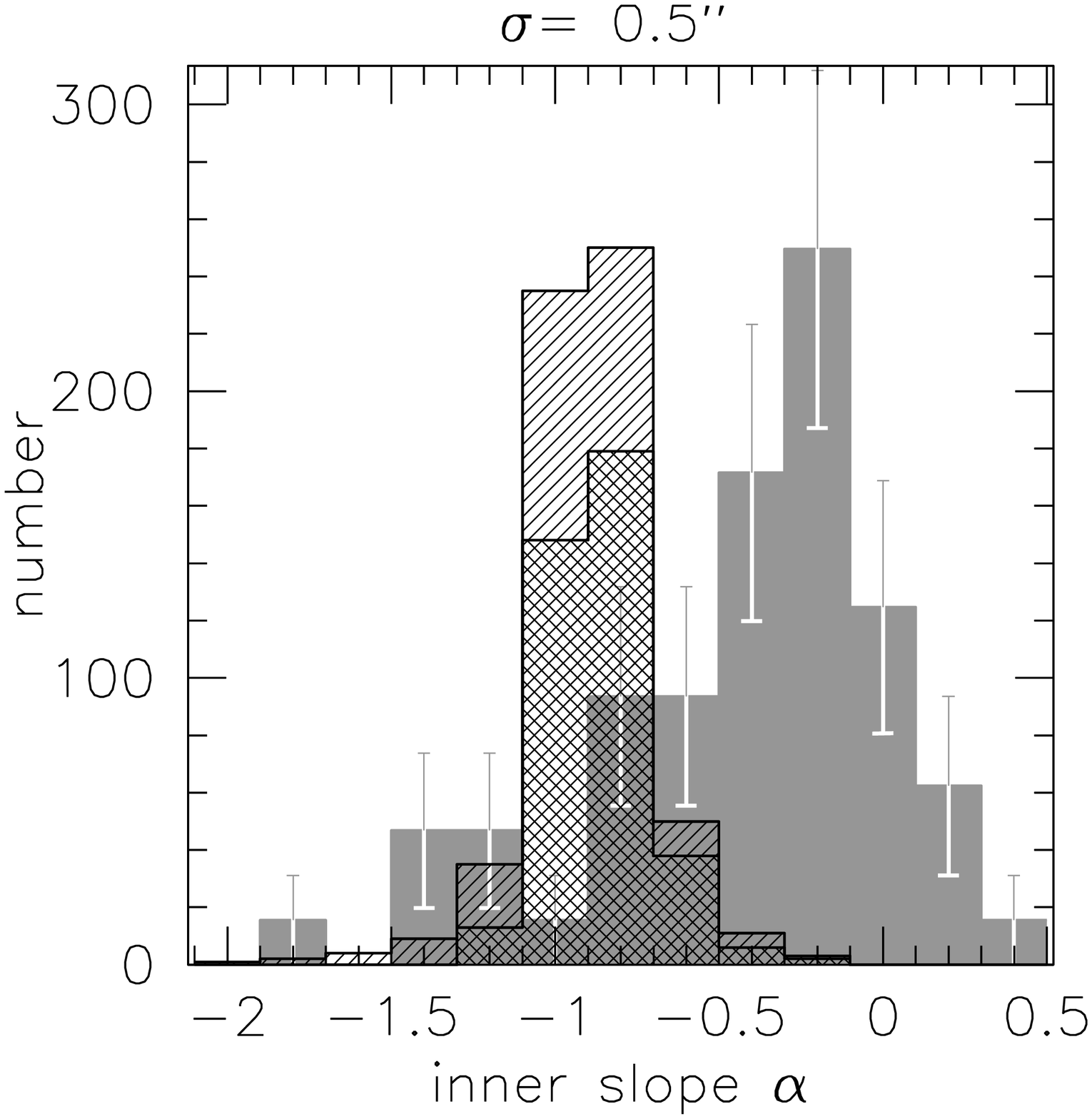}
\epsfxsize=0.2\hsize 
\epsfbox{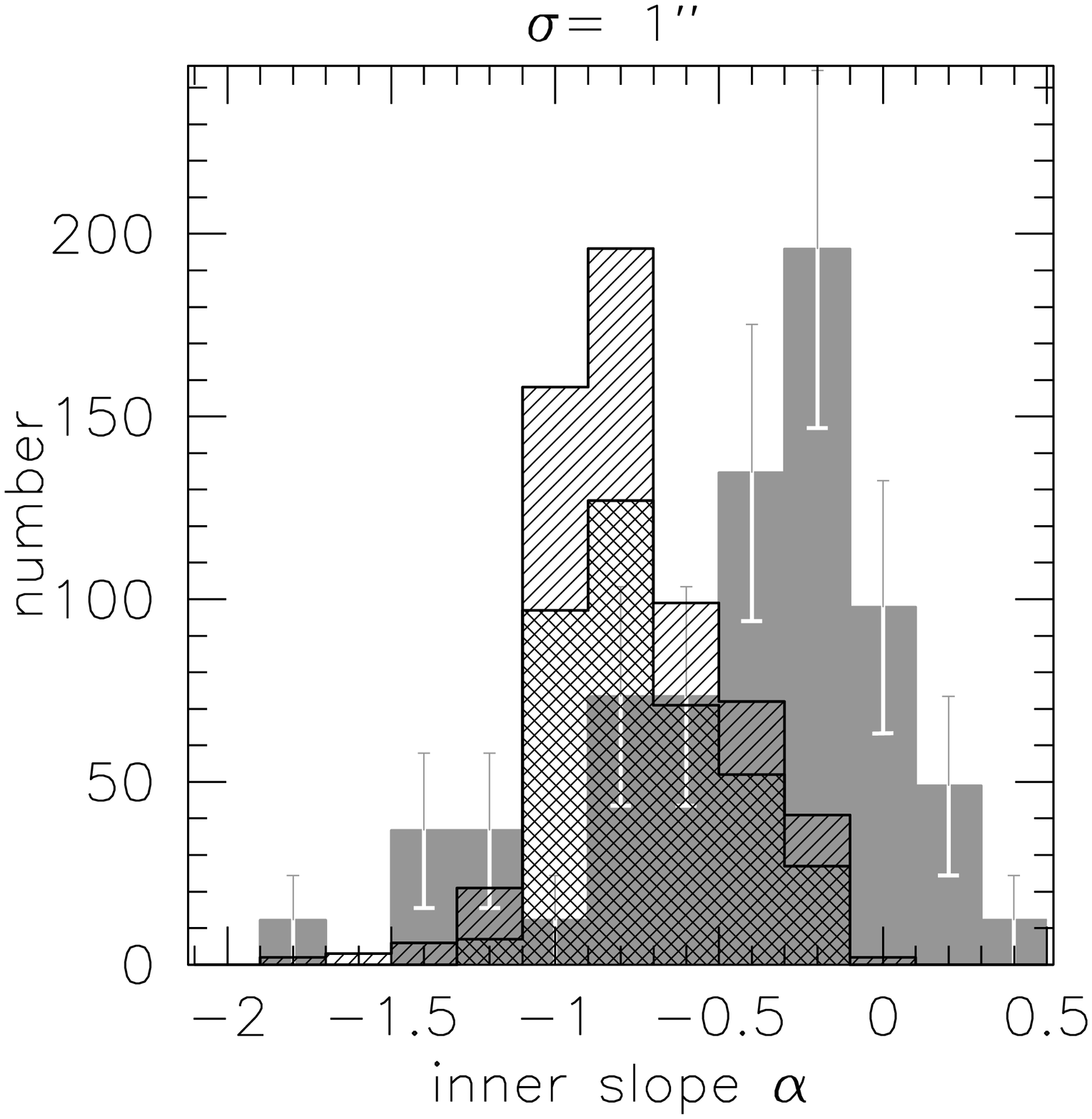}
\epsfxsize=0.2\hsize 
\epsfbox{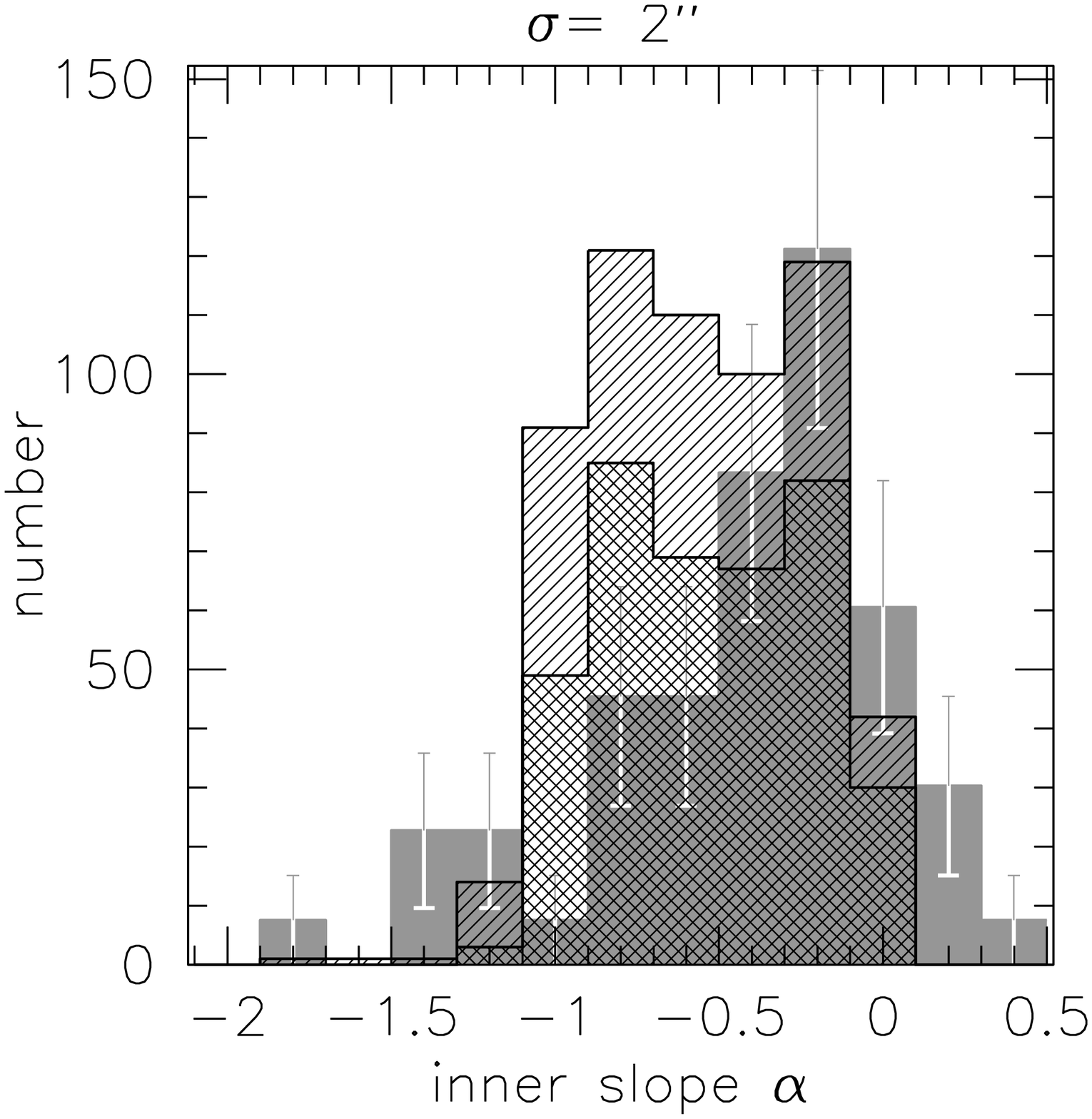}
\epsfxsize=0.2\hsize 
\epsfbox{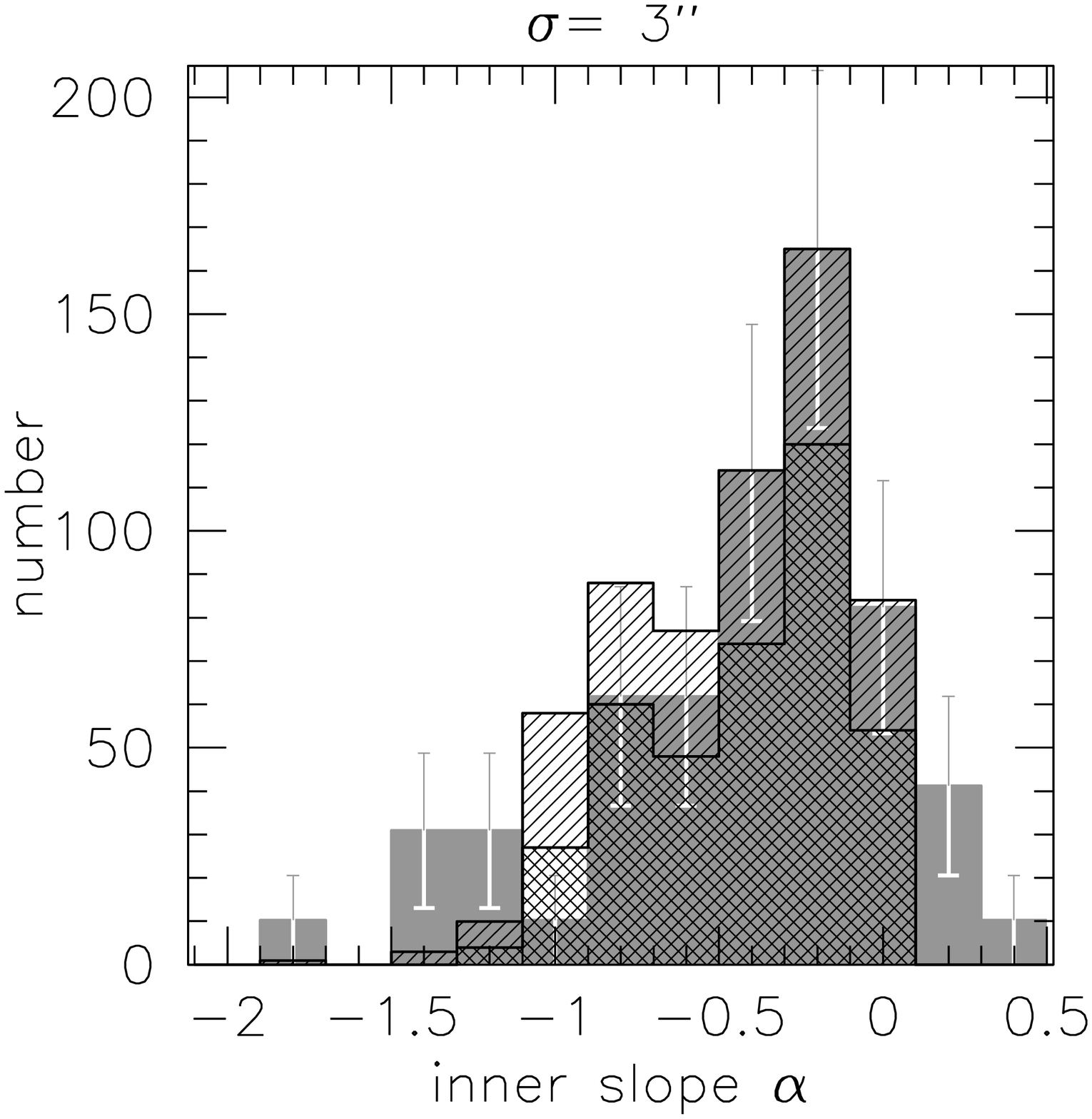}}
\hbox{\epsfxsize=0.2\hsize 
\epsfbox{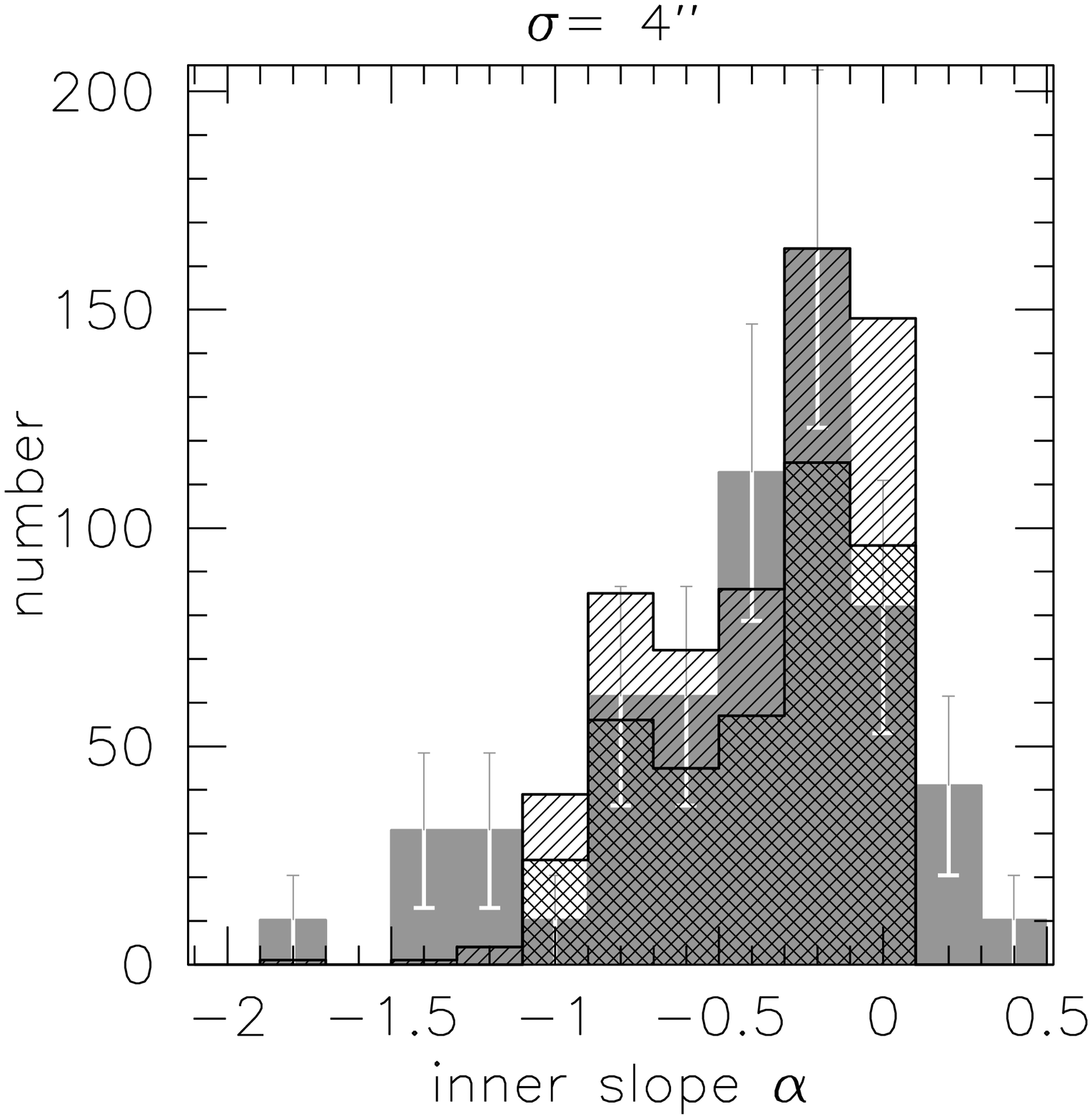}
\epsfxsize=0.2\hsize 
\epsfbox{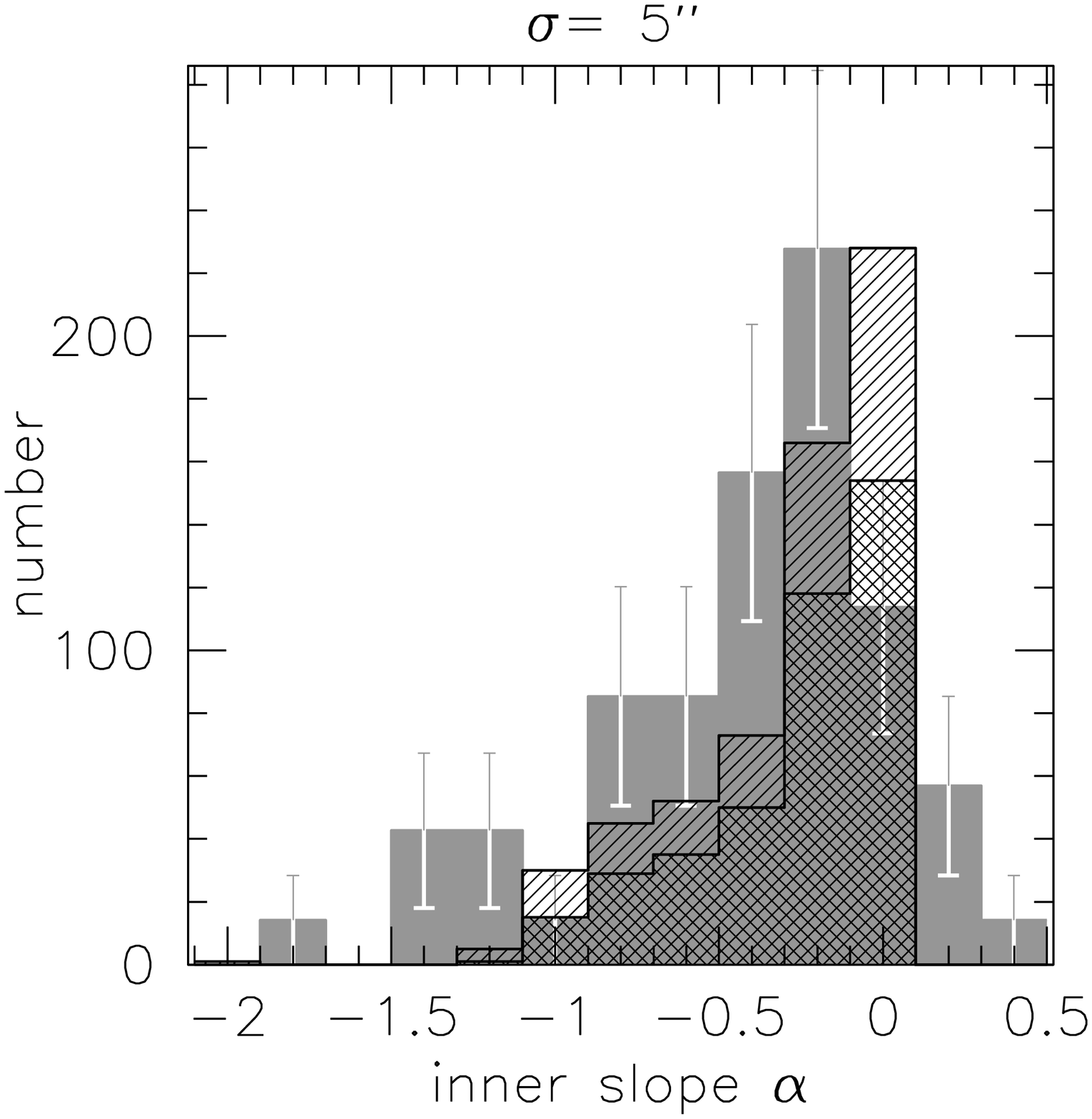}
\epsfxsize=0.2\hsize 
\epsfbox{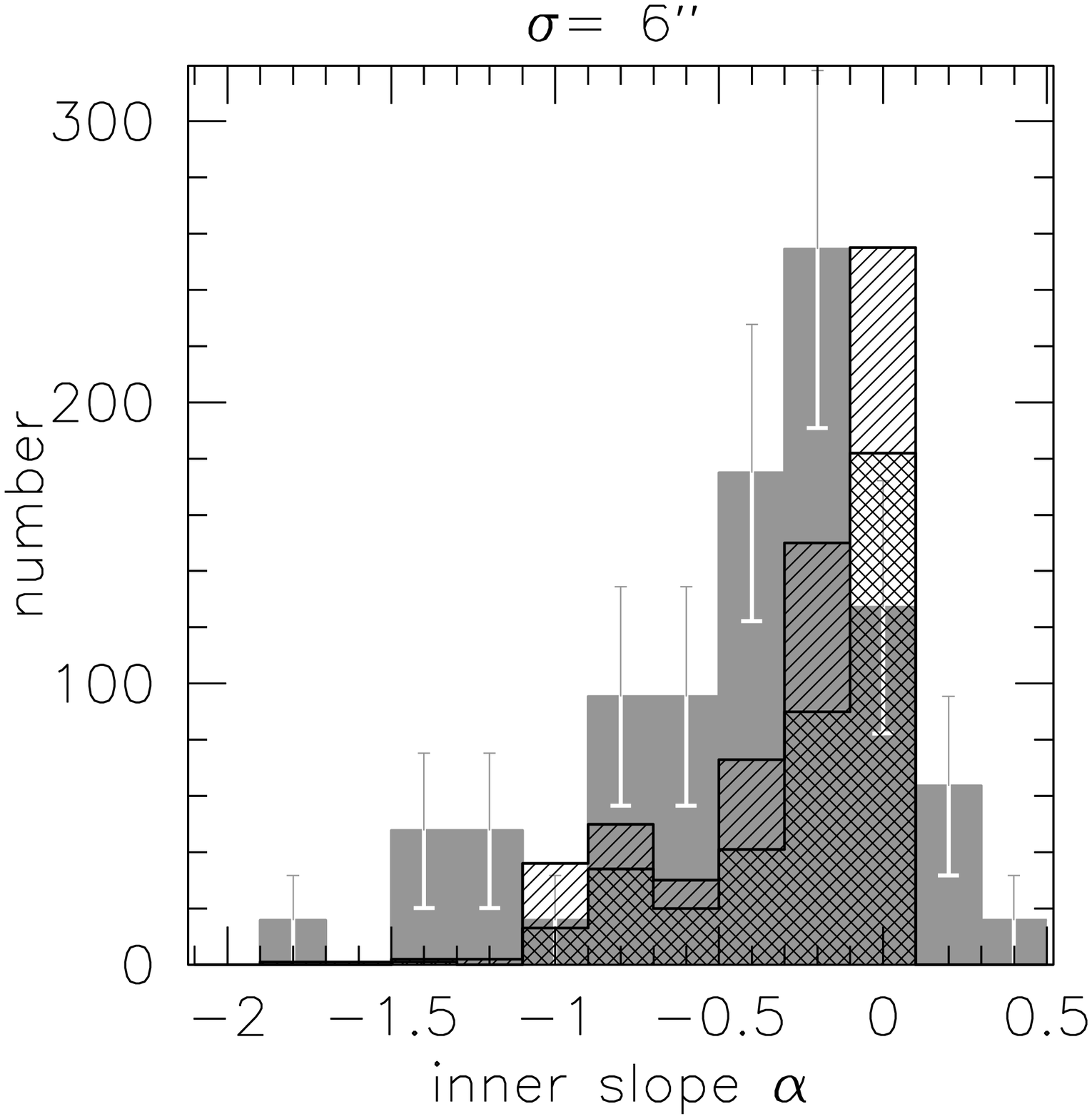}}
\caption[gausshisto.ps]{Comparison of the simulated 
distribution of NFW halos slopes (hatched histogram; low-resolution
galaxies: single-hatched; high-resolution galaxies: cross-hatched)
with that actually observed (grey histogram). The dispersion of the
Gaussian distribution for offsets is given above each sub-panel.
\label{gaussNFWhisto}}
\end{center} 
\end{figure*}

The $r_{\rm in}-\alpha$ comparison is shown in
Fig.~\ref{gaussNFWdata}. The observed and theoretical distributions
are distinctly different, as also seen in the histograms in
Fig.~\ref{gaussNFWhisto}. The NFW models show a pronounced peak at
$\alpha=-1$ not seen in the data.  The high-resolution and
low-resolution histograms look very similar. The only way to
explain the observed distribution with NFW halos and centre offsets is
to assume that for all galaxies observed the position of the dynamical
centre is offset from the optical centre by $\sigma \sim 3-4''$. At
higher dispersions the distributions becomes too biased towards
$\alpha=0$. We can put this in context by comparing with the average
scale length of the F-LSB galaxies in \citet{edb_phot}, which is
$11.0''$. Compared to the sizes of the optical disks, the offsets
needed are thus significant.  For uniform offset distributions similar
conclusions can be derived.

\subsubsection{ISO halos\label{sec:iso2}}

\begin{figure*} 
\begin{center}
\hbox{\epsfxsize=0.9\hsize 
\epsfbox{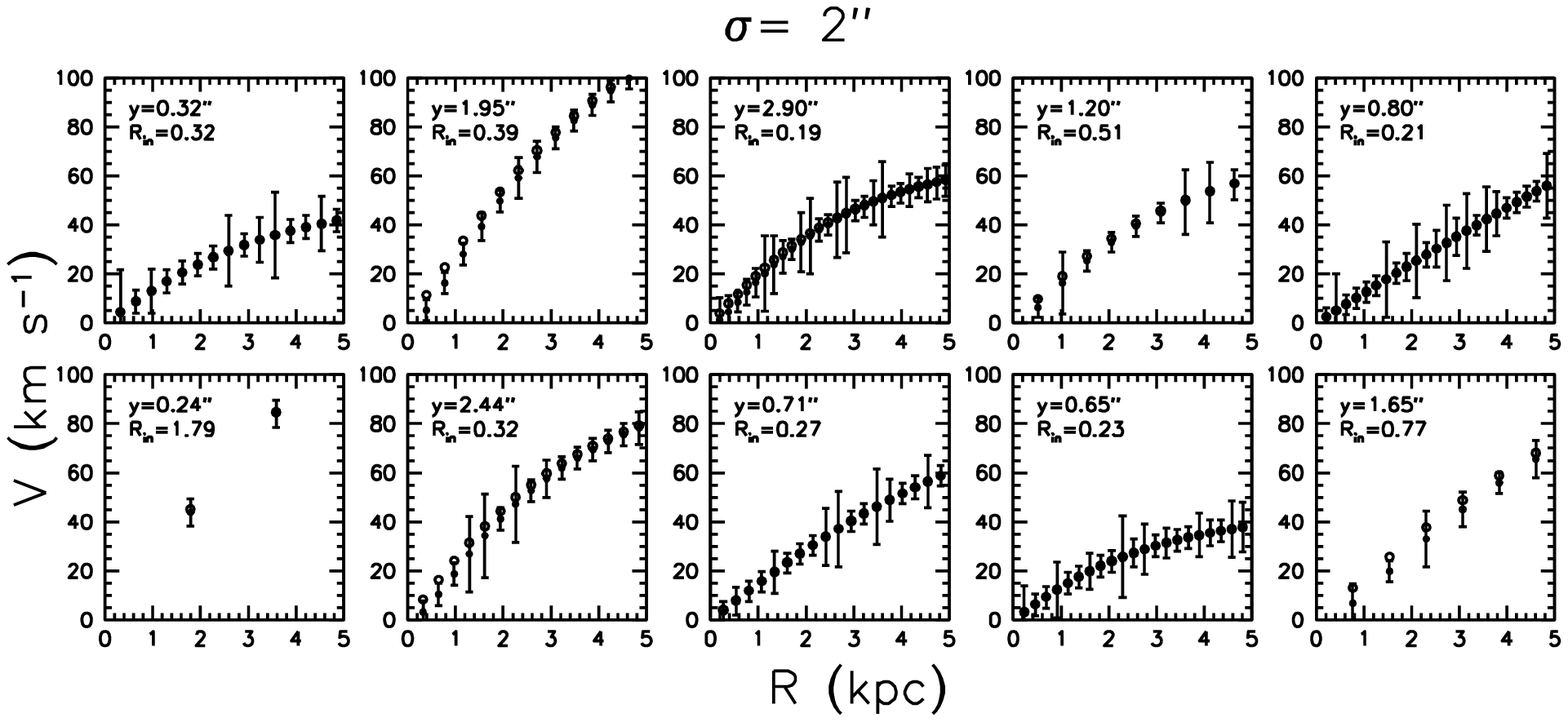}}
\hbox{\epsfxsize=0.9\hsize 
\epsfbox{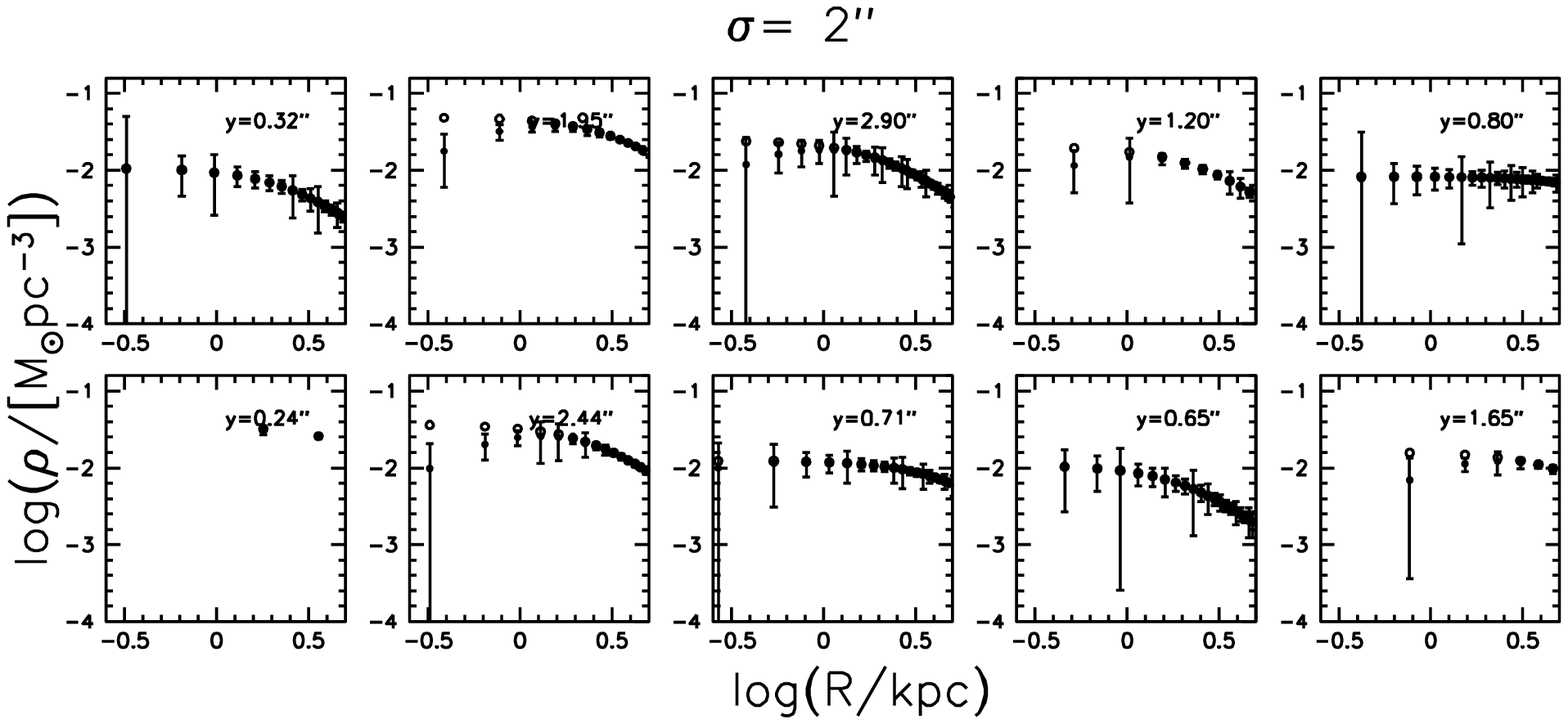}}
\caption[curvesISO.ps]{As Fig.~\ref{NFWcurves}, but showing ISO curves and 
profiles.
\label{ISOcurves}}
\end{center} 
\end{figure*}

\begin{figure*} 
\begin{center}
\hbox{\epsfxsize=0.32\hsize 
\epsfbox{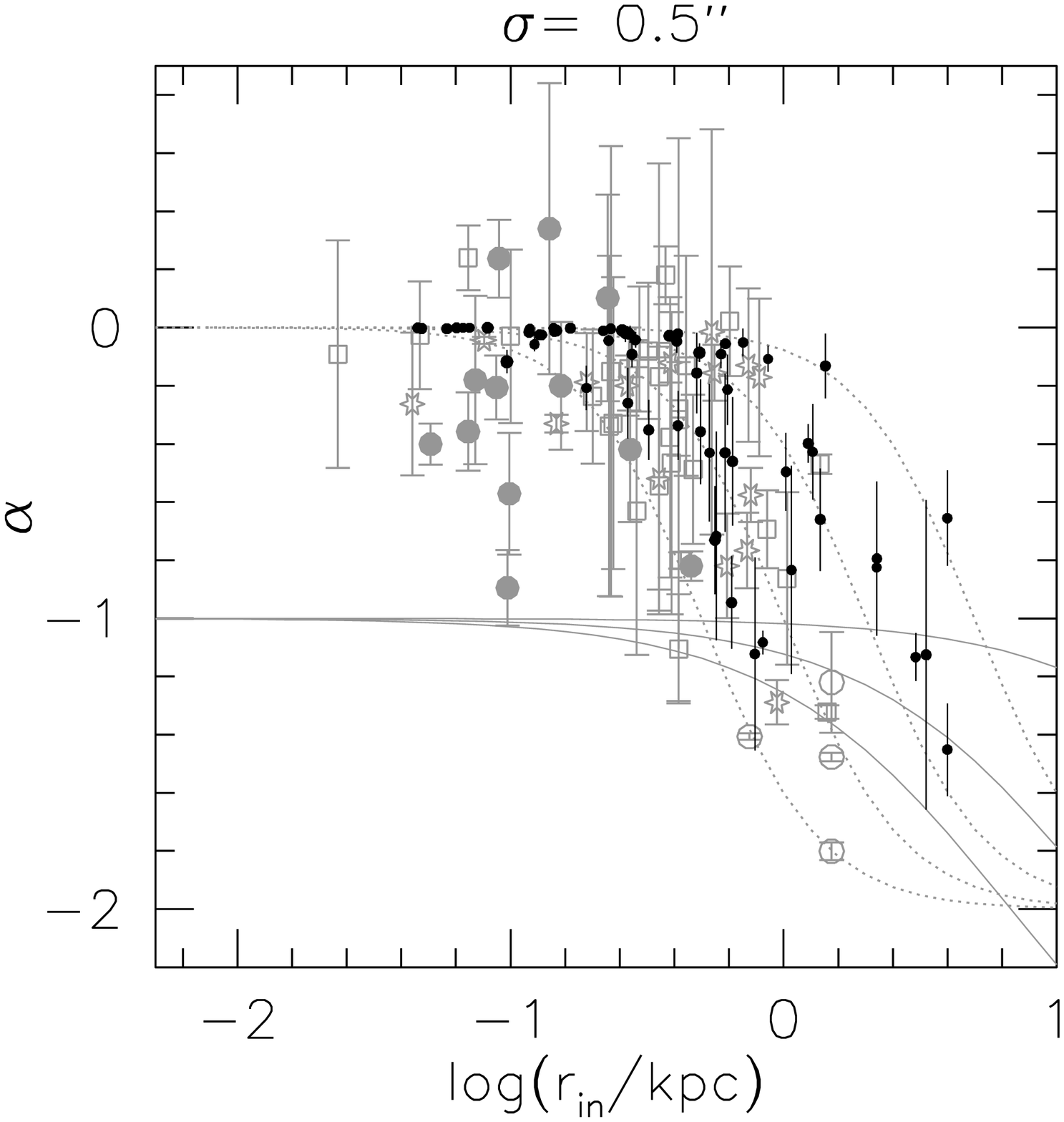}
\epsfxsize=0.32\hsize 
\epsfbox{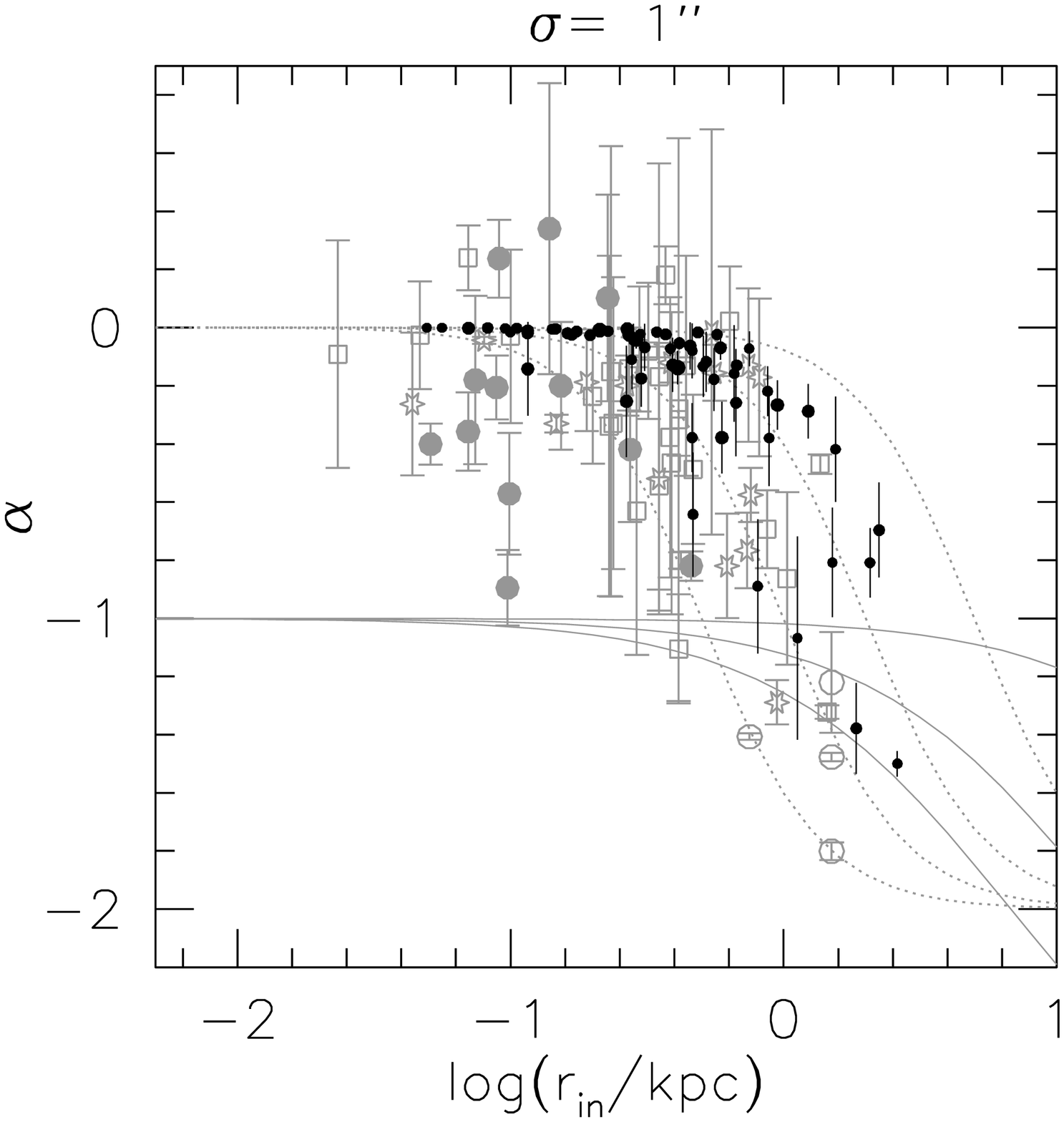}
\epsfxsize=0.32\hsize 
\epsfbox{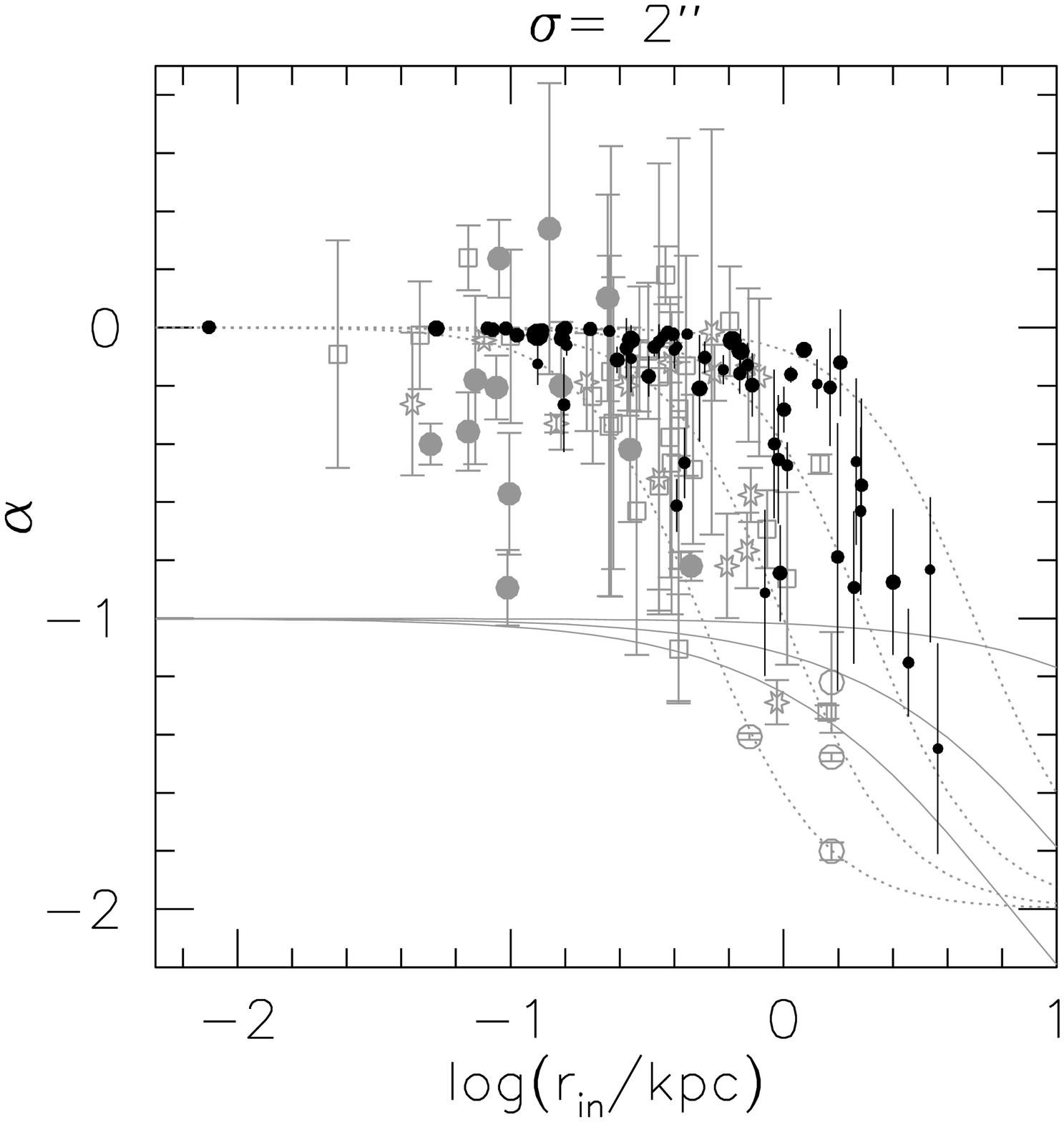}}
\hbox{\epsfxsize=0.32\hsize 
\epsfbox{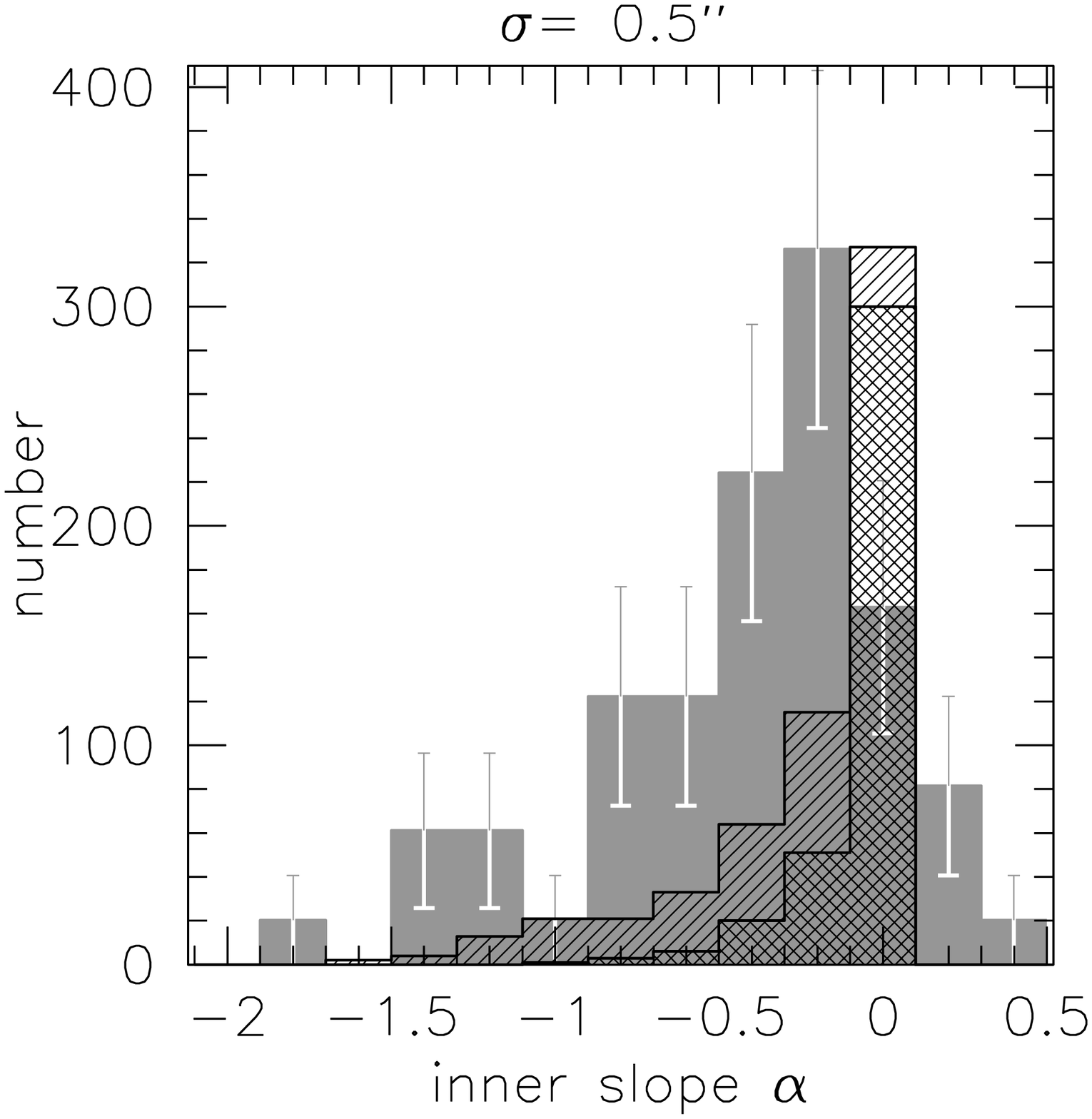}
\epsfxsize=0.32\hsize 
\epsfbox{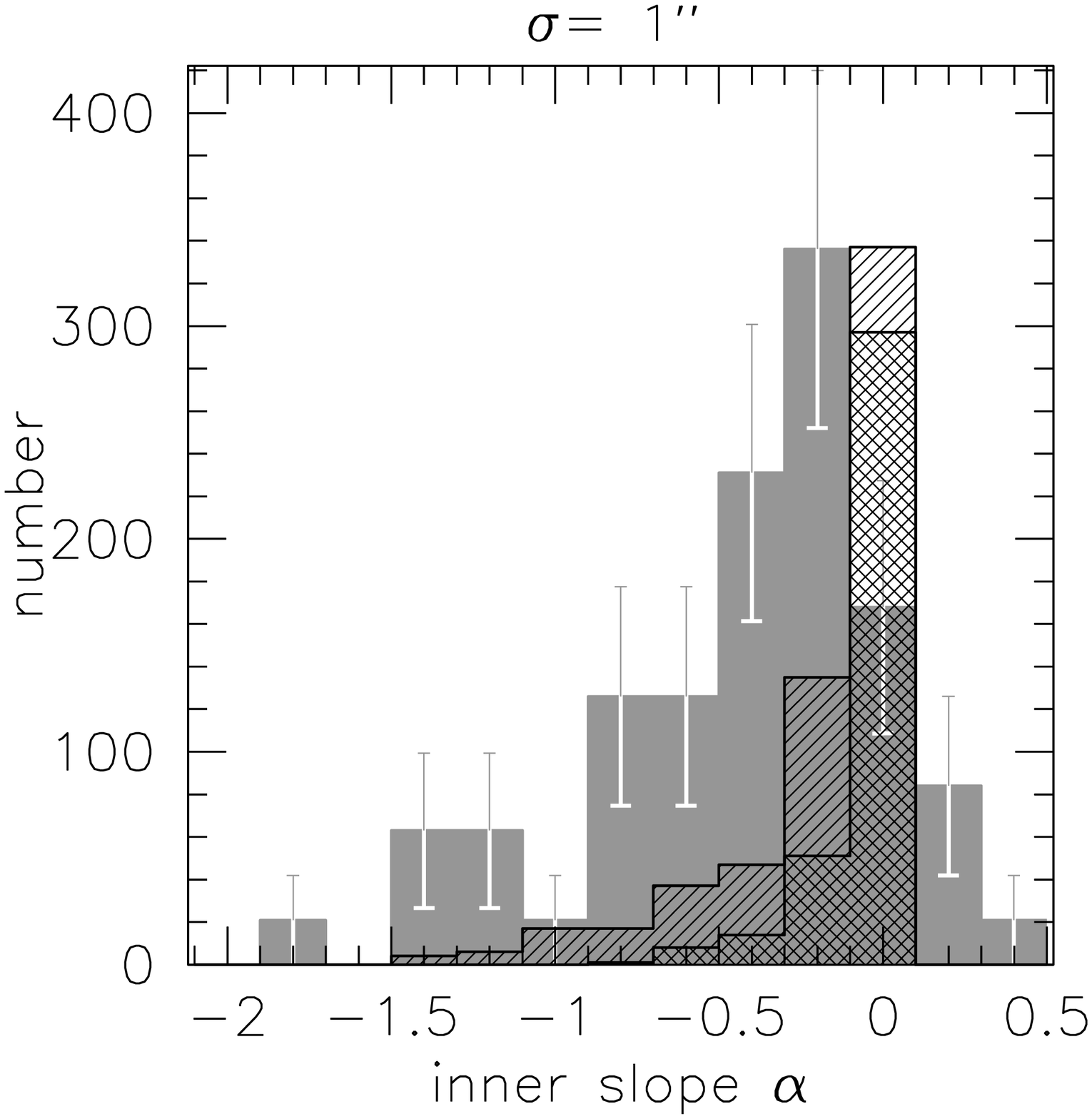}
\epsfxsize=0.32\hsize 
\epsfbox{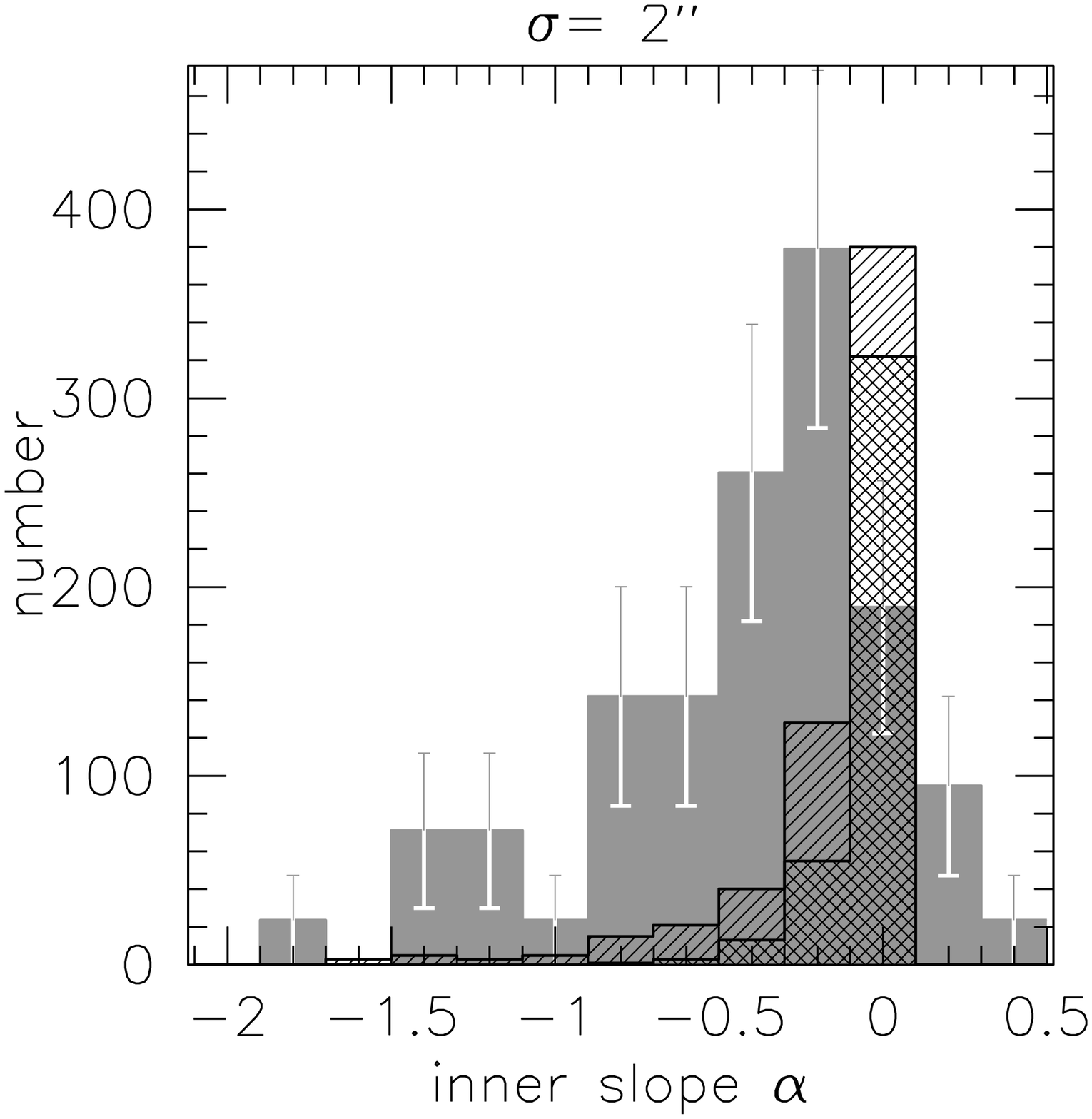}}
\caption[gaussISOhisto.ps]{As Figs.~\ref{gaussNFWdata} 
and \ref{gaussNFWhisto}, but now showing the results for ISO
halos. The bin at $\alpha=0.5$ also contains galaxies with $\alpha >
0.5$.
\label{gaussISOhisto}}
\end{center} 
\end{figure*}

In Fig.~\ref{ISOcurves} a random selection of 10 rotation curves
derived assuming the ISO model and a Gaussian offset distribution with
$\sigma=2''$ is shown in comparison with the corresponding no-offset
rotation curve.  Also shown are the corresponding mass density
profiles.  The $r_{\rm in}-\alpha$ plots and histograms for the ISO model
are presented in Fig.~\ref{gaussISOhisto}. The ISO simulations are a
better (though not perfect) match to the data than the NFW
simulations. ISO halos are fairly insensitive to the effects of moderate
offsets between optical and dynamical centres. We thus only present
the results $\sigma < 2''$.

The peak in the simulated distribution occurs at $\alpha = 0$, offset
from the observed $\alpha = -0.2 \pm 0.2$. There is an excess of
steeper slope galaxies in the observations. The inner parts of real
galaxies clearly have density profiles that are not precisely flat. A
possible explanation is that the minimum disk assumption is slightly
inappropriate, and that the stellar disk causes the steeper observed
slopes. However, as stated before, modeling this effect is beyond the
scope of our simple one-component models.  The values of the slopes
are insensitive to the precise value of the offset, indicating that if
galaxies do have ISO halos, they will always be obvious from the
observations. The simplest explanation is that offsets between optical and dynamical center are small.

\subsubsection{Observational offsets}

A mismatch between the centres has the same effect as
an incorrectly centered slit. The absence of systematic differences
thus also implies an absence of these observational offsets.  This is
to be expected from technical point of view: the optical centres in
LSB galaxies are often visible in the slit camera, and in most cases
used to line up the slit of the spectrograph with the galaxy.  The
accuracy of a typical telescope pointing system and the accuracy of
the offset procedure from nearby stars is generally very good. Offset
and pointing tests performed during the observations generally are
repeatable down to the $\sim 0.3''$ level.  This was
already addressed by \citet{optcur_data} and dBB02 who present
rotation curves of a few LSB galaxies that have been observed multiple
times at different dates on different telescopes by different
observers and show that these agree very well with each other (see
also \citealt{marchesini02}).  Residual observational uncertainties
could be introduced by the process of measuring the positions of the
centres of the galaxies from e.g.\ CCD images, catalogs or \HI
observations.  These offsets can for obvious reasons only be a small
fraction of the size of the galaxy, and are therefore  $\sim
1''$ or less, consistent with the scatter implied by the simulations.

\section{Tests with real data}

To test whether some of the effects described above are observable in
practice, we re-observed one of the galaxies from dBB02.  We
chose UGC 4325, which is a typical late-type dwarf galaxy. It is nearby
($D=10.1$ Mpc), resulting in a linear resolution of $\sim 50$ pc arcsec$^{-1}$.
Its \HI velocity field is regular, symmetric and appears undisturbed
\citep{swaters_phd} with an inclination of 41\degr.

UGC 4325 was observed in early Feb 2002 with the 1.93m telescope at
the Observatoire de Haute Provence using the long-slit Carelec
spectrograph. The set-up, observing procedures and data-reduction were
identical to those described in dBB02.  We obtained H$\alpha$ spectra
of UGC 4325 along its major axis (PA 231$\degr$), parallel to the
major axis but offset by $+5''$ (0.25 kpc), parallel to the major axis
offset by $-5''$, as well as two spectra centered on the galaxy but
with position angle offsets of $+30\degr$ and $-30\degr$.

\begin{figure*} 
\begin{center}
\epsfxsize=\hsize 
\epsfbox{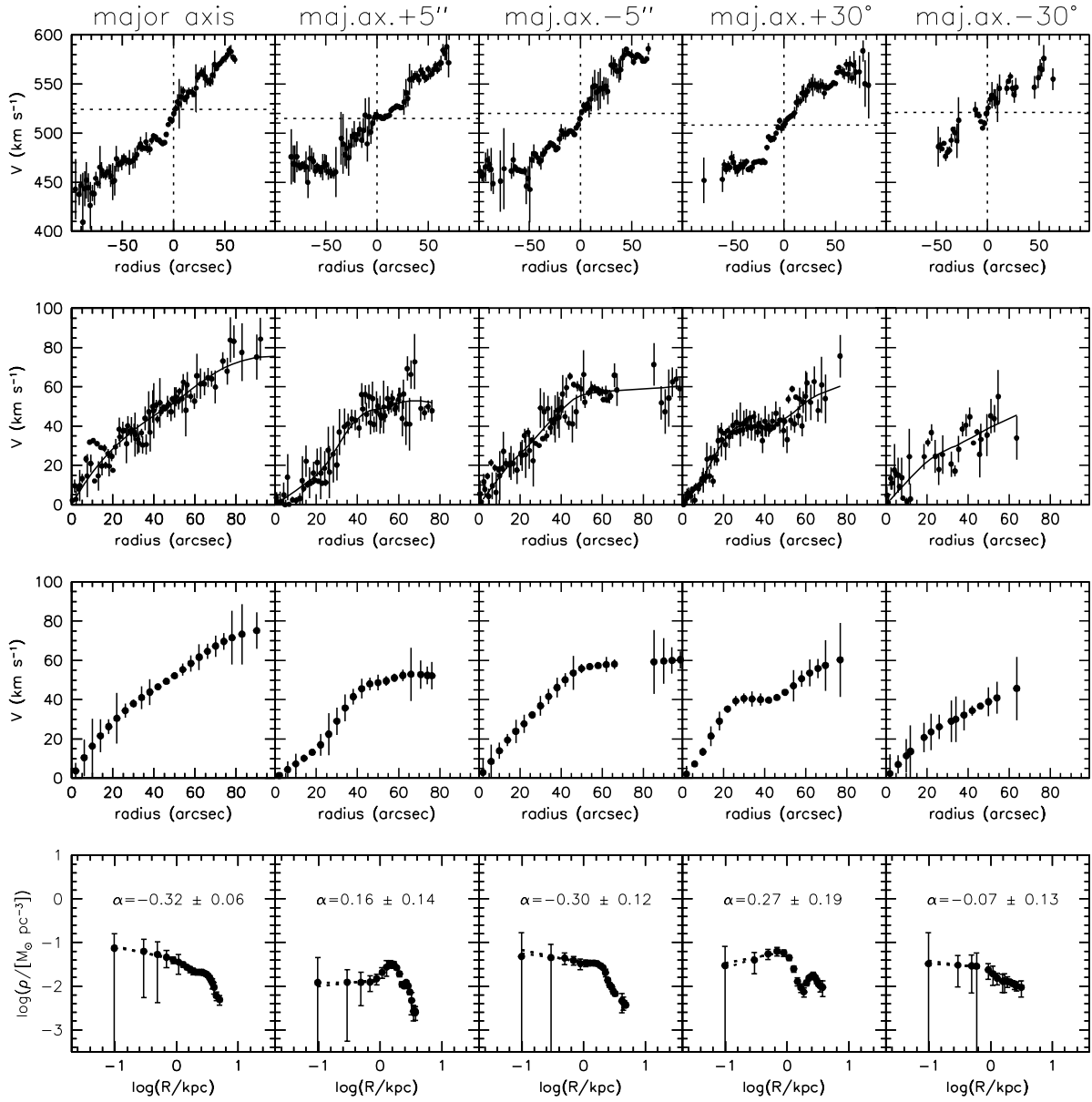}
\caption[u4325.ps]{
Observed rotation curves of UGC 4325, after applying spatial and
position angle offsets. From left to right: major axis rotation curve;
parallel to major axis offset by $+5''$; parallel to major axis offset
by $-5''$; position angle offset of $+30\degr$; position angle offset
of $-30\degr$. Top row shows the raw rotation curves. Second row shows
folded rotation curves. Third row shows re-sampled rotation
curves. Bottom row shows mass-density profiles derived from the
re-sampled rotation curves.
\label{4325}}
\end{center} 
\end{figure*}

The resulting raw rotation curves (not corrected for inclination) are
shown in the top row of Fig.~\ref{4325}. Especially for the offset
curves the positions of the ``centres'' along the slit are difficult
to determine, and we determined the systemic velocity and central
position by looking for the point that gave maximum symmetry between
approaching and receding sides. However, as the curves are almost
linearly rising in the inner parts, the precise choice of the centre
of symmetry does not affect the slopes.  The symmetrized and
uncorrected curves are shown in the second row of Fig.~\ref{4325}.
Following \citet{optcur_pap2} we have performed local polynomial fits
to the folded curves in order to bring out the underlying shape. We
also resampled the curves with a spacing of $4''$. The resulting
curves are shown in the third row of Fig.~\ref{4325}. Also shown in
the bottom row are the mass-density profiles.

The similarity between the major-axis curve and the two off-axis
curves is more consistent with an ISO model and velocity field as
shown in Fig.~\ref{velfi}.    The curves
with position angle offsets only show modest changes in the slope,
consistent with U4325 having a fairly axisymmetric potential, and
negligible streaming motions.  For UGC 4325 the systematic effects
discussed in this paper are not obviously present and likely small enough to be
unobservable.  If UGC 4325 is typical, then the data presented here
show  that systematic effects discussed in Sect.~3 are unlikely to
have a large effect on the observed rotation curves.

\section{Discussion}

If systematics effects indeed play only a small role in
determining the observed mass-density profile, some of the scatter  in
the observed value of the slope could be due to intrinsic differences in the
shapes of the mass-density profiles. This can be used to improve on the
results for the ISO halo.  Figure~\ref{gaussISOhisto} shows that for
this model the peak in the histogram occurs at values of $\alpha$ that
are slightly too large.  Here we discuss two additional
observationally motivated models that were designed as an improvement
on the ISO model. The first one is the  \citet{Bur95} halo
(Sect.~\ref{sec:bur}), the second one the  halo model proposed
by \citet{krav98} based on LSB \HI rotation curves (Sect.~\ref{sec:alpha}).

\subsection{Burkert halos\label{sec:bur}}

The halo model presented in \citet{Bur95} is a compromise between the
ISO model and CDM models: it has the constant density core found in
ISO halos, but shows the steep drop-off in density at large radii
found in CDM simulations.
The density profile of a Burkert halo can be described as
\begin{equation}
\rho(R) = \frac{\rho_0}{(1+R/r_0)(1+(R/r_0)^2)}
\end{equation}
and the corresponding rotation curve is given by
\begin{equation}
V^2(R)= \\\ 2\pi G \rho_0 r_0^3  \frac{1}{R}\Bigl \{ \ln \bigl [
(1+\frac{R}{r_0})\sqrt{1+(\frac{R}{r_0})^2} \bigr ] -
\arctan{\frac{R}{r_0}} \Bigr \},
\end{equation}
where $r_0$ is the equivalent of a core radius and $\rho_0$ is the
central density.  This particular model gives a good description of
the rotation curves of dwarfs and LSB galaxies
\citep{Bur95,blais01,marchesini02}. However, the steep $R^{-3}$
drop-off in density in the outer parts, as implied by the above
equations, has not been observed unambiguously, and therefore fits
using this model mostly constrain the inner part of the curve.
Here we characterise the Burkert halo in terms of its maximum rotation
velocity, making the assumption that the observed rotation curves do
not probe the declining $\rho \sim R^{-3}$ part of the rotation curve.
$V_{\rm max}$ occurs at $R_{\rm max}=3.25\,r_0$, and we can express $r_0$ as
\begin{equation}
r_0 = \sqrt{\frac{V^2_{\rm max}}{0.8606\pi G \rho_0}}.
\label{eq:burkert}
\end{equation}
We describe the selection of halo parameters for the simulated Burkert 
halos in Sec.~\ref{sec:scale}.

\subsection{Kravtsov et al.\ (1998) halos\label{sec:alpha}}

The distribution of observed inner mass-density slopes peaks at a
value of $\alpha=-0.2$ (dBMBR, dBB02). Before the optical rotation
curves used in those analyses became available, \citet{krav98} fitted
power-law models to the \HI rotation curves of dwarfs and LSB galaxies
(the latter from \citealt{edb_bmh96}) and found that the inner
rotation curves could be described with a power-law with a slope
$\alpha_{\rm HI} = -0.2$ (the agreement between the optical and \HI
values is certainly an indication that the effects of beam smearing on
the mjority of the \HI curves must have been minor).

The density profiles described so far are part of a wider set of 
profiles of the form \citep{krav98,zhao96,blais01}
\begin{equation}
\rho(R) = \frac{\rho_0}{[q+(R/r_0)^{-\alpha}][1+(R/r_0)^{\gamma}]^{(\beta+\alpha)/\gamma}}
\label{eq:kravrho}
\end{equation}
where we have changed the order and signs of some of the exponents as
defined in \citet{krav98} and \citet{blais01} to be consistent with
our notation.  The above equation can be used to describe NFW halos
with $(q,\alpha,\beta,\gamma) = (0,-1,3,1)$, ISO halos with
$(1,-2,2,2)$ while the   Burkert halo has parameter set $(1,-1,3,2)$. The
parameter $\alpha$ determines the inner slope, $\gamma$ determines the
rate of turnover in the mass profile, while $\beta$ gives the slope of
the mass distribution in the outer parts of halos. The $q$ parameter indicates
the presence $(q=1)$ or absence $(q=0)$ of a constant-density core.
\citet{krav98} find that the LSB \HI rotation  curves from 
\citet{edb_bmh96} can be best described with the parameter
set $(0,-0.2,3,1.5)$. They note that the $\gamma$ and $\beta$
parameters are difficult to constrain and choose values suggested by
CDM simulations.  Though we will use these values, we note that the
observed mass profiles of LSB galaxies are equally well described by
asymptotically flat rotation curves with $\beta=2$.  As we are here
only interested in the inner parts of the halos, this choice does not
affect the results.

The corresponding rotation curve can be described by a model of the form
\begin{equation}
V(R) = V_{t} \frac{(R/r_0)^g}{[1+(R/r_0)^a]^{(g+b)/a}}.
\label{eq:kravrot}
\end{equation}
The parameter $g$ gives the inner slope of the rotation curve and is
related to $\alpha$ by $g =1+\alpha/2$ (note our definition as
$\alpha$ as a negative number). We thus find $(a,b,g) =$ (1.5, 0.34,
0.9). The maximum rotation velocity is related to $V_t$ by
$V_{max}=V_t (g/a)^{g/a}(1+g/b)^{-(b+g)/a}$.  The parameter sets
$(q,\alpha,\beta,\gamma) = (0, -0.2, 3, 2)$ and $(a,b,g)=(1.5, 0.34, 0.9)$
thus fully describe the mass-density profiles and rotation curves of a
halo with an $\alpha = -0.2$ inner mass-density profile.  These
profiles will be used to optimise the description of the data by our
simulations. 

\subsection{Scaling the models\label{sec:scale}}

The NFW and ISO models each have a well-defined range of allowed halo
parameters, either derived from cosmological simulations or
observations.  These ranges are less well defined for the Burkert and
\alfa\ halos.  In order to simulate these halos with realistic
parameters we assume that the ISO curves modeled here are a good
description of observed rotation curves.  We scale the Burkert and
\alfa\ models to show the same diversity in shapes and amplitudes.

\begin{figure*} 
\begin{center}
\epsfxsize=0.47\hsize 
\epsfbox{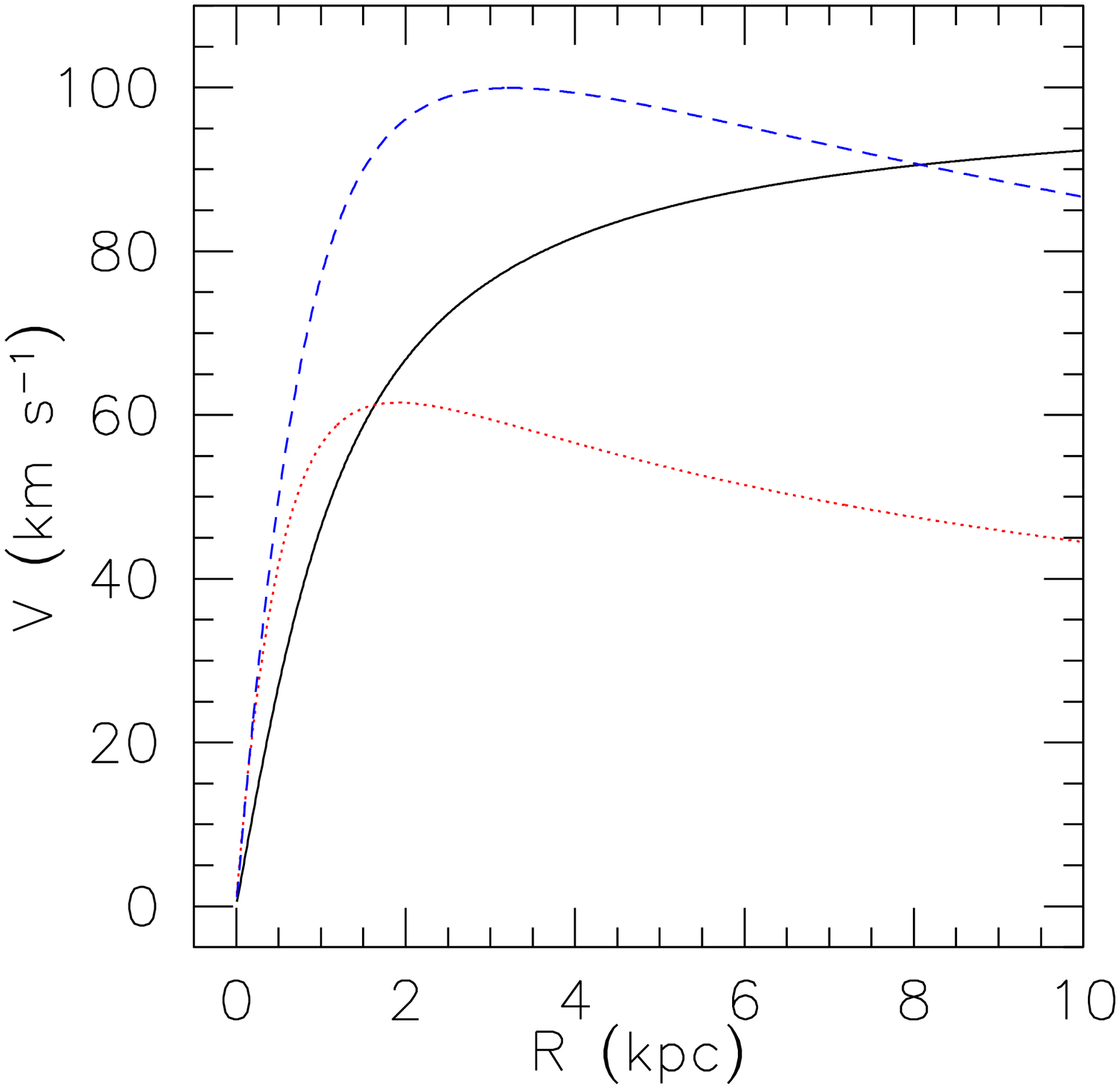}
\epsfxsize=0.47\hsize 
\epsfbox{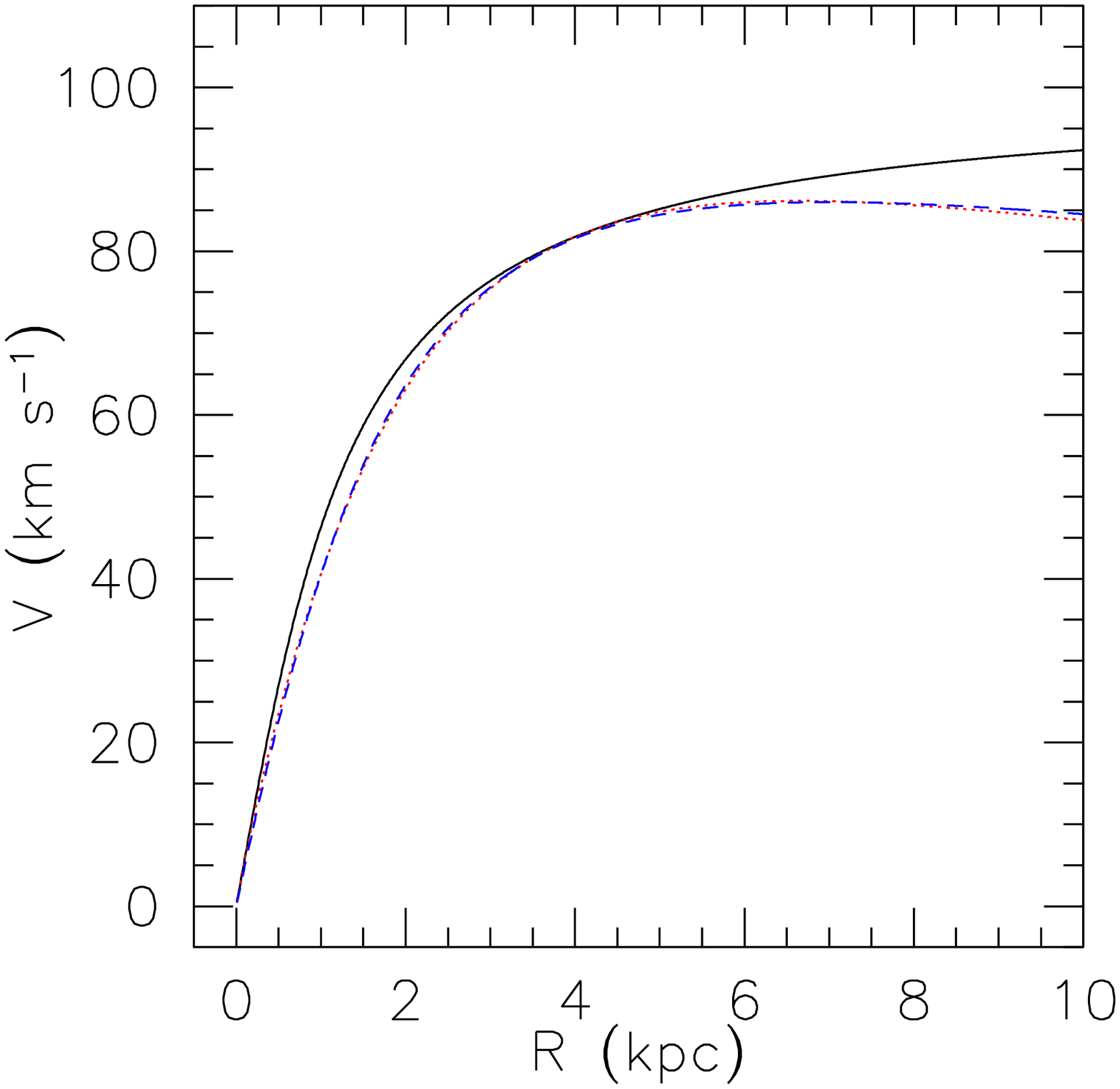}
\caption[comparehalo.ps]{
Comparison of the rotation curves of the ISO, Burkert and \alfa\
models, before and after scaling. Left panel: rotation curves of the
three models, assuming $V_{\infty} = V_{\rm max} = 100$ \kms, and $R_C =
r_0 = 1$ kpc. Full drawn curve: ISO halo; dotted curve: \alfa\ halo;
dashed curve: Burkert halo. Right panel: the three curves after the
scaling described in the text.
\label{comparehalo}}
\end{center} 
\end{figure*}

The left panel in Fig.~\ref{comparehalo} compares the rotation curves
of each of the three models for $V_{\infty} = V_{\rm max} = V_t= 100$
\kms, and $R_C = r_0 = 1$~kpc.  Using these identical numerical
parameters the ISO curve rises more slowly than the Burkert and \alfa\
curves.  The latter has a lower maximum velocity, as for this model
$V_{\rm max} = 0.62\,V_t$ (see Sec.~\ref{sec:alpha}). Simply using
identical halo parameters in the simulations thus makes the Burkert
and \alfa\ halos consistently more compact and more likely to suffer
from systematic resolution effects.  The differences in
Fig.~\ref{comparehalo} are due to the different scalings that
$r_0$ and $R_C$ imply in the various models and we can scale the
curves in order to get a more consistent behaviour.  Equations
\ref{eq:ISO} and \ref{eq:burkert} show that for a given central
density and maximum or asymptotic velocity, the ISO and Burkert radii
scale as $R_C/r_0 = 1/2.156$.  For a given central density and maximum
velocity, one needs to assume a value of $r_0$ that is $\sim 2.2$
times larger than the corresponding value of $R_C$ in order to get a
Burkert curve that matches approximately the shape of the ISO curve.
The Burkert and \alfa\ curves can be scaled in radius by considering
their maximum rotation velocities. For the Burkert halo this occurs at
$r=3.25\, r_0$; for the \alfa\ halo at $r=1.91\,r_0$. A scaling of
$1.69$ in radius will thus line up the
\alfa\ curve with the Burkert one (which implies a total scaling of
3.64 with respect to the ISO halo).

The velocity scales can be matched with those of the ISO halos by
multiplying the velocities of the Burkert halo with $0.86$; that of
the \alfa\ halo by multiplying with $1.38$.  We do not normalise to
the asymptotic velocity of the ISO halo as might perhaps be expected,
as this would have resulted in substantially different shapes.  The
right panel in Fig.~\ref{comparehalo} shows the scaled rotation
curves.  This procedure thus merely serves to get a sample of
simulated Burkert and \alfa\ curves with shapes that match those shown
by the ISO curves (and as observed in real galaxies). This minimises
possible systematic effects in the simulations due to differences in
shape or resolution.  The scaling does not affect the values of the
central slopes.

In order to simulate a realistic range of parameters for the Burkert
and \alfa\ halos we sample the same numerical range of parameters as
for the ISO halos (as described in Sect.~\ref{sec:iso}), and multiply
their radial and velocity scales by the factors described above.

\subsection{Results for Burkert and \alfa\ halos}

\begin{figure*} 
\begin{center}
\epsfxsize=0.48\hsize 
\epsfbox{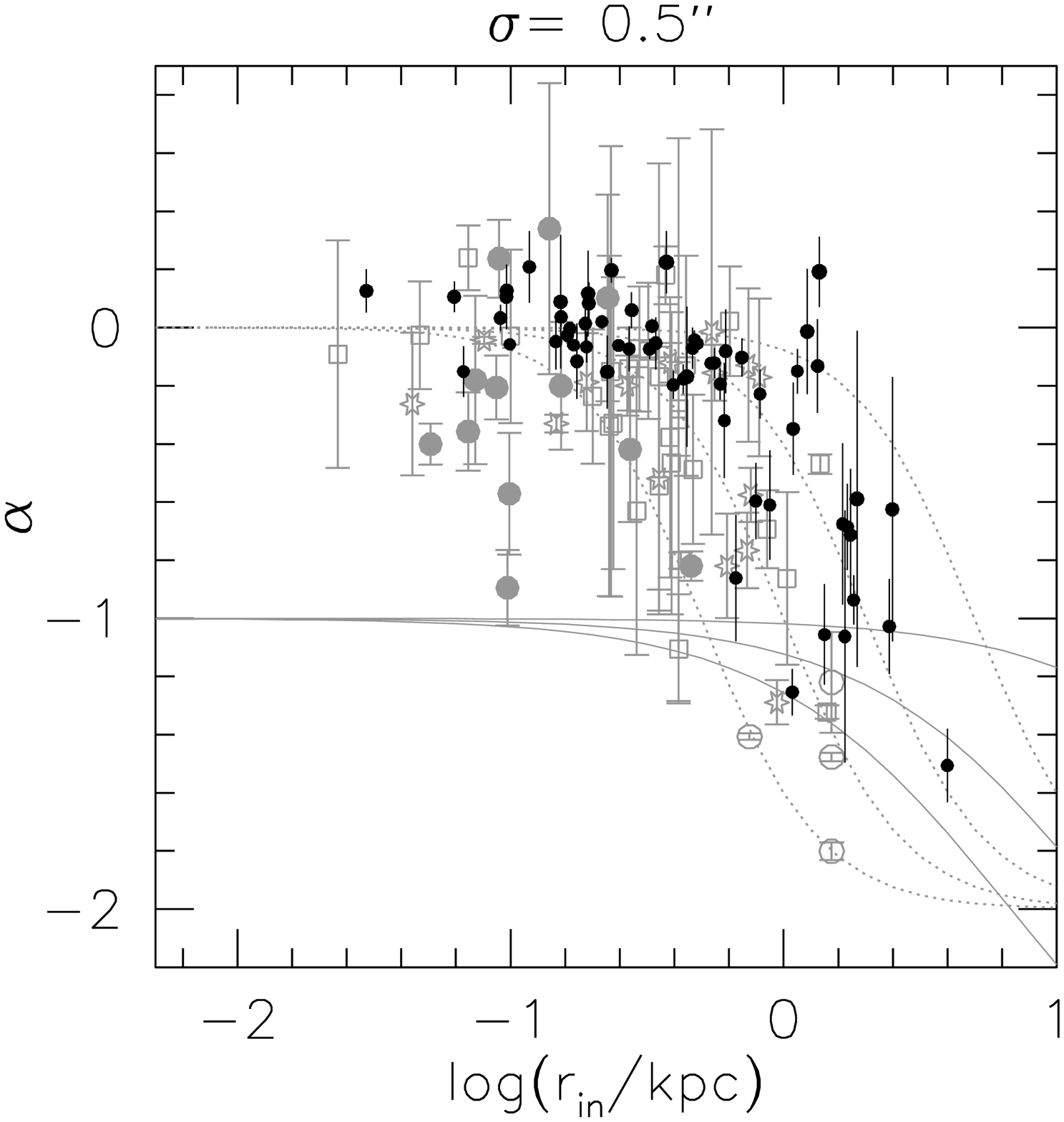}
\epsfxsize=0.48\hsize 
\epsfbox{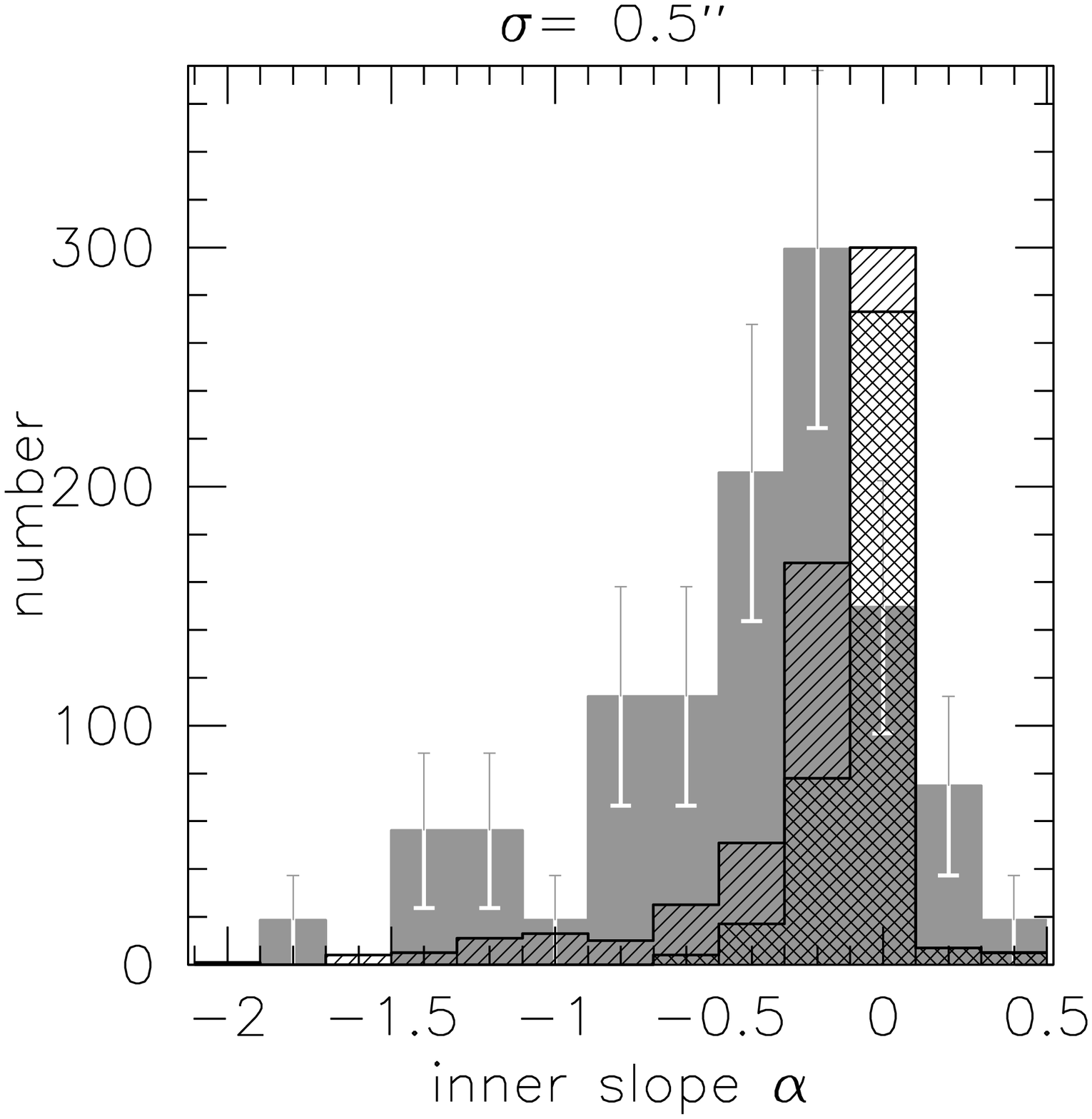}
\caption[gaussBURhisto.ps]{As Figs.~\ref{gaussNFWdata} and 
\ref{gaussNFWhisto}, but now showing the results for Burkert halos. 
The bin at $\alpha=0.5$ also contains galaxies with $\alpha > 0.5$.
\label{gaussBURhisto}}
\end{center} 
\end{figure*}

As we established that the systematic effects for core-dominated
models can only be small, we concentrate here on the simulation
results in the absence of these effects, except for the small $\sigma
= 0.5''$ scatter in the position of the dynamical centre (see
Sect.~\ref{sec:iso2}).  Results for the Burkert halos are expected to
be similar to those of the ISO model, as the Burkert model was
designed to be a compromise between a core model in the inner parts
and a CDM $\rho \sim R^{-3}$ model in the outer parts. As is clear
from Fig.~\ref{gaussBURhisto}, the only real difference is that the
Burkert models show a slightly larger number of steep slopes than the
ISO model, due to the steeper drop-off in the outer parts.

\begin{figure*} 
\begin{center}
\epsfxsize=0.48\hsize 
\epsfbox{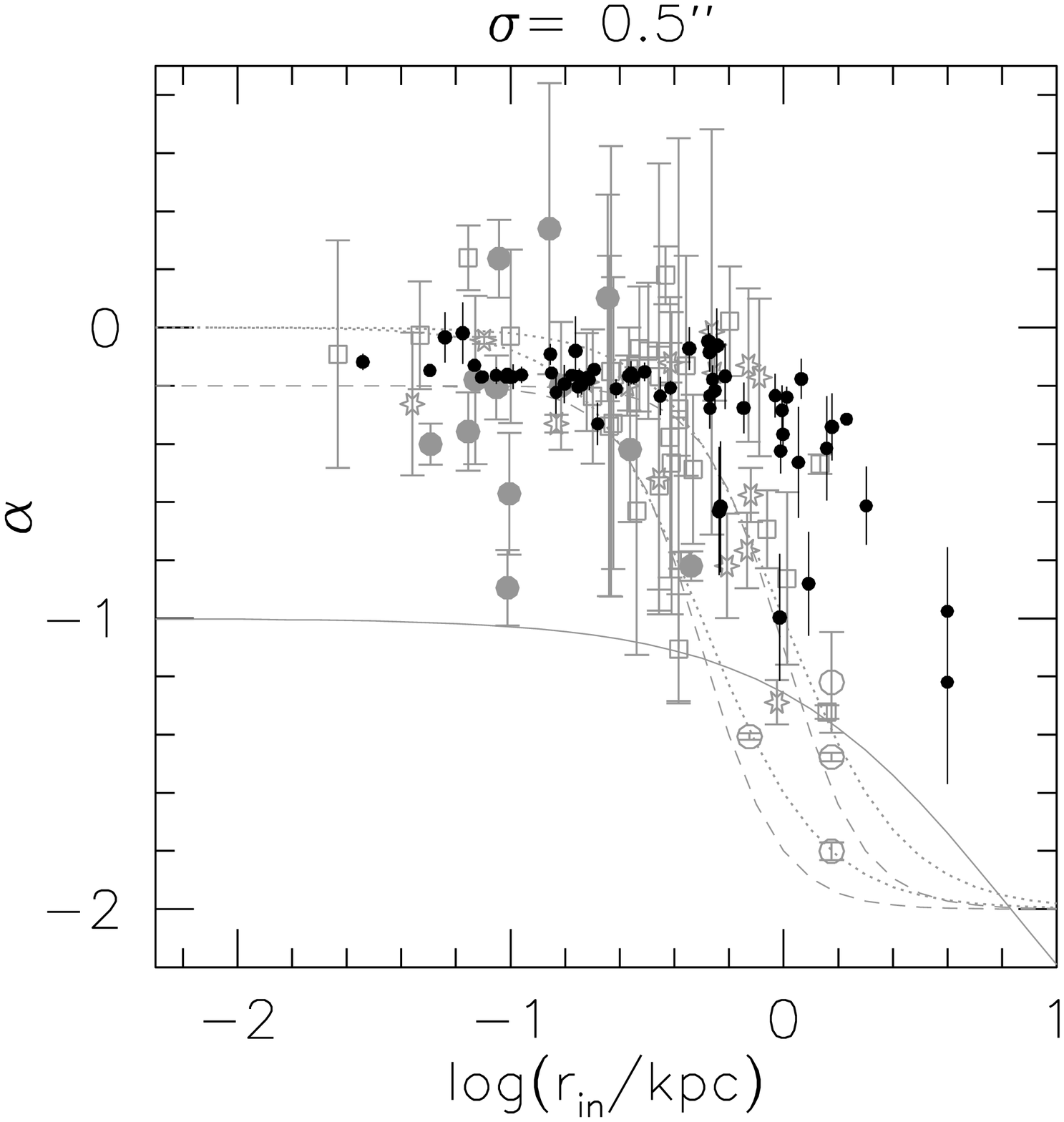}
\epsfxsize=0.48\hsize 
\epsfbox{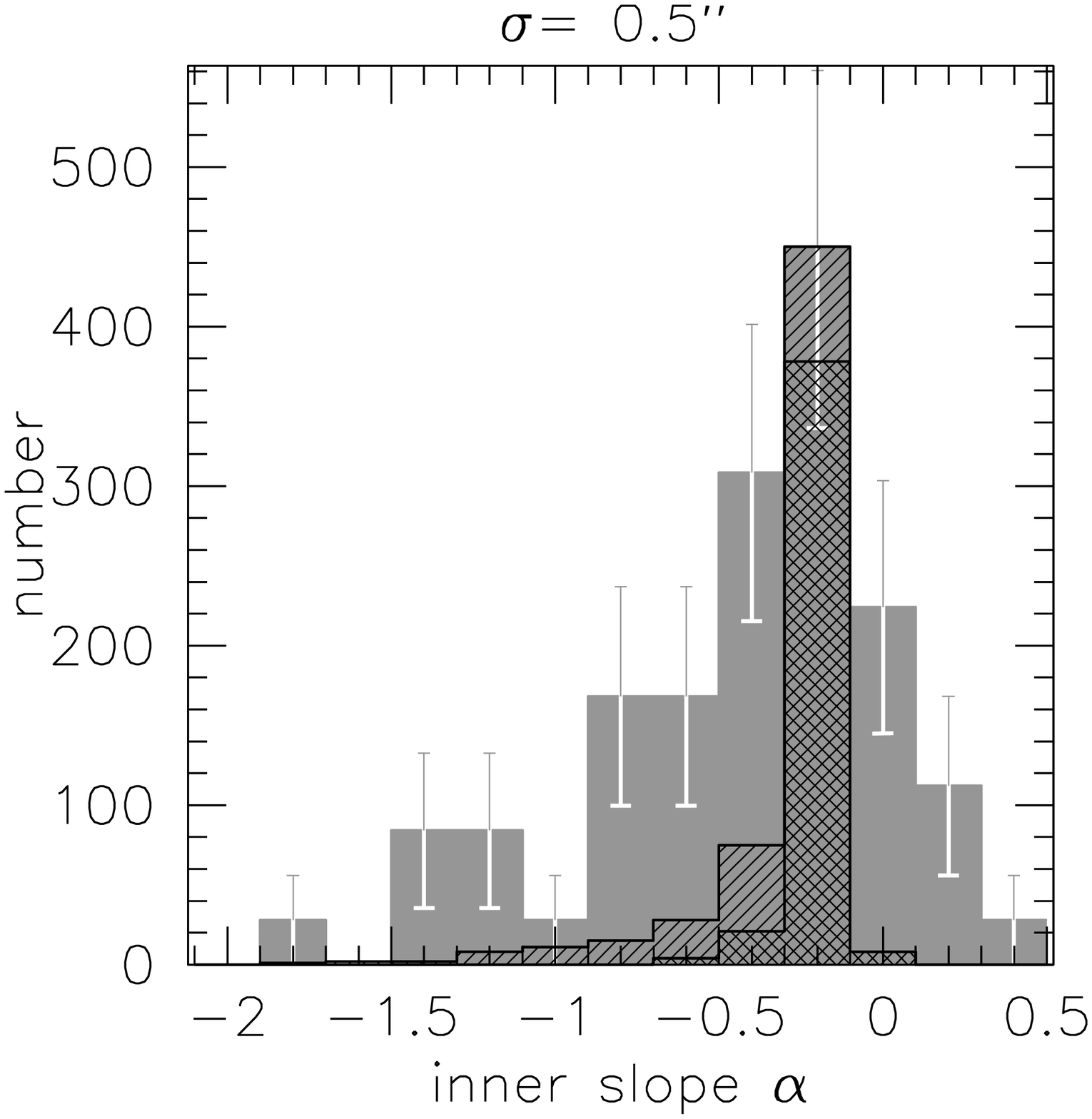}
\caption[gaussKRAhisto.ps]{As Figs.~\ref{gaussNFWdata} and 
\ref{gaussNFWhisto}, but now showing the results for \alfa\ halos. The bin at $\alpha=0.5$ also contains galaxies with $\alpha > 0.5$.
\label{gaussKRAhisto}}
\end{center} 
\end{figure*}

Fig.~\ref{gaussKRAhisto} shows the results for the \alfa\ halos.  The
peak of the histogram distribution now matches well (but was
constructed to do so). The $r_{\rm in} - \alpha$ distributions show slopes
that are slightly too shallow for large values of $r_{\rm in}$.  This is
introduced by the rotation curve scaling we used and the points can be
shifted horizontally by using a different scaling. However, as we are
only interested in the global trends, and are not trying to model
individual galaxies in detail, we retain our simple scaling
relations. The scaling does not affect the inner slopes.

\begin{figure*} 
\begin{center}
\epsfxsize=0.45\hsize 
\epsfbox{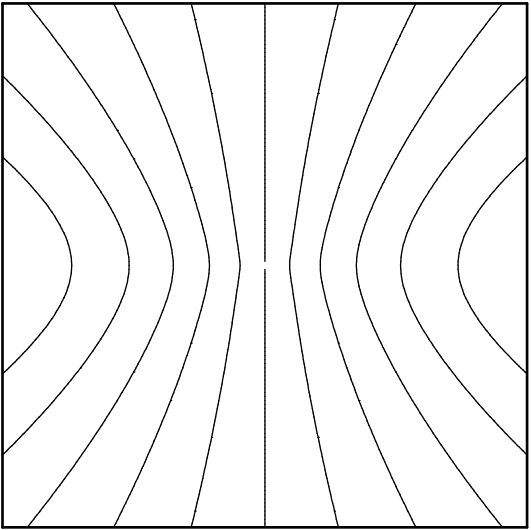}
\caption[velfiKRA.ps]{Velocity fields of the inner parts of massless 
disks embedded in a Kravtsov halo. The velocity field is seen under an
inclination angle of 60\degr, and a position angle of 90\degr. The
boxes measure $5 \times 5$ kpc. The vertical minor axis contour is 0
\kms, increasing in steps of 10 \kms outwards. The halo parameters are
$r_0 = 3.64$ kpc and $V_{\rm max} = 138$ \kms. 
\label{velfikra}}
\end{center} 
\end{figure*}

We show the velocity field of a \alfa\ halo in Fig.~\ref{velfikra}. The halo
parameters were chosen to give a rotation curve similar to the one
used for the ISO velocity field in Fig.~\ref{velfi}. The shape of the
iso-velocity contours is very similar to the ISO velocity field, and
do not show the ``pinching'' that the NFW iso-velocity contours show.

\subsection{\alfa\ halos with scatter}

The \alfa\ model fits the peak in the $\alpha$ histogram best, but was
of course constructed to do so.  One feature of the data that none of
the three core-dominated models can reproduce are the wings towards
steeper slopes found in the data histogram. It is easy to produce
shallower slopes with systematic effects, but very few effects
actually will produce steeper slopes.  Furthermore, the ``monolithic''
nature of the models (for each model all halos have identical slopes)
creates pronounced peaks in the histograms.  In real galaxies we might
expect the contributions of e.g.\ the disk to differ from galaxy to
galaxy, introducing a scatter in the values for the inner total
mass-density slope. We test here the hypothesis that there is a small
scatter in the intrinsic value of the inner slopes which is
responsible for the observed wings in the data histograms.

We use the \alfa\ model and add a small Gaussian scatter to the
slope. We choose an intrinsic dispersion in the mass-density slope
$\sigma_{\alpha} = 0.2$. This choice has no physical motivation, but
is simply suggested by the width of the peak in the data
histogram. Other more fine-tuned choices may very well be possible,
our aim here is simply to describe the data in as few assumptions as
possible with a simple one-component model.  Adding a Gaussian scatter
$\sigma_{\alpha}$ to the $\alpha$-parameter in Eq.~\ref{eq:kravrho},
implies modifying the $g$-parameter in Eq.~\ref{eq:kravrot} with a
Gaussian contribution with dispersion $\sigma_g = \sigma_{\alpha}/2$.

\begin{figure*} 
\begin{center}
\epsfxsize=0.48\hsize 
\epsfbox{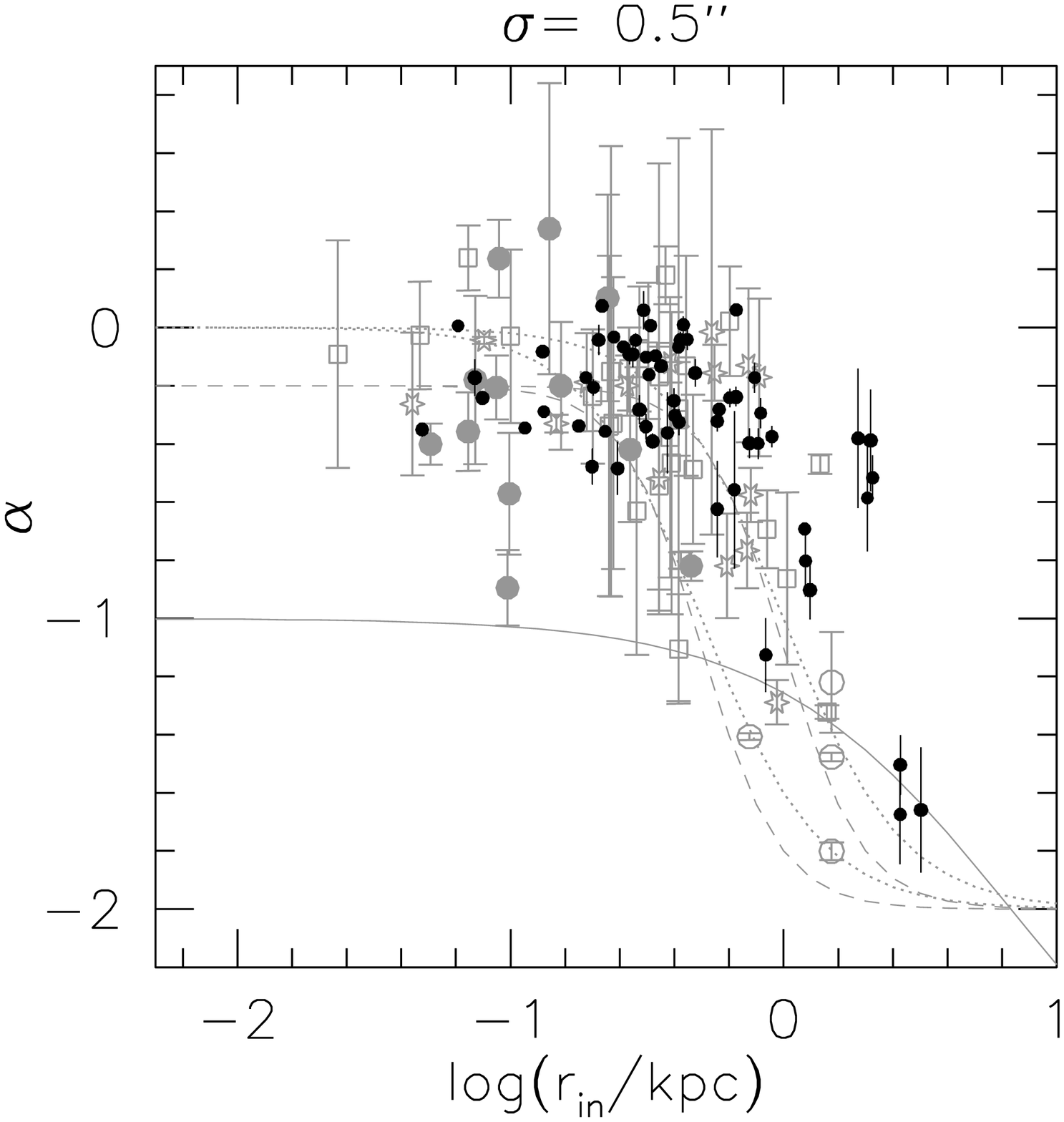}
\epsfxsize=0.48\hsize 
\epsfbox{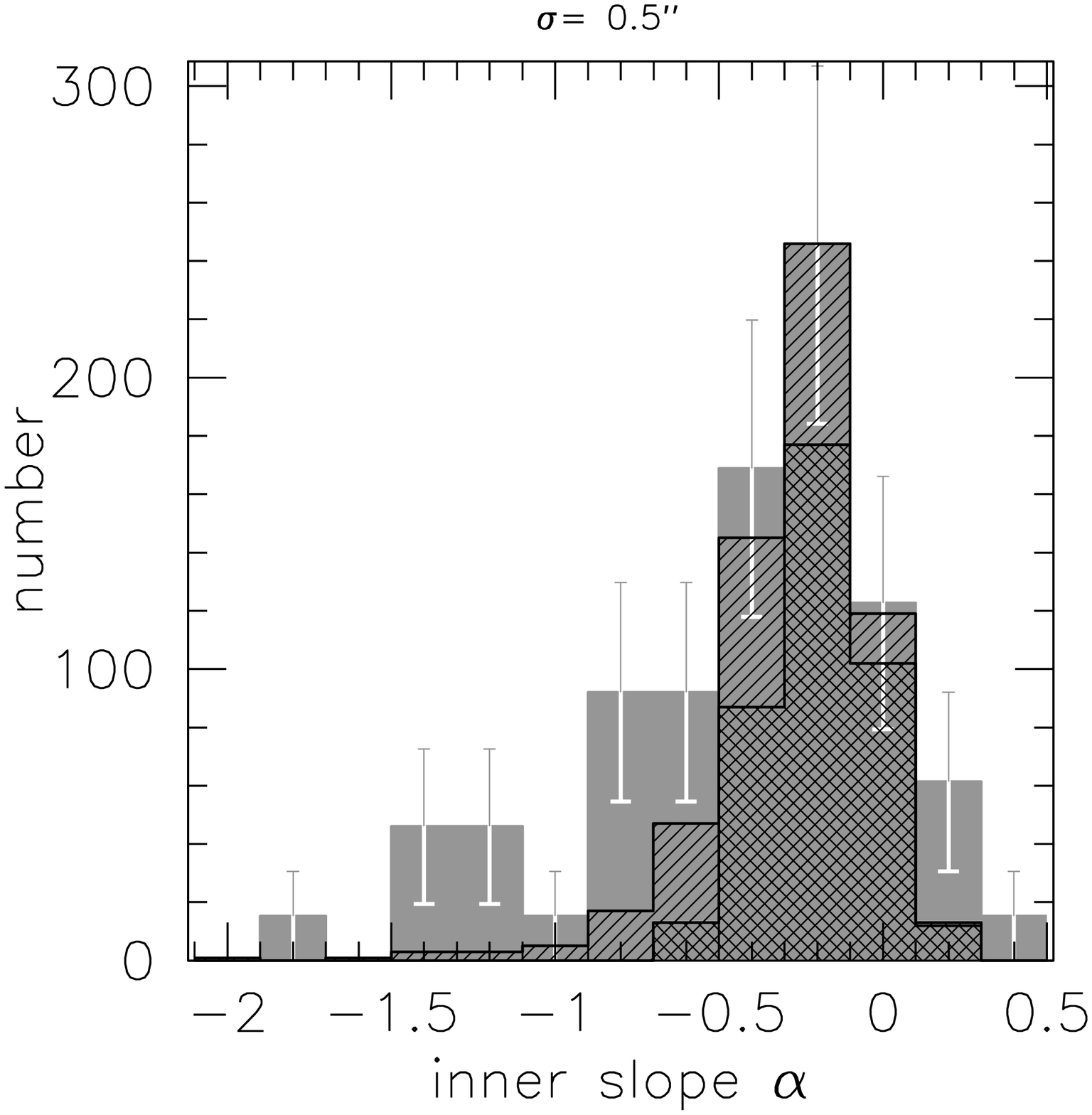}
\caption[gaussSCAhisto.ps]{As Figs.~\ref{gaussNFWdata} and 
\ref{gaussNFWhisto}, but now showing the results for  \alfa\ halos 
with a $\sigma_{\alpha} = 0.2$ Gaussian scatter added to the value for
the inner slope. The bin at $\alpha=0.5$ also contains galaxies with
$\alpha > 0.5$.
\label{gaussSCAhisto}}
\end{center} 
\end{figure*}

The results for the \alfa\ simulations with scatter
$\sigma_{\alpha}=0.2$ are shown in Fig.~\ref{gaussSCAhisto}.  The peak
again matches (naturally). The high- and low-$\alpha$ wings of the
histogram fit the data better than previous models.  The value for the
scatter can probably be fine-tuned, and the distribution of the
scatter is most likely not Gaussian (our assumptions in a small number
of cases lead to hollow halos).  The observed distributions of slopes
is thus consistent with halos with inner slopes $\alpha = -0.2 \pm 0.2
(1\sigma)$ and no systematic effects.

\section{Conclusions}

We have investigated the effect of systematic effects such as
non-circular motions, asymmetries and lopsidedness on the values of
the inner slopes of various halo models.  We find that no realistic
combination of a single systematic effect and NFW models can explain
the observed distribution of slopes.

Core-dominated models, on the other hand, are consistent with the data
in the absence of large systematic effects.  The best match to the
data is obtained by using halos with an inner mass-density slope
$\alpha=-0.2$ and an intrinsic Gaussian scatter in the slope of
$\sigma_{\alpha} \sim 0.2$. As our models assume minimum disk, these
slopes should be interpreted as those of the combined mass-profiles of
stars, gas and dark matter in a galaxy.

It is perhaps possible to construct a combination of multiple
systematic effects to explain the data in the CDM framework.  However,
such a combination of effects would have to completely wipe out the
strong $\alpha = -1$ peak in the $\alpha$ histograms. The resulting
models would be completely dominated by non-cosmological effects,
and not contain any link between properties of dark matter halos and the
visible galaxies that inhabit them.  Yet, they would have to explain and
obey the Tully-Fisher relation, for example, and thus require a
tremendous amount of fine-tuning.

The implied observational signature of CDM halos is strong and if
present should be easily seen.  

We thus conclude that the trends observed in the $r_{\rm in} - \alpha$
plot are mostly resolution effects, in combination with intrinsic
scatter in the inner slopes.  The cusp problem is certainly genuine;
it can not plausibly be attributed to systematic errors in the data.
LSB galaxies have shallow mass-density slopes implying that (i) the
current CDM model, or more precisely the current generation of N-body
models based on the CDM prescription, do not correctly describe
structure at the scale of galaxies, or (ii) non-cosmological effects
destroy the cusp.  The second option reduces the elegance and
predictive power of the CDM model.

\section*{Acknowledgements}
EdB thanks PPARC for Advanced Fellowship support.

\newpage

\clearpage
\newpage 

\clearpage 

\begin{table*}
\begin{minipage}{50mm}
\caption{Restricted sample\label{sample}}
\begin{tabular}{@{}llll}
\hline
Name & Class. & Name & Class. \\
\hline
       F583-1 &U &      UGC 11648&U\\
       F583-4 &U &      UGC 11748&U\\
    ESO-LV 1200211 &U&      DDO 64&U\\
    ESO-LV 2060140 &U&     DDO 185&B\\
    ESO-LV 3020120 &B&     DDO 189&U\\
    ESO-LV 4880490 &B&      NGC 1560&U\\
        UGC 731 &B&      NGC 4395&U\\
       UGC 3371&U&             NGC 4455&B\\
       UGC 4325&U&            UGC 11583&U\\
      UGC 11557&D&&\\
\hline
\end{tabular}

'B' denotes barred; 'U' unbarred; 'D' dubious
\end{minipage}
\end{table*}

\end{document}